\documentclass[preprint,3p]{elsarticle}

\usepackage{float}
\usepackage{amssymb}
\usepackage{amsmath}
\usepackage{graphicx}
\usepackage{dcolumn}
\usepackage{bm}
\usepackage{psfrag}
\usepackage{tikz}
\usetikzlibrary{calc}
\usepackage{pgfplots}
\usepackage{enumitem}
\usepackage{subcaption}
\usepackage{hyperref}
\usepackage[normalem]{ulem}
\allowdisplaybreaks
\pgfplotsset{compat=1.18}

\begin{document}

\begin{frontmatter}



\title{A Kinetic Scheme Based On Positivity Preservation For Multi-component Euler Equations}


\author[1]{Shashi Shekhar Roy\corref{cor1}%
\fnref{fn1}}
\author[2]{S. V. Raghurama Rao\fnref{fn2}}
\cortext[cor1]{Corresponding author}
\fntext[fn1]{E-mail addresses: shashi@iisc.ac.in, shashisroy@gmail.com}
\fntext[fn2]{E-mail address: raghu@iisc.ac.in}
\affiliation[1]{organization={Ph.D. Research Scholar, Department of Aerospace Engineering, Indian Institute of Science},
            city={Bangalore},
            country={India}}
						
\affiliation[2]{organization={Department of Aerospace Engineering, Indian Institute of Science},
            city={Bangalore},
            country={India}}

\begin{abstract}
A kinetic model with flexible velocities is presented for solving the multi-component Euler equations. The model employs a two-velocity formulation in 1D and a three-velocity formulation in 2D. In 2D, the velocities are aligned with the cell-interface to ensure a locally one-dimensional macroscopic normal flux in a finite volume. The velocity magnitudes are defined to satisfy conditions for preservation of positivity of density of each component as well as of overall pressure for first order accuracy under a CFL-like time-step restriction. Additionally, at a stationary contact discontinuity, the velocity definition is modified to achieve exact capture. The basic scheme is extended to third order accuracy using a Chakravarthy-Osher type flux-limited approach along with Strong Stability Preserving Runge-Kutta (SSPRK) method. Benchmark 1D and 2D test cases, including shock-bubble interaction problems, are solved to demonstrate the efficacy of the solver in accurately capturing the relevant flow features.  
\end{abstract}

%


\end{frontmatter}



\section{Introduction}
  Kinetic theory provides a powerful framework for solving the hyperbolic Euler equations numerically. Instead of directly discretizing the macroscopic equations, kinetic schemes discretize the Boltzmann equation and recover the schemes for the macroscopic equations through a moment closure procedure. The key advantages of this approach include simplified upwinding and independence from the underlying eigen-structure of the macroscopic system. Some of the popular early generation schemes are described in \cite{chu1965kinetic,sanders1974possible,pullin1980direct,reitz1981one,deshpande1986second,deshpande1986kinetic,mandal1994kinetic,kaniel1988kinetic,perthame1990boltzmann,prendergast1993numerical,raghurama1995peculiar}. Of particular interest to us are kinetic schemes incorporating flexible velocities, as implemented in \cite{ROY2023106016,roy2024kineticschemebasedpositivity}, due to their accuracy and robustness. However, most existing kinetic schemes have been developed primarily for single-component gas flows, which is a significant limitation given that many real-world applications involve multi-component mixtures.   

Extending numerical schemes developed for the single-component Euler equations to their multi-component counterparts presents unique challenges. As an example, a simple extension of Roe\textquotesingle s scheme (with Roe-averages) to the multi-component Euler equations is not possible without manipulating the Roe-average matrix (see \cite{10.1007/978-3-663-13975-1_1},\cite{Fernandez1989}). More generally, as highlighted by Abgrall \& Karni \cite{ABGRALL2001594}, numerical schemes for multi-component flows must address two major issues. The first challenge is ensuring the positivity of density of each component. Preserving this property enhances the robustness of a numerical scheme. The preservation of overall pressure positivity is an additional advantage, which adds to the robustness of a numerical scheme in handling strong gradients. The second challenge is the suppression of spurious pressure oscillations at material interfaces (contact discontinuities). This issue arises for conservative numerical schemes due to the numerical smearing of material interfaces and the rapid variation in the equation of state caused by a change in gas composition at such discontinuities. Several strategies have been proposed to mitigate these challenges. Larrouturou \cite{LARROUTUROU199159} analyzed the mass fraction positivity issue and modified the discretization of the species conservation equation(s) to ensure the positivity of mass fraction of the species. However, this approach does not prevent pressure oscillations. Karni \cite{KARNI199431} proposed a non-conservative approach, solving the equations in primitive form along with a viscous correction to reduce conservation errors. While effective in reducing oscillations, this method is neither effective over long times nor in handling strong shocks \cite{LeFloch_1994}.  More recently, Gouasmi et al. \cite{GOUASMI2020112912} developed an entropy stable scheme for the multi-component Euler equations, ensuring entropy consistency and robustness even when partial densities vanish; however, like other conservative schemes, it exhibits pressure anomalies at moving material interfaces. These works highlight the inherent difficulties in solving the multi-component Euler equations numerically, motivating further research, including our kinetic scheme based on positivity preservation.  

In this paper, we present a kinetic model with flexible velocities in a vector-kinetic framework for solving the multi-component Euler equations. Our approach employs a two-velocity model ($\lambda$ and -$\lambda$) for each component in 1D.  In 2D, we define a three-velocity model that leads to a locally one-dimensional formulation for the resulting macroscopic normal flux at a cell-interface in a finite volume. The flexible velocities are defined to satisfy positivity preservation conditions for the first order explicit scheme, under a CFL-like limiting condition on the time-step. Furthermore, at a steady contact discontinuity, the velocity definition is adapted to ensure its exact resolution. The basic numerical scheme is extended upto third order accuracy using a Chakravarthy-Osher type flux-limited approach and a higher order Runge-Kutta method. The proposed numerical scheme neither requires the computation of Roe-averages nor is it strongly dependent on the underlying eigen-structure.  Through a series of standard multi-component compressible flow test cases in 1D and 2D, we demonstrate that our scheme is robust, entropic, and capable of accurately capturing key flow features, including those in shock-bubble interaction problems.   

\section{Governing Equations}
The Boltzmann equation describing the evolution of the velocity distribution function $f(t,\textbf{x}, \textbf{v})$ for a single gas is given by 
\begin{equation}
\frac{\partial f}{\partial t}+ \textbf{v} \cdot\frac{\partial f}{\partial \textsl{\textbf{x}}}= Q(f)
\label{eq:ge_1}
\end{equation}
Here, the collision term $Q(f)$ represents the rate of change of $f$ due to collisions. During collisions, the net mass (or number), momentum, and energy of the particles remain conserved. The collision term drives $f$ towards equilibrium and vanishes in the limit. The equilibrium distribution function given by gas-kinetic theory is the Maxwell-Boltzmann equilibrium distribution, defined by 
\begin{equation}
f^{eq}_{Maxwell}= \frac{\rho}{I_{0}} \left(\frac{\beta}{\pi}\right)^{N/2} exp\left(-\beta|\textbf{v}-\textbf{u}|^{2}\right) exp\left(-I/I_{0}\right) 
\label{eq:ge_2}
\end{equation}
where $\beta$= $\frac{1}{2RT}$, $\textbf{u}(t,\textbf{x})$ is the macroscopic velocity, $N$ is the translational degrees of freedom, $I$ is the internal energy variable corresponding to non-translational degrees of freedom, $I_{0}~=~\frac{2- N(\gamma -1)}{2(\gamma -1)}RT$ and $\gamma$= $\frac{c_{p}}{c_{v}}$. A simplification to the collision term was given by Bhatnagar, Gross and Krook (BGK) \cite{bhatnagar1954model}. The BGK model approximates the collision term as 
\begin{equation}
Q(f)= -\frac{1}{\epsilon}\left[f- f^{eq}\right]
\label{eq:ge_3}
\end{equation}
Here $\epsilon$ is the relaxation time. By taking moments of the Boltzmann equation, {\em i.e.}, by integrating the product of the Boltzmann equation and the moment function vector $\bm{\Psi}~=~\left[1, v_{1}, ..,v_{N}, I+ \frac{|\textbf{v}|^{2}}{2}\right]^{T}$ over $\textbf{v}$ and $I$, we derive the macroscopic conservation laws of mass, momentum and energy. Furthermore, in the limit of $\epsilon~\rightarrow$~0, $f$ relaxes instantaneously to $f^{eq}$. In this limit, the moments of the Boltzmann equation yield the inviscid Euler equations, as follows.  
\begin{equation}
\int_{\mathbb{R}^{N}}d\textbf{v} \int_{\mathbb{R}^{+}}dI \ \bm{\Psi}\left(\frac{\partial f}{\partial t}+ \frac{\partial (v_{i}f)}{\partial x_{i}}= 0,f= f^{eq}\right) \Rightarrow \frac{\partial \textbf{U}}{\partial t}+ \frac{\partial \textbf{G}_{i}}{\partial x_{i}}= 0
\label{eq:ge_4}
\end{equation}
Here, $\textbf{U}$ is the vector of conserved variables and $\textbf{G}_{i}$ is the inviscid flux vector in the direction $i$ for the Euler equations. They are given by 
\begin{equation}
\textbf{U}= \begin{bmatrix} \rho \\ \rho u_{j} \\ \rho E \end{bmatrix}, \ \textbf{G}_{i}= \begin{bmatrix} \rho u_{i} \\ \rho u_{i}u_{j}+ p \delta_{ij} \\ (\rho E+ p)u_{i} \end{bmatrix}
\label{eq:ge_5}
\end{equation}
and the total energy $E= e + \frac{|\textbf{u}|^{2}}{2}$. The ideal gas satisfies the equations e= $c_{v}$T (caloric equation of state) and p=$\rho$RT (thermal equation of state), with R being the gas constant. The specific heat capacities satisfy the relations, $ c_{p}-c_{v}$= R, and $\frac{c_{p}}{c_{v}}= \gamma$. The ratio of specific heats, $\gamma$, is constant for an ideal gas.

The kinetic theory for a gas mixture comprises a Boltzmann equation for each species/component. Component gases are assumed to be chemically inert. Thus, during collisions, the mass of each gas component as well as total momentum and energy of the gas mixture remain conserved.  Additionally, we assume local thermal equilibrium, {\em i.e.}, a single temperature is defined for all the components present at any given point in space and time. In the present work, we consider the number of components $N_{c}$ to be $2$.  The extension to a general $N_{c}$ number of components is then straightforward. The 1D Boltzmann-BGK equations for a two-component gas mixture are 
\begin{subequations}
\begin{equation}
\frac{\partial f_{1}}{\partial t}+ v \frac{\partial f_{1}}{\partial x}= -\frac{1}{\epsilon}\left(f_{1}- f_{1}^{eq}\right)
\label{eq:ge_6a}
\end{equation}
\begin{equation}
\frac{\partial f_{2}}{\partial t}+ v \frac{\partial f_{2}}{\partial x}= -\frac{1}{\epsilon}\left(f_{2}- f_{2}^{eq}\right)
\label{eq:ge_6b}
\end{equation}
\end{subequations}
Here, $f_{1}^{eq}$ and $f_{2}^{eq}$ are the Maxwellian equilibrium distributions for components $1$ and $2$ respectively. Given the vectors $\bm{\Psi}_{1}= \left[1, 0, v, I+ \frac{|\textbf{v}|^{2}}{2}\right]^{T}$ and $\bm{\Psi}_{2}= \left[0, 1, v, I+ \frac{|\textbf{v}|^{2}}{2}\right]^{T}$, the moment of $\bm{\Psi}_{1}$*\eqref{eq:ge_6a}+ $\bm{\Psi}_{2}$*\eqref{eq:ge_6b} give us the macroscopic conservation laws. Under the assumption of instantaneous relaxation ($\epsilon \rightarrow$0), the moments give us the the two-component 1D Euler equations, as follows.  
\begin{equation}
\left\langle \bm{\Psi}_{1}\left(\frac{\partial f_{1}}{\partial t}+ \frac{\partial (vf_{1})}{\partial x}= 0,f_{1}= f_{1}^{eq}\right)+ \bm{\Psi}_{2}\left(\frac{\partial f_{2}}{\partial t}+ \frac{\partial (vf_{2})}{\partial x}= 0,f_{2}= f_{2}^{eq}\right) \right\rangle \Rightarrow \frac{\partial \textbf{U}}{\partial t}+ \frac{\partial \textbf{G}}{\partial x}= 0
\label{eq:ge_7}
\end{equation}
Here $\left\langle \right\rangle$ refers to taking moments. The conserved variable vector $\textbf{U}$ and the inviscid flux vector $\textbf{G}$ are given by 
\begin{equation}
\textbf{U}= \begin{bmatrix} \rho_{1} \\ \rho_{2} \\ \rho u \\ \rho E \end{bmatrix}, \textbf{G}= \begin{bmatrix} \rho_{1} u \\\rho_{2} u\\ \rho u^{2}+ p \\ (\rho E+ p)u \end{bmatrix}, 
\label{eq:ge_8}
\end{equation}
Here, the ideal gas component $c$ satisfies the relations, $e_{c}=\left(c_{v}\right)_{c}T$ and $p_{c}=\rho_{c}R_{c}T$. The overall density and pressure are given by 
\begin{equation}
\rho = \sum^{2}_{c=1}\rho_{c}, \ p = \sum_{c}p_{c} \text{(Dalton's law of partial pressures)}= \rho R T
\label{eq:ge_9}
\end{equation}
with 
\begin{equation}
R= \frac{\sum_{c}\rho_{c}R_{c}}{\sum_{c}\rho_{c}}\text{. Also, }c_{p}= \frac{\sum_{c}\rho_{c}\left(c_{p}\right)_{c}}{\sum_{c}\rho_{c}}, \ c_{v}= \frac{\sum_{c}\rho_{c}\left(c_{v}\right)_{c}}{\sum_{c}\rho_{c}}, \ \gamma = \frac{c_{p}}{c_{v}}
\label{eq:ge_10}
\end{equation}
Finally,
\begin{equation}
\rho E = \sum_{c} \rho_{c} E_{c}= \sum_{c} \rho_{c} e_{c}+ \sum_{c} \rho_{c} \frac{u^{2}}{2}= \rho c_{v} T+ \frac{\rho u^{2}}{2}= \frac{p}{(\gamma -1)}+ \frac{\rho u^{2}}{2}
\label{eq:ge_11}
\end{equation}
It has to be mentioned here that the total energy of a component $c$ is actually given by $\rho_{c}\varepsilon_{c}$= $\rho_{c}E_{c}$+ $\rho_{c}h^{0}_{c}$, where $h^{0}_{c}$ is the heat of formation of the component $c$. However, this addition does not change the system of equations. Hence, the heat of formations, $h^{0}_{c}$ can be set to zero without changing the system (see \cite{10.1007/BFb0083869}). Next, the mass conservation equation for one of the components, say component $2$, can be replaced by mass conservation equation for the overall mixture by instead defining the moment vector $\bm{\Psi}_{1}$ to be $\left[1, 1, v, I+ \frac{|\textbf{v}|^{2}}{2}\right]^{T}$. The Euler equations then become 
\begin{equation}
 \frac{\partial \textbf{U}}{\partial t}+ \frac{\partial \textbf{G}}{\partial x}= 0, \ \textbf{U}= \begin{bmatrix} \rho_{1} \\ \rho \\ \rho u \\ \rho E \end{bmatrix}, \textbf{G}= \begin{bmatrix} \rho_{1} u \\\rho u\\ \rho u^{2}+ p \\ (\rho E+ p)u \end{bmatrix},
\label{eq:ge_12}
\end{equation}
Thus the two-component Euler equations can be written as Euler equations for the overall mixture and an additional mass conservation law for one of the components. In general, the $N_{c}$ component Euler equations have $N_{c}-1$ additional species mass conservation equations. These species mass conservation equations can also be written in terms of the corresponding mass fraction, given by $W_{c}= \rho_{c}/\sum_{c} \rho_{c}$. Thus, Equation \eqref{eq:ge_12} can be rewritten as 
\begin{equation}
 \frac{\partial \textbf{U}}{\partial t}+ \frac{\partial \textbf{G}}{\partial x}= 0, \ \textbf{U}= \begin{bmatrix} \rho W \\ \rho \\ \rho u \\ \rho E \end{bmatrix}, \textbf{G}= \begin{bmatrix} \rho W u \\\rho u\\ \rho u^{2}+ p \\ (\rho E+ p)u \end{bmatrix}, W= \frac{\rho_{1}}{\rho}
\label{eq:ge_13}
\end{equation}
In terms of mass fraction,
\begin{equation}
R= \sum_{c}W_{c}R_{c}, \ c_{p}= \sum_{c}W_{c}\left(c_{p}\right)_{c}, \ c_{v}= \sum_{c}W_{c}\left(c_{v}\right)_{c}
\label{eq:ge_14}
\end{equation}
From the mass conservation equations in Equation \eqref{eq:ge_13}, we get,
\begin{equation}
\frac{\partial W}{\partial t}+ u \frac{\partial W}{\partial x}= 0
\label{eq:}
\end{equation}
Thus, a discontinuity in gas composition is a material interface, {\em i.e.}, it propagates with velocity $u$, as a contact discontinuity, with pressure and velocity remaining constant across the discontinuity. On the other hand, across shocks and expansion fans, the composition of the mixture does not change.

\section{Kinetic model for 1D Euler equations}
In the present work, we are modeling only in the velocity $v$ space. Therefore, we introduce the following truncated equilibrium distributions for a component $c$ by integrating w.r.t. the internal energy variable $I$.  
\begin{equation}
\breve{f}_{c}^{eq} = \int^{\infty}_{0} \ f_{c}^{eq} dI, \ \ \hat{f}^{eq}_{c,i} = \Psi_{c,i} \breve{f}_{c}^{eq}
\label{eq:1d_euler_1}
\end{equation} 
Then, the moment relations for the equilibrium distribution function are given by 
\begin{subequations}
\label{eq:1d_euler_2}
\begin{equation}
\int_{-\infty}^{\infty} dv \ \sum_{c}\Psi_{c,i} \breve{f}_{c}^{eq} = \int_{-\infty}^{\infty} dv \ \sum_{c}\hat{f}^{eq}_{c,i}= \left\langle  \sum_{c}\hat{f}^{eq}_{c,i} \right\rangle = U_{i}
\end{equation}
\begin{equation}
\int_{-\infty}^{\infty} dv \sum_{c}v\Psi_{c,i} \breve{f}_{c}^{eq} = \int_{-\infty}^{\infty} dv \sum_{c}v\hat{f}^{eq}_{c,i} = \left\langle \sum_{c}v\hat{f}^{eq}_{c,i} \right\rangle = G_{i}
\end{equation}
\end{subequations}
Here, $U_{i}$ and $G_{i}$ are the $i^{th}$ terms of the conserved variable vector $\textbf{U}$ and flux vector $\textbf{G}$ respectively, as defined in \eqref{eq:ge_13}. In our 1-D approach, we replace $\hat{f}^{eq}_{c,i}$ corresponding to Maxwellian distribution function for each component $c$ with two Dirac-delta distributions ($\delta$) (as in \cite{shrinath2023kinetic}). One dirac-delta function is placed at a non-negative velocity $\lambda$, and the other at a non-positive velocity $-\lambda$. The velocity $\lambda$ is flexible, and is fixed later based on positivity considerations. The equilibrium distribution $\hat{f}^{eq}_{c,i}$ for our kinetic model can, thus, be written as 
\begin{equation}
\hat{f}^{eq}_{c,i}= f^{eq}_{1,c,i}\delta(v- \lambda) + f^{eq}_{2,c,i}\delta(v+ \lambda), \ \lambda \geq 0
\label{eq:1d_euler_3}
\end{equation}  
This leads to a vector kinetic framework (similar to that introduced by Natalini \cite{natalini1998discrete} and Aregba-Driollet and Natalini \cite{aregba2000discrete}). In the resulting Flexible Velocity Boltzmann Equations, $f_{1,c,i}$ and $f_{2,c,i}$, are being advected by velocities $\lambda$ and $-\lambda$ respectively. This vector Boltzmann equation for a component $c$, which corresponds to the $i^{th}$ macroscopic equation, is given by
\begin{equation}
\frac{\partial \textbf{f}_{c,i}}{\partial t}+ \frac{\partial (\Lambda\textbf{f}_{c,i})}{\partial x}= 
- \frac{1}{\epsilon} \left[\textbf{f}_{c,i}- \textbf{f}_{c,i}^{eq}\right] 
\label{eq:1d_euler_4}
\end{equation}
Here 
\begin{equation}
\textbf{f}_{c,i}^{eq}= \begin{bmatrix} f^{eq}_{1,c,i} \\ f^{eq}_{2,c,i} \end{bmatrix} \ \textrm{and} \ \Lambda= \begin{bmatrix} \lambda & 0\\ 0 & -\lambda \end{bmatrix} 
\label{eq:1d_euler_5}
\end{equation}
This vector kinetic framework simplifies moment relations compared to the classical framework, replacing complex integrals and the Gaussian distribution based equilibrium functions with straightforward summations. These moment relations for \eqref{eq:1d_euler_4} are given by 
\begin{subequations}
\label{eq:1d_euler_6}
\begin{equation}
\textbf{P}_{i}\sum_{c}\textbf{f}_{c,i}^{eq}= \sum_{c}f^{eq}_{1,c,i}+ \sum_{c}f^{eq}_{2,c,i}= U_{i}
\end{equation}
\begin{equation}
\textbf{P}_{i}\sum_{c}\Lambda\textbf{f}_{c,i}^{eq}= \lambda\sum_{c}f^{eq}_{1,c,i}- \lambda\sum_{c}f^{eq}_{2,c,i}= G_{i}
\end{equation}
\end{subequations}
Here, the row vector $\textbf{P}_{i}= \begin{bmatrix}1 & 1\end{bmatrix}$. From the moment relations in \eqref{eq:1d_euler_6}, we get 
\begin{equation}
\sum_{c}f^{eq}_{1,c,i}= \frac{U_{i}}{2}+ \frac{G_{i}}{2\lambda}, \ \ \sum_{c}f^{eq}_{2,c,i}= \frac{U_{i}}{2}- \frac{G_{i}}{2\lambda}
\label{eq:1d_euler_7}
\end{equation}
Using a finite volume approach, we numerically solve the Boltzmann equations \eqref{eq:1d_euler_4} in their conservation form for each cell $j$. In 1D, we assume a constant cell width, $\Delta x$. The operator-splitting technique is applied, where at the end of the $n^{th}$ time step, the distribution function undergoes instantaneous relaxation to the corresponding equilibrium state. Following this, the advective components of the Boltzmann equations are discretized and solved numerically to update the distribution function for the next time step, as follows.

\begin{subequations}
\label{eq:1d_euler_8}
\[\text{Relaxation step: Instantaneous, {\em i.e.},} \epsilon\rightarrow 0\text{. Thus,} \]
\begin{equation}
(\textbf{f}_{c,i})^{n}_{j}=  (\textbf{f}_{c,i}^{eq})^{n}_{j}
\end{equation}
\[\text{Advection step: The advective parts of Boltzmann equations are given by,}\]
\begin{equation}
\frac{\partial (\textbf{f}_{c,i})_{j}}{\partial t}+ \frac{\partial (\textbf{h}_{c,i})_{j}}{\partial x}= 0; \textbf{h}_{c,i}= \Lambda\textbf{f}_{c,i}^{eq}\text{. In integral form,}\nonumber
\end{equation}
\begin{equation}
\frac{d (\textbf{f}_{c,i})_{j}}{dt} = -\frac{1}{\Delta x}\left[ (\textbf{h}_{c,i})_{j+1/2}^{n}- (\textbf{h}_{c,i})_{j-1/2}^{n}\right]
\end{equation}
\end{subequations}
In the present work, the interface kinetic flux $(\textbf{h}_{c,i})_{j+\frac{1}{2}}$ is defined using a flux difference splitting approach. The temporal derivative is approximated using forward Euler method for the first order scheme. The discretized equations thus become 
\begin{subequations}
\begin{equation}
(\textbf{f}_{c,i})_{j}^{n+1} = (\textbf{f}_{c,i}^{eq})_{j}^{n} -\frac{\Delta t}{\Delta x}\left[ (\textbf{h}_{c,i})_{j+\frac{1}{2}}^{n}- (\textbf{h}_{c,i})_{j-\frac{1}{2}}^{n}\right]
\label{eq:1d_euler_9a}
\end{equation}
\begin{equation}
(\textbf{h}_{c,i})_{j+\frac{1}{2}} = \frac{1}{2}\left\{(\textbf{h}_{c,i})_{j}+ (\textbf{h}_{c,i})_{j+1}\right\}-\frac{1}{2}\left\{(\Delta \textbf{h}_{c,i}^{+})_{j+\frac{1}{2}}- (\Delta \textbf{h}_{c,i}^{-})_{j+\frac{1}{2}}\right\}
\label{eq:1d_euler_9b}
\end{equation}
\begin{equation}
(\Delta \textbf{h}_{c,i}^{+})_{j+ \frac{1}{2}}= \left(\Lambda^{+}\Delta \textbf{f}_{c,i}^{eq}\right)_{j+ \frac{1}{2}}= \begin{bmatrix}\left(\lambda\Delta f^{eq}_{1,c,i}\right)_{j+ \frac{1}{2}} \\ 0 \end{bmatrix}= \begin{bmatrix}(\lambda)_{j+\frac{1}{2}}\left\{(f^{eq}_{1,c,i})_{j+1} -(f^{eq}_{1,c,i})_{j}\right\} \\ 0 \end{bmatrix}
\label{eq:1d_euler_9c}
\end{equation}
\begin{equation}
(\Delta \textbf{h}_{c,i}^{-})_{j+ \frac{1}{2}}= \left(\Lambda^{-}\Delta \textbf{f}_{c,i}^{eq}\right)_{j+ \frac{1}{2}}= \begin{bmatrix} 0 \\ \left(-\lambda\Delta f^{eq}_{2,c,i}\right)_{j+ \frac{1}{2}} \end{bmatrix}= \begin{bmatrix} 0 \\ (-\lambda)_{j+\frac{1}{2}}\left\{(f^{eq}_{2,c,i})_{j+1} -(f^{eq}_{2,c,i})_{j}\right\} \end{bmatrix}
\label{eq:1d_euler_9d}
\end{equation}
\begin{equation}
\Lambda^{\pm}= \frac{\Lambda \pm |\Lambda|}{2}
\label{eq:1d_euler_9e}
\end{equation}
\end{subequations}
Summing the discretized equations \eqref{eq:1d_euler_9a} over all $c$ components, we get 
\begin{subequations}
\begin{equation}
\left(\sum_{c}\textbf{f}_{c,i}\right)_{j}^{n+1} = \left(\sum_{c}\textbf{f}_{c,i}^{eq}\right)_{j}^{n} -\frac{\Delta t}{\Delta x}\left[ \left(\sum_{c}\textbf{h}_{c,i}\right)_{j+\frac{1}{2}}^{n}- \left(\sum_{c}\textbf{h}_{c,i}\right)_{j-\frac{1}{2}}^{n}\right]
\label{eq:1d_euler_10a}
\end{equation}
\begin{equation}
\left(\sum_{c}\textbf{h}_{c,i}\right)_{j+\frac{1}{2}} = \frac{1}{2}\left\{\left(\sum_{c}\textbf{h}_{c,i}\right)_{j}+ \left(\sum_{c}\textbf{h}_{c,i}\right)_{j+1}\right\}-\frac{1}{2}\left\{\left(\sum_{c}\Delta \textbf{h}_{c,i}^{+}\right)_{j+\frac{1}{2}}- \left(\sum_{c}\Delta \textbf{h}_{c,i}^{-}\right)_{j+\frac{1}{2}}\right\}
\label{eq:1d_euler_10b}
\end{equation}
\begin{equation}
\left(\sum_{c}\Delta \textbf{h}_{c,i}^{+}\right)_{j+\frac{1}{2}}= \begin{bmatrix}\left(\lambda\sum_{c}\Delta f^{eq}_{1,c,i}\right)_{j+ \frac{1}{2}} \\ 0 \end{bmatrix}= \begin{bmatrix}(\lambda)_{j+\frac{1}{2}}\left\{\left(\sum_{c}f^{eq}_{1,c,i}\right)_{j+1} -\left(\sum_{c}f^{eq}_{1,c,i}\right)_{j}\right\} \\ 0 \end{bmatrix}
\label{eq:1d_euler_10c}
\end{equation}
\begin{equation}
\left(\sum_{c}\Delta \textbf{h}_{c,i}^{-}\right)_{j+\frac{1}{2}}= \begin{bmatrix} 0 \\ \left(-\lambda\sum_{c}\Delta f^{eq}_{2,c,i}\right)_{j+ \frac{1}{2}} \end{bmatrix}= \begin{bmatrix} 0 \\ (-\lambda)_{j+\frac{1}{2}}\left\{\left(\sum_{c}f^{eq}_{2,c,i}\right)_{j+1} -\left(\sum_{c}f^{eq}_{2,c,i}\right)_{j}\right\} \end{bmatrix}
\label{eq:1d_euler_10d}
\end{equation}
\end{subequations}
The macroscopic update formula, obtained by taking moment of equation \eqref{eq:1d_euler_10a}, is given by 
\begin{subequations}
\begin{equation}
(U_{i})_{j}^{n+1} = (U_{i})_{j}^{n} -\frac{\Delta t}{\Delta x}\left[ (G_{i})_{j+\frac{1}{2}}^{n}- (G_{i})_{j-\frac{1}{2}}^{n}\right]
\label{eq:1d_euler_11a}
\end{equation}
\begin{equation}
(G_{i})_{j+\frac{1}{2}}= \textbf{P}_{i} \left(\sum_{c}\textbf{h}_{c,i}\right)_{j+\frac{1}{2}}= \frac{1}{2}\left\{(G_{i})_{j}+ (G_{i})_{j+1}\right\}-\frac{1}{2}\left\{(\Delta G_{i}^{+})_{j+\frac{1}{2}}- (\Delta G_{i}^{-})_{j+\frac{1}{2}}\right\}
\label{eq:1d_euler_11b}
\end{equation}
\begin{equation}
(\Delta G_{i}^{+})_{j+\frac{1}{2}} = \textbf{P}_{i}\left(\sum_{c}\Delta \textbf{h}_{c,i}^{+}\right)_{j+ \frac{1}{2}}= \left(\lambda\sum_{c}\Delta f^{eq}_{1,c,i}\right)_{j+ \frac{1}{2}} = \frac{1}{2} \left\{ (G_{i})_{j+1}- (G_{i})_{j}\right\}+ \frac{\lambda_{j+\frac{1}{2}}}{2} \left\{ (U_{i})_{j+1}- (U_{i})_{j} \right\}
\label{eq:1d_euler_11c}
\end{equation}
\begin{equation}
(\Delta G_{i}^{-})_{j+\frac{1}{2}} = \textbf{P}_{i}\left(\sum_{c}\Delta \textbf{h}_{c,i}^{-}\right)_{j+ \frac{1}{2}}= \left(-\lambda\sum_{c}\Delta f^{eq}_{2,c,i}\right)_{j+ \frac{1}{2}} = \frac{1}{2} \left\{ (G_{i})_{j+1}- (G_{i})_{j}\right\}- \frac{\lambda_{j+\frac{1}{2}}}{2} \left\{ (U_{i})_{j+1}- (U_{i})_{j} \right\}
\label{eq:1d_euler_11d}
\end{equation}
\end{subequations}
The macroscopic flux vector at the interface can then be rewritten as 
\begin{equation}
\textbf{G}_{j+\frac{1}{2}}= \frac{1}{2} \left( \textbf{G}_{j}+ \textbf{G}_{j+1} \right) - \frac{\lambda_{j+\frac{1}{2}}}{2}\left(\textbf{U}_{j+1}-\textbf{U}_{j}\right)
\label{eq:1d_euler_12}
\end{equation}
In the following subsections, we will determine the optimal value of $\lambda_{j+\frac{1}{2}}$ based on considerations of positivity and exact capture of a steady contact discontinuity.

\subsection{Positivity analysis}
We consider a numerical scheme for the multi-component Euler equations to be positivity preserving if it ensures the positivity of: (i) density of each component and (ii) the overall pressure. Thus, given $\rho_{c}(x,t_{0})>$0 $\forall$ $c$ $\in$ [1,$N_{c}$] and $p(x,t_{0})>$0, positivity preservation requires that $\rho_{c}(x,t)>$0 and $p(x,t)>$0 $\forall$ $t>t_{0}$. Let $\textbf{W}$ be the set of all physically admissible conserved variable vectors $\textbf{U}$, {\em i.e.}, states having positive density of each component as well as overall pressure. Then, for positivity preservation,
\begin{equation}
\textbf{U}(x,t_{0})\in \textbf{W} \Rightarrow \textbf{U}(x,t)\in \textbf{W} \ \forall \ t>t_{0}
\label{eq:1d_pos_1}
\end{equation}
To perform the positivity analysis for our first order accurate numerical scheme, we first rewrite the vector form of the macroscopic update formula as follows.
\begin{equation}
\begin{split}
&\textbf{U}^{n+1}_{j}= \textbf{U}^{n}_{j}- \frac{\Delta t}{\Delta x}\left(\textbf{G}^{n}_{j+\frac{1}{2}}- \textbf{G}^{n}_{j-\frac{1}{2}}\right) \\
&= \textbf{U}^{n}_{j}- \frac{\Delta t}{2\Delta x} \left[ \left\{ \left( \lambda\right)^{n}_{j+\frac{1}{2}}\textbf{U}^{n}_{j} + \textbf{G}^{n}_{j} \right\}  + \left\{ -\left( \lambda\right)^{n}_{j+\frac{1}{2}}\textbf{U}^{n}_{j+1} + \textbf{G}^{n}_{j+1} \right\} \right] \\
& + \frac{\Delta t}{2\Delta x} \left[ \left\{ \left( \lambda\right)^{n}_{j-\frac{1}{2}}\textbf{U}^{n}_{j-1} + \textbf{G}^{n}_{j-1} \right\} + \left\{ -\left( \lambda\right)^{n}_{j-\frac{1}{2}}\textbf{U}^{n}_{j} + \textbf{G}^{n}_{j} \right\} \right] \\
&= \frac{\Delta t}{2\Delta x} \underbrace{\left\{ \left(\lambda\right)^{n}_{j+\frac{1}{2}}\textbf{U}^{n}_{j+1} - \textbf{G}^{n}_{j+1} \right\}}_{\text{Term 1}} +
 \frac{\Delta t}{2\Delta x}\underbrace{\left\{ \left(\lambda\right)^{n}_{j-\frac{1}{2}}\textbf{U}^{n}_{j-1} + \textbf{G}^{n}_{j-1} \right\}}_{\text{Term 2}}  + \underbrace{\textbf{U}^{n}_{j}- \frac{\Delta t}{2\Delta x} \left[\left\{ \left( \lambda\right)^{n}_{j+\frac{1}{2}}+ \left( \lambda\right)^{n}_{j-\frac{1}{2}}\right\} \textbf{U}^{n}_{j}  \right]}_{\text{Term 3}}
\end{split}
\label{eq:1d_pos_2}
\end{equation}   
Let $\textbf{U}^{n}_{j} \in \textbf{W}$, $\forall$ j. Then, the numerical scheme is positivity preserving, {\em i.e.}, $\textbf{U}^{n+1}_{j} \in \textbf{W}$,  if terms 1, 2 and 3 in equation \eqref{eq:1d_pos_2} are all positive. Thus, our numerical method is positivity preserving if the following conditions are all satisfied.  
\begin{enumerate}
	\item $\left\{\left(\lambda\right)_{j+\frac{1}{2}}\textbf{U}_{j+1} - \textbf{G}_{j+1}\right\} \in \textbf{W}$ . This condition gives us (derivation in \ref{appendix:a0}) \label{condition1}
\begin{equation}
\left(\lambda\right)_{j+\frac{1}{2}} \geq \left(u_{j+1}+ \sqrt{\frac{\gamma_{j+1} -1}{2\gamma_{j+1}}} a_{j+1}\right) 
\label{eq:1d_pos_3}
\end{equation}
\item $\left\{\left(\lambda\right)_{j-\frac{1}{2}}\textbf{U}_{j-1} + \textbf{G}_{j-1}\right\} \in \textbf{W}$. Similarly, $\left\{\left(\lambda\right)_{j+\frac{1}{2}}\textbf{U}_{j} + \textbf{G}_{j}\right\} \in \textbf{W}$. From this condition, we get \label{condition2}
\begin{equation}
\left(\lambda\right)_{j+\frac{1}{2}} \geq \left(-u_{j}+ \sqrt{\frac{\gamma_{j} -1}{2\gamma_{j}}} a_{j}\right)
\label{eq:1d_pos_4}
\end{equation}
The positivity conditions \eqref{eq:1d_pos_3} and \eqref{eq:1d_pos_4} are combined by taking the following maximum.
\begin{equation}
\left(\lambda\right)_{j+\frac{1}{2}} \geq max\left( -u_{j}+ \sqrt{\frac{\gamma_{j} -1}{2\gamma_{j}}} a_{j}, u_{j+1}+ \sqrt{\frac{\gamma_{j+1} -1}{2\gamma_{j+1}}} a_{j+1}\right)
\label{eq:1d_pos_5}
\end{equation}
	\item $\left[1- \frac{\Delta t}{2\Delta x}\left\{ \left( \lambda\right)^{n}_{j+\frac{1}{2}}+ \left( \lambda\right)^{n}_{j-\frac{1}{2}}\right\}\right]\geq$0. Thus, we get the following limit on the global time step based on positivity preservation requirement.
	\begin{equation}
	\Delta t \leq  \Delta t_{p}= min_{j}\left( \frac{2 \Delta x}{\lambda_{j+\frac{1}{2}}+ \lambda_{j-\frac{1}{2}}} \right)
	\label{eq:1d_pos_6}
	\end{equation}	
	\end{enumerate}
	
	\subsection{Time step}
	\subsubsection{Time step restriction based on stability}
	The equilibrium distributions, or rather the sum of equilibrium distributions for all components for our kinetic model, given by Equation \eqref{eq:1d_euler_7}, can be written in vector form as,
	\begin{equation}
	\sum_{s}\textbf{f}^{eq}_{1,c}= \frac{\textbf{U}}{2}+ \frac{\textbf{G}}{2\lambda}, \ \sum_{s}\textbf{f}^{eq}_{2,c}= \frac{\textbf{U}}{2}- \frac{\textbf{G}}{2\lambda}
	\label{eq:1d_dt_1}
	\end{equation}
	Then, the Bouchut's criterion \cite{bouchut1999construction} requires the following condition to be satisfied for stability.
	\begin{equation}
	eig\left(\frac{\partial \sum_{c} \textbf{f}^{eq}_{j,c}}{\partial \textbf{U}}\right) \subset\left[0,\infty\right), \ j=1,2
	\label{eq:1d_dt_2}
	\end{equation}
	Here $eig$ refers to the eigenspectrum. Substituting \eqref{eq:1d_dt_1} into \eqref{eq:1d_dt_2} and simplifying, we get 
	\begin{equation}
	\lambda \geq max \left( \left|u-a\right|, \left|u\right|, \left|u+a\right|\right)
	\label{eq:1d_dt_3}
	\end{equation}
Based on the stability criterion in \eqref{eq:1d_dt_3}, we impose the following limit on the global time step.  
	\begin{equation}
	\Delta t \leq \Delta t _{s}= min_{j}\left(\frac{\Delta x}{\lambda_{max,j}}\right), \ \lambda_{max,j}= max\left(\left|u-a\right|, \left|u\right|, \left|u+a\right|\right)_{j}
	\label{eq:1d_dt_4}
	\end{equation}
	\textbf{Global time step }: To ensure that the restrictions imposed by both the positivity and stability criteria are satisfied, we define the global time step as follows.
\begin{equation}
	\Delta t= \sigma \ min (\Delta t_{p}, \Delta t_{s}),  \ 0< \sigma \leq 1
	\label{eq:1d_dt_5}
	\end{equation}
	That is, the global time step is the minimum of $\Delta t_{p}$ (defined in \eqref{eq:1d_pos_6}) and $\Delta t_{s}$ (defined in \eqref{eq:1d_dt_4}), multiplied by the CFL no., $\sigma$. Thus, our time-step computation accounts for both positivity and stability constraints.  
	
	\subsection{Fixing $\lambda$}
\begin{figure}[!h] 
\centering
\resizebox{\textwidth}{!}{
\includegraphics[width=15cm]{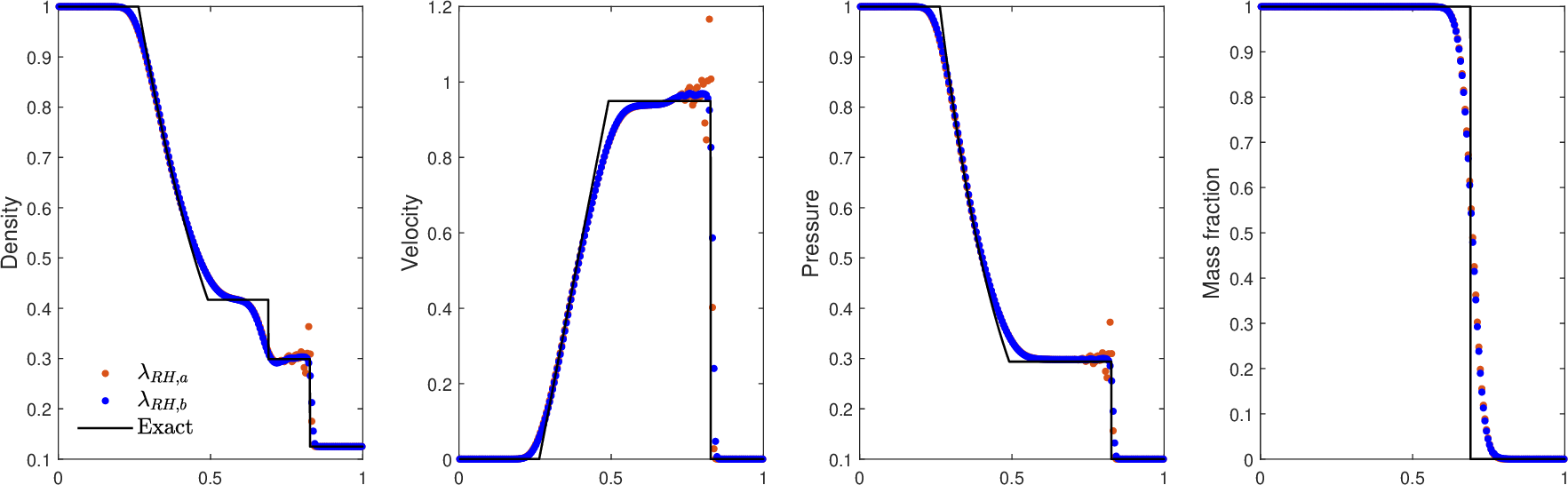}}
\caption{First order accurate numerical results for the multi-component Sod's shock tube problem for $\lambda_{RH}$=$\lambda_{RH,a}$\eqref{eq:1d_lambda_2} and $\lambda_{RH,b}$\eqref{eq:1d_lambda_3}. $\lambda$ is given by \ref{eq:1d_lambda_1}. Results with $\lambda_{RH,b}$ are improved over $\lambda_{RH,a}$}
\label{fig:1d_sod_compare}
\end{figure}

	We aim to fix $\lambda$ ($\geq$0) for our kinetic model such that it leads to an accurate and robust numerical scheme. Therefore, we define $\lambda$ such that it satisfies the positivity preservation condition \eqref{eq:1d_pos_5}, as follows.  
	\begin{equation}
	\left(\lambda\right)_{j+\frac{1}{2}} = max\left(\lambda_{RH}, -u_{j}+ \sqrt{\frac{\gamma_{j} -1}{2\gamma_{j}}} a_{j}, u_{j+1}+ \sqrt{\frac{\gamma_{j+1} -1}{2\gamma_{j+1}}} a_{j+1}\right)
	\label{eq:1d_lambda_1}
	\end{equation}
	Here, $\lambda_{RH}$ is a non-negative scalar numerical wave speed satisfying the Rankine-Hugoniot conditions at the interface (implemented in \cite{ROY2023106016}). We have considered two possible definitions of $\lambda_{RH}$ for the two-component Euler equations, as given below.  
	\begin{equation}
	\left(\lambda_{RH,a}\right)_{j+\frac{1}{2}}= min^{4}_{i=1}\left\{\frac{\left|\Delta G_{i}\right|}{\left|\Delta U_{i}\right|+ \epsilon_{0}}\right\}, \ \Delta= ()_{j+1}- ()_{j}
	\label{eq:1d_lambda_2}
	\end{equation}
	\begin{equation}
	\left(\lambda_{RH,b}\right)_{j+\frac{1}{2}}= min^{4}_{i=2}\left\{\frac{\left|\Delta G_{i}\right|}{\left|\Delta U_{i}\right|+ \epsilon_{0}}\right\}= min\left\{\frac{\left|\Delta \left(\rho u\right)\right|}{\left|\Delta \left(\rho\right)\right|+ \epsilon_{0}},\frac{\left|\Delta \left(\rho u^{2}+p\right)\right|}{\left|\Delta \left(\rho u\right)\right|+ \epsilon_{0}}, \frac{\left|\Delta \left( \rho Eu+ pu\right)\right|}{\left|\Delta \left(\rho E\right)\right|+ \epsilon_{0}}\right\}
	\label{eq:1d_lambda_3}
	\end{equation}
In the present work, $\epsilon_{0}= 10^{-10}$.  $\epsilon_{0}$ added to a non-negative denominator prevents division by zero. In Figure \ref{fig:1d_sod_compare}, we have compared the first order accurate numerical results for the Sod's shock tube problem (details of the test in section \ref{sod_diff_y}) for the two choices of $\lambda_{RH}$, with $\lambda$ given by \ref{eq:1d_lambda_1}. The observation for this test case as well as other standard test cases has been that numerical oscillations are lower for $\lambda_{RH,b}$ than for $\lambda_{RH,a}$. Thus, $\lambda_{RH}$ for our numerical scheme is chosen as in equation \eqref{eq:1d_lambda_3}.

\begin{figure}
\centering
\resizebox{\textwidth}{!}{
\includegraphics[width=15cm]{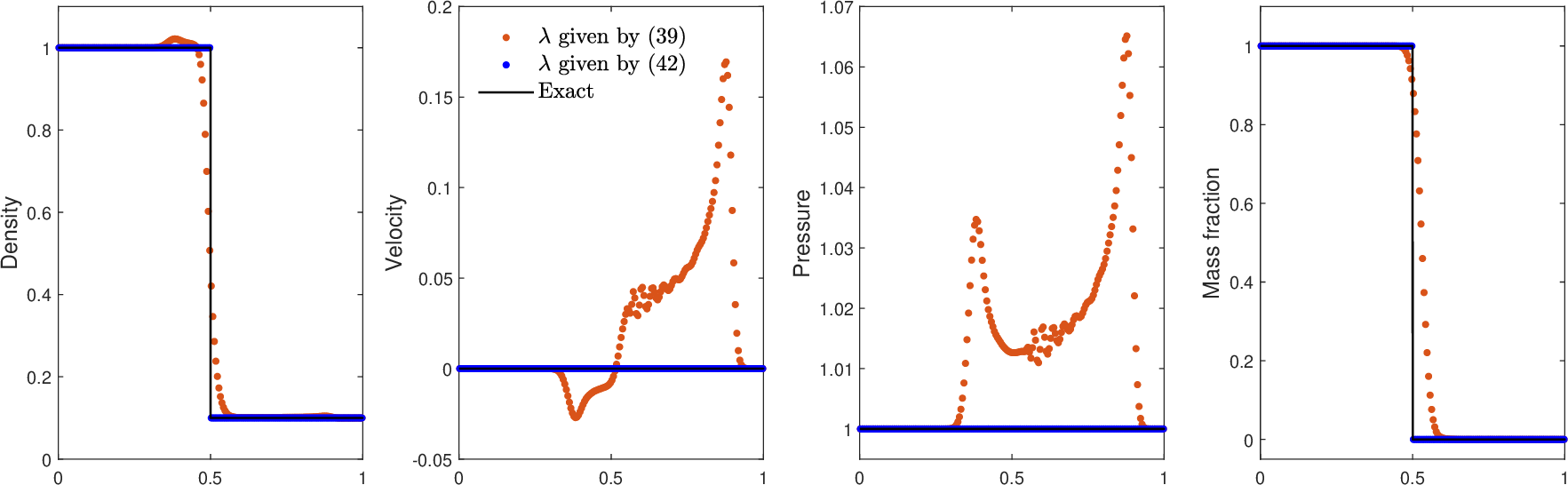}}
\caption{First order accurate numerical results for the multi-component steady contact discontinuity problem, for $\lambda$ given by \ref{eq:1d_lambda_1} and \ref{eq:1d_lambda_4}. $\lambda_{RH}$ is given by \eqref{eq:1d_lambda_3}. The contact discontinuity is resolved exactly using \ref{eq:1d_lambda_4}}
\label{fig:1d_steadyc_compare}
\end{figure}
 Designing a numerical scheme that accurately captures a material contact discontinuity is crucial for accuracy. However, our scheme with $\lambda$ defined by Equation \eqref{eq:1d_lambda_1} along with Equation \eqref{eq:1d_lambda_3} fails to exactly capture a steady material contact discontinuity (see Section \ref{steady_contact} for details of the test case). Moreover, as shown in Figure \ref{fig:1d_steadyc_compare}, it introduces unphysical pressure oscillations across the discontinuity. To address this, we have refined the definition of $\lambda$ to ensure the exact resolution of steady material contact discontinuities while maintaining positivity preservation elsewhere. Utilizing the fact that a steady contact discontinuity exhibits a density jump but no pressure jump, we conducted extensive numerical experimentation and arrived at the following definition of $\lambda$ that meets both the above mentioned objectives.  
\begin{subequations}
\label{eq:1d_lambda_4}
\begin{equation} 
\nonumber
\text{ If }\frac{|\rho_{j+1}-\rho_{j}|}{\left(\frac{\rho_{j}+ \rho_{j+1}}{2}\right)}> 0.1 \text{ and }  \frac{|p_{j+1}-p_{j}|}{\left(\frac{p_{j}+ p_{j+1}}{2}\right)}< 0.1 \text{, then} : 
\end{equation} 
\begin{equation} 
\left(\lambda\right)_{j+\frac{1}{2}} = abs\_sign\left(u_{j}+u_{j+1}\right)* \ max\left(\lambda_{RH}, -u_{j}+ \sqrt{\frac{\gamma_{j} -1}{2\gamma_{j}}} a_{j}, u_{j+1}+ \sqrt{\frac{\gamma_{j+1} -1}{2\gamma_{j+1}}} a_{j+1}\right)
\end{equation}
\begin{equation}
\text{ Else }: \ \ \left(\lambda\right)_{j+\frac{1}{2}} = max\left(\lambda_{RH}, -u_{j}+ \sqrt{\frac{\gamma_{j} -1}{2\gamma_{j}}} a_{j}, u_{j+1}+ \sqrt{\frac{\gamma_{j+1} -1}{2\gamma_{j+1}}} a_{j+1}\right)\text{, where}
\end{equation}
\begin{equation}
\left(\lambda_{RH}\right)_{j+\frac{1}{2}}= min\left\{\frac{\left|\Delta \left(\rho u\right)\right|}{\left|\Delta \left(\rho\right)\right|+ \epsilon_{0}},\frac{\left|\Delta \left(\rho u^{2}+p\right)\right|}{\left|\Delta \left(\rho u\right)\right|+ \epsilon_{0}}, \frac{\left|\Delta \left( \rho Eu+ pu\right)\right|}{\left|\Delta \left(\rho E\right)\right|+ \epsilon_{0}}\right\}, \ \epsilon_{0}= 10^{-10}
\end{equation}
\begin{equation}
abs\_sign(x)= \left\{\begin{array}{cc}
1,\text{ if }|x|>\epsilon_{0}\\
0,\text{ otherwise }\end{array} \right\}
\end{equation}
\end{subequations}
 Thus, at a steady contact discontinuity, the numerical diffusion coefficient $\lambda$ dynamically switches to zero, satisfying the Rankine-Hugoniot conditions and enabling exact capture of the contact discontinuity. Figure \ref{fig:1d_steadyc_compare} demonstrates that, with the modified definition \eqref{eq:1d_lambda_4}, our scheme captures a multi-component steady contact discontinuity exactly. This approach is in contrast to our kinetic model for a single gas, described in \cite{roy2024kineticschemebasedpositivity}. In \cite{roy2024kineticschemebasedpositivity}, both asymmetrical as well as symmetrical distribution of velocities are utilized, and the natural supersonic upwinding for the asymmetrical model leads to exact capture of a steady shock.

\subsection{Extension to higher order accuracy}
While our first order scheme can capture a steady contact discontinuity exactly, pressure oscillations still exist across a moving contact discontinuity (see section \ref{moving_contact_diff_gamma}). Thus, there is a need to increase the order of accuracy of the numerical scheme to reduce these numerical pressure oscillations and to obtain a more accurate solution. We have used a Chakravarthy-Osher type flux-limited approach \cite{10.1007/978-1-4613-8689-6_9}, together with a higher order Runge Kutta method, to extend our kinetic scheme up to third order accuracy. The third order ($3O$) kinetic flux summed over $c$, obtained by adding anti-diffusion terms to the first order kinetic flux, is given by 
\begin{equation}
\begin{split}
&\left(\sum_{c}\textbf{h}_{c,i}\right)_{j+\frac{1}{2}, 3O}= \left(\sum_{c}\textbf{h}_{c,i}\right)_{j+\frac{1}{2}}\\
&+ \frac{1}{6}\Phi\left( \frac{b\left(\sum_{c}\Delta \textbf{h}_{c,i}^{+}\right)_{j+ \frac{1}{2}}}{\left(\sum_{c}\Delta \textbf{h}_{c,i}^{+}\right)_{j- \frac{1}{2}}}\right)\left(\sum_{c}\Delta \textbf{h}_{c,i}^{+}\right)_{j- \frac{1}{2}}- \frac{1}{6}\Phi\left(\frac{b\left(\sum_{c}\Delta \textbf{h}_{c,i}^{-}\right)_{j+ \frac{1}{2}}}{\left(\sum_{c}\Delta \textbf{h}_{c,i}^{-}\right)_{j+ \frac{3}{2}}}\right)\left(\sum_{c}\Delta \textbf{h}_{c,i}^{-}\right)_{j+ \frac{3}{2}} \\
&+ \frac{1}{3}\Phi\left( \frac{b\left(\sum_{c}\Delta \textbf{h}_{c,i}^{+}\right)_{j- \frac{1}{2}}}{\left(\sum_{c}\Delta \textbf{h}_{c,i}^{+}\right)_{j+ \frac{1}{2}}}\right)\left(\sum_{c}\Delta \textbf{h}_{c,i}^{+}\right)_{j+ \frac{1}{2}}- \frac{1}{3}\Phi\left(\frac{b\left(\sum_{c}\Delta \textbf{h}_{c,i}^{-}\right)_{j+ \frac{3}{2}}}{\left(\sum_{c}\Delta \textbf{h}_{c,i}^{-}\right)_{j+ \frac{1}{2}}}\right)\left(\sum_{c}\Delta \textbf{h}_{c,i}^{-}\right)_{j+ \frac{1}{2}}
\end{split}
\label{eq:1d_HO_1}
\end{equation}
The definition in \eqref{eq:1d_HO_1} is applicable term-wise. $\left(\sum_{c}\textbf{h}_{c,i}\right)_{j+\frac{1}{2}}$ is the first order flux defined in \eqref{eq:1d_euler_10b}. The flux differences $\sum_{c}\Delta \textbf{h}_{c,i}^{\pm}$ are given by \eqref{eq:1d_euler_10c} and \eqref{eq:1d_euler_10d}. $\Phi$ is a diagonal matrix of the minmod limiter function. The constant compression parameter $b$ lies in the range $1<b\leq 4$. Setting all the limiters in \eqref{eq:1d_HO_1} to 1, we get the following fully third order flux.
\begin{equation}
\begin{split}
\left(\sum_{c}\textbf{h}_{c,i}\right)_{j+\frac{1}{2}, 3O}= \left(\sum_{c}\textbf{h}_{c,i}\right)_{j+\frac{1}{2}} &+ \frac{1}{6}\left(\sum_{c}\Delta \textbf{h}_{c,i}^{+}\right)_{j- \frac{1}{2}}- \frac{1}{6}\left(\sum_{c}\Delta \textbf{h}_{c,i}^{-}\right)_{j+ \frac{3}{2}} \\
&+ \frac{1}{3}\left(\sum_{c}\Delta \textbf{h}_{c,i}^{+}\right)_{j+ \frac{1}{2}}- \frac{1}{3}\left(\sum_{c}\Delta \textbf{h}_{c,i}^{-}\right)_{j+ \frac{1}{2}}
\end{split}
\label{eq:1d_HO_2a}
\end{equation}
If we substitute $b$=1 in Equation \eqref{eq:1d_HO_1}, then, owing to the symmetric nature of the minmod function, the equation simplifies to the following second order flux formulation.
\begin{equation}
\begin{split}
&\left(\sum_{c}\textbf{h}_{c,i}\right)_{j+\frac{1}{2}, 2O}= \left(\sum_{c}\textbf{h}_{c,i}\right)_{j+\frac{1}{2}}\\
&+ \frac{1}{2}\Phi\left( \frac{\left(\sum_{c}\Delta \textbf{h}_{c,i}^{+}\right)_{j+ \frac{1}{2}}}{\left(\sum_{c}\Delta \textbf{h}_{c,i}^{+}\right)_{j- \frac{1}{2}}}\right)\left(\sum_{c}\Delta \textbf{h}_{c,i}^{+}\right)_{j- \frac{1}{2}}- \frac{1}{2}\Phi\left(\frac{\left(\sum_{c}\Delta \textbf{h}_{c,i}^{-}\right)_{j+ \frac{1}{2}}}{\left(\sum_{c}\Delta \textbf{h}_{c,i}^{-}\right)_{j+ \frac{3}{2}}}\right)\left(\sum_{c}\Delta \textbf{h}_{c,i}^{-}\right)_{j+ \frac{3}{2}}
\end{split}
\label{eq:1d_HO_2}
\end{equation}
In the present work, we have taken $b$=4. Next, we rewrite Equation \eqref{eq:1d_HO_1} in expanded form, as follows.
\begin{equation}
\begin{split}
&\left(\sum_{c}\textbf{h}_{c,i}\right)_{j+\frac{1}{2}, 3O}= \begin{bmatrix} \sum_{c}h_{1,c,i} \\ \sum_{c}h_{2,c,i}\end{bmatrix}_{j+ \frac{1}{2}, 3O}= \begin{bmatrix} \sum_{c}h_{1,c,i} \\ \sum_{c}h_{2,c,i}\end{bmatrix}_{j+ \frac{1}{2}}\\
&+ \frac{1}{6} \begin{bmatrix}\phi\left\{\frac{b\left(\lambda\sum_{c}\Delta f^{eq}_{1,c,i}\right)_{j+ \frac{1}{2}}}{\left(\lambda\sum_{c}\Delta f^{eq}_{1,c,i}\right)_{j- \frac{1}{2}}}\right\}\left(\lambda\sum_{c}\Delta f^{eq}_{1,c,i}\right)_{j- \frac{1}{2}} \\0 \end{bmatrix}- \frac{1}{6} \begin{bmatrix}0 \\ \phi\left\{\frac{b\left(-\lambda\sum_{c}\Delta f^{eq}_{2,c,i}\right)_{j+ \frac{1}{2}}}{\left(-\lambda\sum_{c}\Delta f^{eq}_{2,c,i}\right)_{j+ \frac{3}{2}}}\right\}\left(-\lambda\sum_{c}\Delta f^{eq}_{2,c,i}\right)_{j+ \frac{3}{2}} \end{bmatrix} \\
&+ \frac{1}{3} \begin{bmatrix}\phi\left\{\frac{b\left(\lambda\sum_{c}\Delta f^{eq}_{1,c,i}\right)_{j- \frac{1}{2}}}{\left(\lambda\sum_{c}\Delta f^{eq}_{1,c,i}\right)_{j+ \frac{1}{2}}}\right\}\left(\lambda\sum_{c}\Delta f^{eq}_{1,c,i}\right)_{j+ \frac{1}{2}} \\0 \end{bmatrix}- \frac{1}{3} \begin{bmatrix}0 \\ \phi\left\{\frac{b\left(-\lambda\sum_{c}\Delta f^{eq}_{2,c,i}\right)_{j+ \frac{3}{2}}}{\left(-\lambda\sum_{c}\Delta f^{eq}_{2,c,i}\right)_{j+ \frac{1}{2}}}\right\}\left(-\lambda\sum_{c}\Delta f^{eq}_{2,c,i}\right)_{j+ \frac{1}{2}} \end{bmatrix}
\end{split}
\label{eq:1d_HO_3}
\end{equation}
Here $\phi(r)$ is the minmod limiter function of ratio $r$. To prevent the need for division, we express the limiter function as a function of two variables and write as $\phi(1,r)$. The minmod limiter function satisfies the following multiplication property.
\begin{equation}
\alpha \phi(x,y)= \phi(\alpha x, \alpha y)
\label{eq:eq:1d_HO_4}
\end{equation}
Equation \eqref{eq:1d_HO_3} can thus be rewritten as 
\begin{subequations}
\label{eq:1d_HO_5}
\begin{equation}
\begin{split}
&\left(\sum_{c}\textbf{h}_{c,i}\right)_{j+\frac{1}{2}, 3O}= \begin{bmatrix} \sum_{c}h_{1,c,i} \\ \sum_{c}h_{2,c,i}\end{bmatrix}_{j+ \frac{1}{2}, 3O}= \begin{bmatrix} \sum_{c}h_{1,c,i} \\ \sum_{c}h_{2,c,i}\end{bmatrix}_{j+ \frac{1}{2}}\\
&+ \frac{1}{6} \begin{bmatrix}\phi\left\{b\left(\lambda\sum_{c}\Delta f^{eq}_{1,c,i}\right)_{j+ \frac{1}{2}},\left(\lambda\sum_{c}\Delta f^{eq}_{1,c,i}\right)_{j- \frac{1}{2}}\right\} \\0 \end{bmatrix}- \frac{1}{6} \begin{bmatrix}0 \\ \phi\left\{b\left(-\lambda\sum_{c}\Delta f^{eq}_{2,c,i}\right)_{j+ \frac{1}{2}},\left(-\lambda\sum_{c}\Delta f^{eq}_{2,c,i}\right)_{j+ \frac{3}{2}}\right\} \end{bmatrix} \\
&+ \frac{1}{3} \begin{bmatrix}\phi\left\{b\left(\lambda\sum_{c}\Delta f^{eq}_{1,c,i}\right)_{j- \frac{1}{2}},\left(\lambda\sum_{c}\Delta f^{eq}_{1,c,i}\right)_{j+ \frac{1}{2}}\right\} \\0 \end{bmatrix}- \frac{1}{3} \begin{bmatrix}0 \\ \phi\left\{b\left(-\lambda\sum_{c}\Delta f^{eq}_{2,c,i}\right)_{j+ \frac{3}{2}},\left(-\lambda\sum_{c}\Delta f^{eq}_{2,c,i}\right)_{j+ \frac{1}{2}}\right\} \end{bmatrix}
\end{split}
\end{equation}
\begin{equation}
\phi(x,y)= minmod(x,y)= \left\{\begin{array}{cc}
x,\text{ if }|x|<|y|\text{ and }xy>0\\
y,\text{ if }|x|>|y|\text{ and }xy>0\\
0,\text{ if }xy<0\end{array} \right\}
\end{equation}
\end{subequations}
The macroscopic flux at the interface then becomes
\begin{equation}
\begin{split}
&(G_{i})_{j+\frac{1}{2}, 3O}= \textbf{P}_{i}\left(\sum_{c}\textbf{h}_{c,i}\right)_{j+\frac{1}{2}, 3O}= (G_{i})_{j+\frac{1}{2}}\\
&+ \frac{1}{6} \phi\left\{b\left(\lambda\sum_{c}\Delta f^{eq}_{1,c,i}\right)_{j+ \frac{1}{2}},\left(\lambda\sum_{c}\Delta f^{eq}_{1,c,i}\right)_{j- \frac{1}{2}}\right\} -\frac{1}{6}\phi\left\{b\left(-\lambda\sum_{c}\Delta f^{eq}_{2,c,i}\right)_{j+ \frac{1}{2}},\left(-\lambda\sum_{c}\Delta f^{eq}_{2,c,i}\right)_{j+ \frac{3}{2}}\right\}\\
&+ \frac{1}{3}\phi\left\{b\left(\lambda\sum_{c}\Delta f^{eq}_{1,c,i}\right)_{j- \frac{1}{2}},\left(\lambda\sum_{c}\Delta f^{eq}_{1,c,i}\right)_{j+ \frac{1}{2}}\right\}- \frac{1}{3}\phi\left\{b\left(-\lambda\sum_{c}\Delta f^{eq}_{2,c,i}\right)_{j+ \frac{3}{2}},\left(-\lambda\sum_{c}\Delta f^{eq}_{2,c,i}\right)_{j+ \frac{1}{2}}\right\}
\end{split}
\label{eq:1d_HO_6}
\end{equation}
or 
\begin{eqnarray}
(G_{i})_{j+\frac{1}{2}, 3O}= (G_{i})_{j+\frac{1}{2}}&+& \frac{1}{6}\phi\left\{b(\Delta G^{+}_{i})_{j+\frac{1}{2}},(\Delta G^{+}_{i})_{j-\frac{1}{2}}\right\}- \frac{1}{6}\phi\left\{b(\Delta G^{-}_{i})_{j+\frac{1}{2}},(\Delta G^{-}_{i})_{j+\frac{3}{2}}\right\} \nonumber\\
&+& \frac{1}{3}\phi\left\{b(\Delta G^{+}_{i})_{j-\frac{1}{2}},(\Delta G^{+}_{i})_{j+\frac{1}{2}}\right\}- \frac{1}{3}\phi\left\{b(\Delta G^{-}_{i})_{j+\frac{3}{2}},(\Delta G^{-}_{i})_{j+\frac{1}{2}}\right\}
\label{eq:1d_HO_7}
\end{eqnarray}
The temporal derivative is approximated using Strong Stability Preserving Runge Kutta \cite{gottlieb2001strong} (SSPRK) method. The update formula is 
\begin{subequations}
\label{eq:1d_HO_8}
\begin{equation}
(U_{i})^{1}_{j}= (U_{i})^{n}_{j}- \Delta t \ \text{R}((U_{i})^{n}_{j})
\end{equation}
\begin{equation}
(U_{i})^{2}_{j}= \frac{1}{4}(U_{i})^{1}_{j}+ \frac{3}{4}(U_{i})^{n}_{j} -\frac{1}{4}\Delta t \ \text{R}((U_{i})^{1}_{j})
\end{equation}
\begin{equation}
(U_{i})^{n+1}_{j}= \frac{2}{3}(U_{i})^{2}_{j}+ \frac{1}{3}(U_{i})^{n}_{j} -\frac{2}{3}\Delta t \ \text{R}((U_{i})^{2}_{j})
\end{equation}
\end{subequations} 
Here, R is the residual, {\em i.e.}, $\text{R}((U_{i})^{n}_{j})= \frac{1}{\Delta x}\left[ (G_{i})_{j+\frac{1}{2}}^{n}- (G_{i})_{j-\frac{1}{2}}^{n}\right]$.  The time step $\Delta t$ for our third order (as well as for second order) method is approximated by 
\begin{equation}
\left(\Delta t_{3O}\right)= \sigma \ min(\frac{\Delta t_{p}}{2}, \Delta t_{s})
\label{eq:1d_HO_9}
\end{equation}
At this point, we note that we haven't yet achieved positivity preservation with the chosen higher order flux-limited approach (unlike in the first order case), and addressing this issue is beyond the scope of the present work.  We restrict ourselves here to demonstrating that (while the basic scheme is also positivity preserving), the higher order extension accurately captures the relevant flow features.

\section{Kinetic model for 2D Euler equations}
Our kinetic model in two dimensions (N= 2) consists of three flexible velocities. Thus, $N_{d}$= 3, which meets the requirement of minimum number of velocities (given by $N_{d}\geq$N+1). Then, the Boltzmann equations for a component $c$, which correspond to the $i^{th}$ macroscopic equation, are given by 
\begin{equation}
\frac{\partial \textbf{f}_{c,i}}{\partial t}+ \frac{\partial (\Lambda_{1}\textbf{f}_{c,i})}{\partial x_{1}}+ \frac{\partial (\Lambda_{2}\textbf{f}_{c,i})}{\partial x_{2}}= - \frac{1}{\epsilon} \left[\textbf{f}_{c,i}- \textbf{f}_{c,i}^{eq}\right] 
\label{eq:2d_euler_1}
\end{equation}
Here 
\begin{equation}
\textbf{f}_{c,i}^{eq}= \begin{bmatrix} f^{eq}_{1,c,i} \\ f^{eq}_{2,c,i} \\ f^{eq}_{3,c,i} \end{bmatrix}, \ \Lambda_{1}= \begin{bmatrix} \lambda_{1,1} & 0 &0\\ 0 & \lambda_{2,1} &0 \\ 0 &0 & \lambda_{3,1}\end{bmatrix}, \ \Lambda_{2}= \begin{bmatrix} \lambda_{1,2} & 0 &0\\ 0 & \lambda_{2,2} &0 \\ 0 &0 & \lambda_{3,2}\end{bmatrix} 
\label{eq:2d_euler_2}
\end{equation}
Given the the row vector $\textbf{P}_{i}= \begin{bmatrix}1 & 1 & 1\end{bmatrix}$, following are the moment relations for the equilibrium distribution function.
\begin{subequations}
\label{eq:2d_euler_3}
\begin{equation}
\textbf{P}_{i}\sum_{c}\textbf{f}_{c,i}^{eq}= \sum_{c}f^{eq}_{1,c,i}+ \sum_{c}f^{eq}_{2,c,i}+ \sum_{c}f^{eq}_{3,c,i}= U_{i}
\end{equation}
\begin{equation}
\textbf{P}_{i} \sum_{c}\Lambda_{1}\textbf{f}_{c,i}^{eq}= \lambda_{1,1}\sum_{c}f^{eq}_{1,c,i}+ \lambda_{2,1}\sum_{c}f^{eq}_{2,c,i}+ \lambda_{3,1}\sum_{c}f^{eq}_{3,c,i}= G_{1,i}
\end{equation}
\begin{equation}
\textbf{P}_{i} \sum_{c}\Lambda_{2}\textbf{f}_{c,i}^{eq}= \lambda_{1,2}\sum_{c}f^{eq}_{1,c,i}+ \lambda_{2,2}\sum_{c}f^{eq}_{2,c,i}+ \lambda_{3,2}\sum_{c}f^{eq}_{3,c,i}= G_{2,i}
\end{equation}
\end{subequations}
Here, $U_{i}$, $G_{1,i}$ and $G_{2,i}$ are the $i^{th}$ conserved variable and flux terms of the 2D Euler equations. The two- component 2D Euler equations are given by,
\begin{equation}
\frac{\partial \textbf{U}}{\partial t} + \frac{\partial \textbf{G}_{1}}{\partial x_{1}} + \frac{\partial \textbf{G}_{2}}{\partial x_{2}} = 0, \ \textrm{with} \ \textbf{U} = \left[ \begin{array}{c} \rho W \\\rho \\ \rho u_{1} \\ \rho u_{2} \\ \rho E \end{array} \right], \ 
    \textbf{G}_{1} = \left[ \begin{array}{c} \rho W u_{1} \\ \rho u_{1} \\ \rho u_{1}^{2}+ p \\ \rho u_{1} u_{2} \\ (\rho  E +p)u_{1}  \end{array} \right] \ \textrm{and} \ 
    \textbf{G}_{2} = \left[ \begin{array}{c} \rho W u_{2} \\ \rho u_{2} \\  \rho u_{2} u_{1} \\ \rho u_{2}^{2}+ p \\(\rho  E +p)u_{2} \end{array} \right]
\label{eq:2d_euler_4}
\end{equation}
Next, we solve the Boltzmann equations \eqref{eq:2d_euler_1} numerically in a finite volume framework, for a structured grid. We use the operator-splitting strategy, leading to instantaneous relaxation and advection steps for a $(j,k)^{th}$ cell, as follows.  
\begin{subequations}
\[\text{Relaxation step: Instantaneous, {\em i.e.}, }\epsilon\rightarrow 0\text{. Thus,} \]
\begin{equation}
(\textbf{f}_{c,i})^{n}_{j,k}=  (\textbf{f}_{c,i}^{eq})^{n}_{j,k}
\label{eq:2d_euler_5a}
\end{equation}
\[\text{Advection step: The advective part of the Boltzmann equations are given by}\]
\begin{equation}
\frac{\partial \textbf{f}_{c,i}}{\partial t}+ \frac{\partial \textbf{h}_{1,c,i}}{\partial x_{1}}+ \frac{\partial \textbf{h}_{2,c,i}}{\partial x_{2}}= 0; \ \textbf{h}_{1,c,i}= \Lambda_{1}\textbf{f}_{c,i}^{eq}, \ \textbf{h}_{2,c,i}= \Lambda_{2}\textbf{f}_{c,i}^{eq}
\label{eq:2d_euler_5b}
\end{equation}
\end{subequations}
Rewriting \eqref{eq:2d_euler_5b} in integral form for $(j,k)^{th}$ cell, we get,
\begin{subequations}
\begin{equation}
A_{j,k}\frac{d \left(\textbf{f}_{c,i}\right)_{j,k}}{dt}+  \oint \textbf{h}_{\perp,c,i}dl =0; \ \textbf{h}_{\perp,c,i}= \Lambda_{\perp}\textbf{f}_{c,i}^{eq}, \ \Lambda_{\perp}= \Lambda_{1}n_{1}+ \Lambda_{2}n_{2}
\label{eq:2d_euler_6a}
\end{equation}
\begin{equation}
\Rightarrow A_{j,k}\frac{d \left(\textbf{f}_{c,i}\right)_{j,k}}{dt}+  \sum_{s=1}^{4} (\textbf{h}_{\perp,c,i})_{s}l_{s} =0\text{ (mid-point quadrature)}
\label{eq:2d_euler_6b}
\end{equation}
\end{subequations} 
Here, the normal and tangential unit vectors for a cell interface $s$ between the left (L) and right (R) cells are given by $\widehat{e}_{\perp}$= ($n_{1}$, $n_{2}$) and $\widehat{e}_{\parallel}$= (-$n_{2}$, $n_{1}$), respectively. For first order accuracy, we discretize equations \eqref{eq:2d_euler_6b} as follows.
\begin{subequations}
\begin{equation}
\left(\textbf{f}_{c,i}\right)^{n+1}_{j,k}= \left(\textbf{f}^{eq}_{c,i}\right)^{n}_{j,k} -\frac{\Delta t}{A_{j,k}}\sum_{s=1}^{4} (\textbf{h}_{\perp,c,i})_{s}l_{s}
\label{eq:2d_euler_7a}
\end{equation}
\begin{equation}
(\textbf{h}_{\perp,c,i})_{s}= \frac{1}{2}\left\{(\textbf{h}_{\perp,c,i})_{L}+ (\textbf{h}_{\perp,c,i})_{R}\right\}-\frac{1}{2}\left\{(\Delta \textbf{h}_{\perp,c,i}^{+})_{s}- (\Delta \textbf{h}_{\perp,c,i}^{-})_{s}\right\}
\label{eq:2d_euler_7b}
\end{equation}
\begin{equation}
(\Delta \textbf{h}_{\perp,c,i}^{\pm})_{s} = \left(\Lambda_{\perp}^{\pm}\Delta\textbf{f}_{c,i}^{eq}\right)_{s}
\label{eq:2d_euler_7c}
\end{equation}
\end{subequations}
Summing equations \eqref{eq:2d_euler_7a} over all $c$ components, we get 
\begin{subequations}
\begin{equation}
\left(\sum_{c}\textbf{f}_{c,i}\right)^{n+1}_{j,k}= \left(\sum_{c}\textbf{f}^{eq}_{c,i}\right)^{n}_{j,k} -\frac{\Delta t}{A_{j,k}}\sum_{s=1}^{4} \left(\sum_{c}\textbf{h}_{\perp,c,i}\right)_{s}l_{s}
\label{eq:2d_euler_8a}
\end{equation}
\begin{equation}
\left(\sum_{c}\textbf{h}_{\perp,c,i}\right)_{s}= \frac{1}{2}\left\{\left(\sum_{c}\textbf{h}_{\perp,c,i}\right)_{L}+ \left(\sum_{c}\textbf{h}_{\perp,c,i}\right)_{R}\right\}-\frac{1}{2}\left\{\left(\sum_{c}\Delta \textbf{h}_{\perp,c,i}^{+}\right)_{s}- \left(\sum_{c}\Delta \textbf{h}_{\perp,c,i}^{-}\right)_{s}\right\}
\label{eq:2d_euler_8b}
\end{equation}
\begin{equation}
\left(\sum_{c}\Delta \textbf{h}_{\perp,c,i}^{\pm}\right)_{s} = \left(\Lambda_{\perp}^{\pm}\sum_{c}\Delta\textbf{f}_{c,i}^{eq}\right)_{s}
\label{eq:2d_euler_8c}
\end{equation}
\end{subequations}
In two dimensions, the kinetic normal flux at an interface $\left(\sum_{c}\textbf{h}_{\perp,c,i}\right)$ has three components, given the three velocities considered. Summing these components yields the macroscopic normal flux, whose general formulation differs from the 1D case. However, to simplify the positivity analysis, we aim to maintain a 1D-like flux structure. Therefore, to ensure that the resulting macroscopic normal flux retains a locally 1D form, we define our three velocities, ($\lambda_{1,1}$,$\lambda_{1,2}$), ($\lambda_{2,1}$,$\lambda_{2,2}$) and ($\lambda_{3,1}$,$\lambda_{3,2}$) as shown in Figure \ref{fig:2d_eq} and detailed in equation \eqref{eq:2d_euler_9} below.

\begin{figure}[h!] 
\centering
\begin{tikzpicture}
\small\begin{axis}
 [every axis plot post/.append style={
  mark=none,domain=0:5,samples=50,smooth},
  xmin=0,xmax=5,ymin=0,ymax= 4,
	axis x line*=bottom, 
  axis y line*=left,
	axis line style={draw=none},
	xtick={},
	unit vector ratio*=1 1 1,
	xticklabels={\empty},
	yticklabel={\empty},
	tick style={draw=none}
	]

\draw [ultra thick](axis cs:1.8,3.2) --(axis cs:2.8, 1) node [anchor= north west]{$s$} ;
\draw [ultra thick,-stealth](axis cs:2.3, 2.1) --(axis cs:3.966,2.857)node [anchor=south east]{$(\textbf{$\lambda$}_{1,1},\textbf{$\lambda$}_{1,2})$};
\draw [ultra thick,-stealth](axis cs:2.3, 2.1) --(axis cs:0.5,2.3)node [anchor=south]{$(\textbf{$\lambda$}_{2,1},\textbf{$\lambda$}_{2,2})$};
\draw [ultra thick,-stealth](axis cs:2.3, 2.1) --(axis cs:1.26712,0.61232)node [anchor=north west]{$(\textbf{$\lambda$}_{3,1},\textbf{$\lambda$}_{3,2})$};
\draw [ultra thick](axis cs:2.24, 2.232) --(axis cs:2.46,2.332);
\draw [ultra thick](axis cs:2.46,2.332) --(axis cs:2.52,2.2);

\draw[black,thick,dashed,<->] (axis cs:2.4, 1.88) --(axis cs:4.066,2.637)node[midway,fill=white]{$\textbf{$\lambda$}_{\perp}$};
\draw[black,thick,dashed,<->] (axis cs:0.88356, 1.45616) --(axis cs:2.3,2.1)node[midway,fill=white]{$\textbf{$-\lambda$}_{\perp}$};
\draw[black,thick,dashed,<->] (axis cs:2, 1.963636) --(axis cs:1.6,2.843636)node[midway,fill=white]{$\textbf{$\lambda$}_{\parallel}$};
\draw[black,thick,dashed,<->] (axis cs:2, 1.963636) --(axis cs:2.4,1.083636)node[midway,fill=white]{$\textbf{$\lambda$}_{\parallel}$};
\draw [ultra thick,-stealth](axis cs:2.6,1.44) --(axis cs:2.8,1.5309)node [anchor=west] {$\widehat{e}_{\perp}$};
\end{axis}
\end{tikzpicture}
\caption{Velocities for 2D equilibrium distribution}
\label{fig:2d_eq}%
\end{figure}
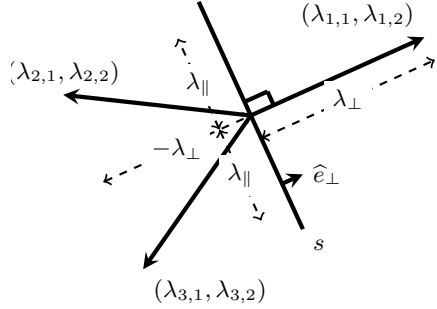

\begin{subequations}
\label{eq:2d_euler_9}
\begin{equation}
\lambda_{1,1}= \lambda_{\perp}n_{1},\hspace{0.5cm} \lambda_{2,1}= (-\lambda_{\perp}n_{1}-\lambda_{\parallel}n_{2}),\hspace{0.5cm} \lambda_{3,1}= (-\lambda_{\perp}n_{1}+\lambda_{\parallel}n_{2})
\end{equation}
\begin{equation}
\lambda_{1,2}= \lambda_{\perp}n_{2},\hspace{0.5cm} \lambda_{2,2}= (-\lambda_{\perp}n_{2}+\lambda_{\parallel}n_{1}),\hspace{0.5cm} \lambda_{3,2}= (-\lambda_{\perp}n_{2}-\lambda_{\parallel}n_{1})
\end{equation}
\end{subequations}
where $\lambda_{\perp}\geq$0. Now, substituting the $\lambda$s defined in \eqref{eq:2d_euler_9} into the moment relations \eqref{eq:2d_euler_3} and solving for the equilibrium distributions, we get,
\begin{equation}
\begin{bmatrix}\sum_{c}f^{eq}_{1,c,i}\\ \sum_{c}f^{eq}_{2,c,i}\\\sum_{c}f^{eq}_{3,c,i}\end{bmatrix}= \frac{1}{2\lambda_{\perp}}\begin{bmatrix} \lambda_{\perp} & n_{1} & n_{2}\\ \frac{\lambda_{\perp}}{2} & \frac{-2\lambda_{\perp}n_{2}- \lambda_{\parallel}n_{1}}{2\lambda_{\parallel}} & \frac{2\lambda_{\perp}n_{1}- \lambda_{\parallel}n_{2}}{2\lambda_{\parallel}} \\ \frac{\lambda_{\perp}}{2} & \frac{2\lambda_{\perp}n_{2}- \lambda_{\parallel}n_{1}}{2\lambda_{\parallel}} & \frac{-2\lambda_{\perp}n_{1}- \lambda_{\parallel}n_{2}}{2\lambda_{\parallel}}\end{bmatrix} \begin{bmatrix}U_{i} \\ G_{1,i} \\ G_{2,i} \end{bmatrix}
\label{eq:2d_euler_10}
\end{equation}
Further,
\begin{subequations}
\label{eq:2d_euler_11}
\begin{equation}
\Lambda_{\perp}= \Lambda_{1}n_{1}+ \Lambda_{2}n_{2}= \begin{bmatrix} \lambda_{\perp} &0 &0\\ 0& -\lambda_{\perp} &0\\ 0& 0& -\lambda_{\perp}\end{bmatrix}
\end{equation}
\begin{equation}
\left(\sum_{c}\Delta \textbf{h}_{\perp,c,i}^{+}\right)_{s} = \left(\Lambda_{\perp}^{+}\sum_{c}\Delta\textbf{f}_{c,i}^{eq}\right)_{s}= \begin{bmatrix} \left(\lambda_{\perp}\sum_{c} \Delta f^{eq}_{1,c,i}\right)_{s} \\0 \\0 \end{bmatrix}= \begin{bmatrix} (\lambda_{\perp})_{s}\left\{ \left(\sum_{c}f^{eq}_{1,c,i}\right)_{R}- \left(\sum_{c}f^{eq}_{1,c,i}\right)_{L}\right\} \\0 \\0 \end{bmatrix}
\end{equation}
\begin{equation}
\left(\sum_{c}\Delta \textbf{h}_{\perp,c,i}^{-}\right)_{s} = \left(\Lambda_{\perp}^{-}\sum_{c}\Delta\textbf{f}_{c,i}^{eq}\right)_{s}= \begin{bmatrix} 0 \\ \left(-\lambda_{\perp}\sum_{c}\Delta f^{eq}_{2,c,i}\right)_{s} \\  \left(-\lambda_{\perp}\sum_{c}\Delta f^{eq}_{3,c,i}\right)_{s} \end{bmatrix}= \begin{bmatrix} 0 \\ (-\lambda_{\perp})_{s}\left\{ \left(\sum_{c}f^{eq}_{2,c,i}\right)_{R}- \left(\sum_{c}f^{eq}_{2,c,i}\right)_{L}\right\} \\ (-\lambda_{\perp})_{s}\left\{ \left(\sum_{c}f^{eq}_{3,c,i}\right)_{R}- \left(\sum_{c}f^{eq}_{3,c,i}\right)_{L}\right\}  \end{bmatrix}
\end{equation}
\end{subequations}
The macroscopic update formula, obtained by taking moments of equation \eqref{eq:2d_euler_8a}, is given by 
\begin{subequations}
\begin{equation}
\left(U_{i}\right)^{n+1}_{j,k}= \left(U_{i}\right)^{n}_{j,k} -\frac{\Delta t}{A_{j,k}}\sum_{s=1}^{4} (G_{\perp,i})_{s}l_{s}; \ G_{\perp,i}= G_{1,i}n_{1}+ G_{2,i}n_{2}
\label{eq:2d_euler_12a}
\end{equation}
\begin{equation}
(G_{\perp,i})_{s}= \textbf{P}_{i}\left(\sum_{c}\textbf{h}_{\perp,c,i}\right)_{s} =\frac{1}{2}\left\{(G_{\perp,i})_{L}+ (G_{\perp,i})_{R}\right\}-\frac{1}{2}\left\{(\Delta G_{\perp,i}^{+})_{s}- (\Delta G_{\perp,i}^{-})_{s}\right\}
\label{eq:2d_euler_12b}
\end{equation}
\begin{eqnarray}
(\Delta G_{\perp,i}^{+})_{s}&=& \textbf{P}_{i}\left(\sum_{c}\Delta \textbf{h}_{\perp,c,i}^{+}\right)_{s}= \left(\lambda_{\perp} \sum_{c}\Delta f^{eq}_{1,c,i}\right)_{s}\nonumber\\
&=& \frac{1}{2}\left\{ (G_{\perp,i})_{R}- (G_{\perp,i})_{L}\right\} +\frac{\left(\lambda_{\perp}\right)_{s}}{2} \left\{ (U_{i})_{R}- (U_{i})_{L}\right\}
\label{eq:2d_euler_12c}
\end{eqnarray}
\begin{eqnarray}
(\Delta G_{\perp,i}^{-})_{s}&=& \textbf{P}_{i}\left(\sum_{c}\Delta \textbf{h}_{\perp,c,i}^{-}\right)_{s}= \left(-\lambda_{\perp} \sum_{c}\Delta \left(f^{eq}_{2,c,i}+ f^{eq}_{3,c,i}\right)\right)_{s} \nonumber\\
&=& \frac{1}{2}\left\{ (G_{\perp,i})_{R}- (G_{\perp,i})_{L}\right\} -\frac{\left(\lambda_{\perp}\right)_{s}}{2} \left\{ (U_{i})_{R}- (U_{i})_{L}\right\}
\label{eq:2d_euler_12d}
\end{eqnarray}
\end{subequations}
The macroscopic normal flux in \eqref{eq:2d_euler_12b} can be rewritten in vector form as,
\begin{equation}
(\textbf{G}_{\perp})_{s}= \frac{1}{2}\left\{(\textbf{G}_{\perp})_{L}+ (\textbf{G}_{\perp})_{R}\right\} -\frac{\left(\lambda_{\perp}\right)_{s}}{2}\left(\textbf{U}_{R}-\textbf{U}_{L}\right)
\label{eq:2d_euler_13}
\end{equation}

\subsection{Positivity analysis}
In our structured grid framework, the normals at interfaces point towards +$\xi$ and +$\eta$ directions respectively. For first order accuracy, the macroscopic update formula in a cell $(j,k)$ is then given by 
\begin{equation}
\begin{split}
&\textbf{U}^{n+1}_{j,k}= \textbf{U}^{n}_{j,k}- \frac{\Delta t}{A_{j,k}}\left[(\textbf{G}_{\perp})^{n}_{j+\frac{1}{2},k}l_{j+\frac{1}{2},k}- (\textbf{G}_{\perp})^{n}_{j-\frac{1}{2},k}l_{j-\frac{1}{2},k}+ (\textbf{G}_{\perp})^{n}_{j,k+\frac{1}{2}}l_{j,k+\frac{1}{2}}- (\textbf{G}_{\perp})^{n}_{j,k-\frac{1}{2}}l_{j,k-\frac{1}{2}}\right]\\
&= \frac{\Delta t}{2 A_{j,k}}\left[\lambda_{\perp}l\left\{-(\textbf{G}_{\perp})_{R}+ \lambda_{\perp} (\textbf{U})_{R}\right\}\right]^{n}_{j+\frac{1}{2},k}
+\frac{\Delta t}{2 A_{j,k}}\left[\lambda_{\perp}l\left\{(\textbf{G}_{\perp})_{L}+ \lambda_{\perp} (\textbf{U})_{L}\right\}\right]^{n}_{j-\frac{1}{2},k}\\
& + \frac{\Delta t}{2 A_{j,k}}\left[\lambda_{\perp}l\left\{-(\textbf{G}_{\perp})_{R}+ \lambda_{\perp} (\textbf{U})_{R}\right\}\right]^{n}_{j,k+\frac{1}{2}}
+\frac{\Delta t}{2 A_{j,k}}\left[\lambda_{\perp}l\left\{(\textbf{G}_{\perp})_{L}+ \lambda_{\perp} (\textbf{U})_{L}\right\}\right]^{n}_{j,k-\frac{1}{2}}\\
&+\underbrace{\textbf{U}^{n}_{j,k}-\frac{\Delta t}{2 A_{j,k}} \left[\left[\lambda_{\perp}l\left\{(\textbf{G}_{\perp})_{L}+ \lambda_{\perp} (\textbf{U})_{L}\right\}\right]^{n}_{j+\frac{1}{2},k}+ \left[\lambda_{\perp}l\left\{-(\textbf{G}_{\perp})_{R}+ \lambda_{\perp} (\textbf{U})_{R}\right\}\right]^{n}_{j-\frac{1}{2},k}\right. }\\
&\underbrace{\left.+ \left[\lambda_{\perp}l\left\{(\textbf{G}_{\perp})_{L}+ \lambda_{\perp} (\textbf{U})_{L}\right\}\right]^{n}_{j,k+\frac{1}{2}}+ \left[\lambda_{\perp}l\left\{-(\textbf{G}_{\perp})_{R}+ \lambda_{\perp} (\textbf{U})_{R}\right\}\right]^{n}_{j,k-\frac{1}{2}}\right]}_{\text{Term 5}}
\end{split}
\label{eq:2d_pos_1}
\end{equation}
Now, for the numerical scheme to be positivity preserving, {\em i.e.}, for $\textbf{U}^{n+1}_{j,k} \in \textbf{W}$, the following conditions have to be satisfied.  
\begin{enumerate}
	\item $\left\{-(\textbf{G}_{\perp})_{R}+ \lambda_{\perp} (\textbf{U})_{R}\right\}_{s} \in \textbf{W}$. This condition gives us,
	\begin{equation}
	\left(\lambda_{\perp}\right)_{s} \geq \left((u_{\perp})_{R}+ \sqrt{\frac{\gamma_{R} -1}{2\gamma_{R}}} a_{R}\right)_{s}, \ u_{\perp}= u_{1}n_{1}+ u_{2}n_{2}
	\label{eq:2d_pos_2}
	\end{equation}
	\item $\left\{(\textbf{G}_{\perp})_{L}+ \lambda_{\perp} (\textbf{U})_{L}\right\}_{s} \in \textbf{W}$. From this condition, we get,
	\begin{equation}
	\left(\lambda_{\perp}\right)_{s} \geq \left(-(u_{\perp})_{L}+ \sqrt{\frac{\gamma_{L} -1}{2\gamma_{L}}} a_{L}\right)_{s}
	\label{eq:2d_pos_3}
	\end{equation}
	Positivity conditions \eqref{eq:2d_pos_2} and \eqref{eq:2d_pos_3} are satisfied when the following condition is satisfied.
	\begin{equation}
	\left(\lambda_{\perp}\right)_{s} \geq max \left(-(u_{\perp})_{L}+ \sqrt{\frac{\gamma_{L} -1}{2\gamma_{L}}} a_{L}, (u_{\perp})_{R}+ \sqrt{\frac{\gamma_{R} -1}{2\gamma_{R}}} a_{R} \right)
	\label{eq:2d_pos_4}
	\end{equation}
	\item Term 5 can be written as a positive term multiplied with $\textbf{U}^{n}_{j,k}$. This leads to the following condition.
	\end{enumerate}
	\begin{equation}
	1 - \frac{\Delta t}{2 A_{j,k}}\left[ \left( \lambda l\right)^{n}_{j+\frac{1}{2},k}+ \left( \lambda l\right)^{n}_{j,k+\frac{1}{2}}+ \left( \lambda l\right)^{n}_{j-\frac{1}{2},k}+ \left( \lambda l\right)^{n}_{j,k-\frac{1}{2}}\right] \geq 0
	\label{eq:2d_pos_5}
	\end{equation}
	The positivity requirement thus leads to the following limit on the global time step.
	\begin{equation}
	\Delta t\leq \Delta t_{p}=min_{j,k}\left[\frac{2 A_{j,k}}{\left[ \left( \lambda l\right)_{j+\frac{1}{2},k}+ \left( \lambda l\right)_{j,k+\frac{1}{2}}+ \left( \lambda l\right)_{j-\frac{1}{2},k}+ \left( \lambda l\right)_{j,k-\frac{1}{2}}\right]}\right]
	\label{eq:2d_pos_6}
	\end{equation}
	
	\subsection{Time step}
	
	\subsubsection{Time step restriction based on stability}
	For our 2D kinetic model, the equilibrium distributions given by equation \eqref{eq:2d_euler_10} can be written in vector form, as follows.
	\begin{subequations}
\label{eq:2d_dt_1}
\begin{equation}
\sum_{c}\textbf{f}^{eq}_{1,c}=\frac{\textbf{U}}{2}+ \frac{\textbf{G}_{\perp}}{2\lambda_{\perp}} 
\end{equation}
\begin{equation}
\sum_{c}\textbf{f}^{eq}_{2,c}= \frac{\textbf{U}}{4}- \frac{\lambda_{\parallel}n_{1}+2\lambda_{\perp}n_{2}}{4\lambda_{\perp}\lambda_{\parallel}}\textbf{G}_{1}- \frac{\lambda_{\parallel}n_{2}-2\lambda_{\perp}n_{1}}{4\lambda_{\perp}\lambda_{\parallel}}\textbf{G}_{2}
\end{equation}
\begin{equation}
\sum_{c}\textbf{f}^{eq}_{3,c}= \frac{\textbf{U}}{4}- \frac{\lambda_{\parallel}n_{1}-2\lambda_{\perp}n_{y}}{4\lambda_{\perp}\lambda_{\parallel}}\textbf{G}_{1}- \frac{\lambda_{\parallel}n_{2}+2\lambda_{\perp}n_{1}}{4\lambda_{\perp}\lambda_{\parallel}}\textbf{G}_{2}
\end{equation}
\end{subequations}
Then, as per Bouchut's stability criterion,
\begin{subequations}
\begin{equation}
eig\left(\frac{\partial \sum_{c}\textbf{f}^{eq}_{1,c}}{\partial \textbf{U}}\right) \subset\left[0,\infty\right)
\label{eq:2d_dt_2a}
\end{equation}
\begin{equation}
eig\left(\frac{\partial \sum_{c}\textbf{f}^{eq}_{2,c}}{\partial \textbf{U}}\right) \subset\left[0,\infty\right)
\label{eq:2d_dt_2b}
\end{equation}
\begin{equation}
eig\left(\frac{\partial \sum_{c}\textbf{f}^{eq}_{3,c}}{\partial \textbf{U}}\right) \subset\left[0,\infty\right)
\label{eq:2d_dt_2c}
\end{equation}
\end{subequations}
Now, $\sum_{c}\textbf{f}^{eq}_{2,c}+\sum_{c}\textbf{f}^{eq}_{3,c}=\frac{\textbf{U}}{2}- \frac{\textbf{G}_{\perp}}{2\lambda_{\perp}}$, which is independent of $\lambda_{\parallel}$. So, we make a simplifying approximation by replacing \eqref{eq:2d_dt_2b} and \eqref{eq:2d_dt_2c} by the following condition.
\begin{equation}
eig\left(\frac{\partial \sum_{c}\left(\textbf{f}^{eq}_{2,c}+ \textbf{f}^{eq}_{3,c}\right)}{\partial \textbf{U}}\right) \subset\left[0,\infty\right)
\label{eq:2d_dt_3}
\end{equation}
The criteria \eqref{eq:2d_dt_2a} and \eqref{eq:2d_dt_3} then lead to 
	\begin{equation}
	\lambda_{\perp} \geq \lambda_{max}= max\left(\left|u_{\perp}-a\right|, \left|u_{\perp}\right|, \left|u_{\perp}+a\right|\right)
	\label{eq:2d_dt_4}
	\end{equation}
	Based on the stability criterion, the global time step is estimated as follows.
	\begin{equation}
	\Delta t \leq \Delta t _{s}= min_{j,k}\left[\frac{A_{j,k}}{(\lambda_{max})_{\xi}l_{\xi}+ (\lambda_{max})_{\eta}l_{\eta}}\right]
	\label{eq:2d_dt_5}
	\end{equation}
	\textbf{Global time step }: The global time step is then given by
	\begin{equation}
	\Delta t= \sigma \ min(\Delta t_{p}, \Delta t_{s}), \ 0<\sigma\leq1
	\label{eq:2d_dt_6}
	\end{equation}
Here, $\Delta t _{p}$ and $\Delta t _{s}$ are defined in \eqref{eq:2d_pos_6} and \eqref{eq:2d_dt_5} respectively.

\subsection{Fixing $\lambda_{\perp}$}
We determine $\lambda_{\perp}$ at an interface $s$ in 2D using the same approach that we followed in 1D. That is, $\left(\lambda_{\perp}\right)_{s}$ is defined to capture a grid-aligned steady contact discontinuity exactly, where as everywhere else it is defined such that it satisfies the positivity condition \eqref{eq:2d_pos_4}, as follows.
\begin{subequations}
\label{eq:2d_lambda}
\begin{equation} 
\nonumber
\text{ If }\frac{|\rho_{R}-\rho_{L}|}{\frac{\rho_{L}+ \rho_{R}}{2}}> 0.1 \text{ and }  \frac{|p_{R}-p_{L}|}{\left(\frac{p_{L}+ p_{R}}{2}\right)}< 0.1 \text{, then}: 
\end{equation} 
\begin{equation} 
\left(\lambda_{\perp}\right)_{s} = abs\_sign\left((u_{\perp})_{L}+(u_{\perp})_{R}\right)* \ max\left(\lambda_{RH}, -(u_{\perp})_{L}+ \sqrt{\frac{\gamma_{L} -1}{2\gamma_{L}}} a_{L}, (u_{\perp})_{R}+ \sqrt{\frac{\gamma_{R} -1}{2\gamma_{R}}} a_{R}\right)
\end{equation}
\begin{equation}
\text{ Else }: \ \ \left(\lambda_{\perp}\right)_{s} = max\left(\lambda_{RH}, -(u_{\perp})_{L}+ \sqrt{\frac{\gamma_{L} -1}{2\gamma_{L}}} a_{L}, (u_{\perp})_{R}+ \sqrt{\frac{\gamma_{R} -1}{2\gamma_{R}}} a_{R}\right)\text{, where}
\end{equation}
\begin{equation}
\left(\lambda_{RH}\right)_{s}= min\left\{\frac{\left|\Delta \left(\rho u_{\perp}\right)\right|}{\left|\Delta \left(\rho\right)\right|+ \epsilon_{0}},\frac{\left|\Delta \left(\rho u_{\perp}u_{1}+pn_{1}\right)\right|}{\left|\Delta \left(\rho u_{1}\right)\right|+ \epsilon_{0}},\frac{\left|\Delta \left(\rho u_{\perp}u_{2}+pn_{2}\right)\right|}{\left|\Delta \left(\rho u_{2}\right)\right|+ \epsilon_{0}}, \frac{\left|\Delta \left( \rho Eu_{\perp}+ pu_{\perp}\right)\right|}{\left|\Delta \left(\rho E\right)\right|+ \epsilon_{0}}\right\}  
\end{equation}
\end{subequations}

\subsection{Extension to higher order accuracy}
To extend our 2D scheme to higher-order accuracy, we adopt the same approach as in 1D. We use Chakravarthy-Osher type flux-limited approach to add anti-diffusion terms to the first order kinetic flux (summed over $c$) at an interface ($j+\frac{1}{2},k$), as follows.
\begin{equation}
\begin{split}
&\left(\sum_{c}\textbf{h}_{\perp,c,i}\right)_{j+\frac{1}{2},k, 3O}= \left(\sum_{c}\textbf{h}_{\perp,c,i}\right)_{j+\frac{1}{2},k}\\
&+ \frac{1}{6}\Phi\left\{ b\left(\sum_{c}\Delta \textbf{h}_{\perp,c,i}^{+}\right)_{j+ \frac{1}{2},k},\left(\sum_{c}\Delta \textbf{h}_{\perp,c,i}^{+}\right)_{j- \frac{1}{2},k}\right\}- \frac{1}{6}\Phi\left\{b\left(\sum_{c}\Delta \textbf{h}_{\perp,c,i}^{-}\right)_{j+ \frac{1}{2},k},\left(\sum_{c}\Delta \textbf{h}_{\perp,c,i}^{-}\right)_{j+ \frac{3}{2},k}\right\} \\
&+ \frac{1}{3}\Phi\left\{ b\left(\sum_{c}\Delta \textbf{h}_{\perp,c,i}^{+}\right)_{j- \frac{1}{2},k},\left(\sum_{c}\Delta \textbf{h}_{\perp,c,i}^{+}\right)_{j+ \frac{1}{2},k}\right\}- \frac{1}{3}\Phi\left\{b\left(\sum_{c}\Delta \textbf{h}_{\perp,c,i}^{-}\right)_{j+ \frac{3}{2},k},\left(\sum_{c}\Delta \textbf{h}_{\perp,c,i}^{-}\right)_{j+ \frac{1}{2},k}\right\}
\end{split}
\label{eq:2d_HO_1}
\end{equation}
Here, $\left(\sum_{c}\textbf{h}_{\perp,c,i}\right)$ is the first order flux, given by Equation \eqref{eq:2d_euler_8b}. The flux differences $\left(\sum_{c}\Delta \textbf{h}_{\perp,c,i}^{\pm}\right)$ are given by \eqref{eq:2d_euler_8c}. Now, Equation \eqref{eq:2d_HO_1} can be rewritten in expanded form as 
\begin{equation}
\begin{split}
&\left(\sum_{c}\textbf{h}_{\perp,c,i}\right)_{j+\frac{1}{2},k, 3O}= \begin{bmatrix} \sum_{c}h_{\perp,1,c,i} \\ \sum_{c}h_{\perp,2,c,i} \\ \sum_{c}h_{\perp,3,c,i} \end{bmatrix}_{j+ \frac{1}{2},k, 3O}= \begin{bmatrix} \sum_{c}h_{\perp,1,c,i} \\ \sum_{c}h_{\perp,2,c,i} \\ \sum_{c}h_{\perp,3,c,i} \end{bmatrix}_{j+ \frac{1}{2},k} \\
+& \frac{1}{6} \begin{bmatrix} \phi\left\{ b\left(\lambda_{\perp}\sum_{c}\Delta f_{1,c,i}^{eq}\right)_{j+ \frac{1}{2},k},\left(\lambda_{\perp}\sum_{c}\Delta f_{1,c,i}^{eq}\right)_{j- \frac{1}{2},k}\right\} \\0 \\0 \end{bmatrix} \\ 
-& \frac{1}{6} \begin{bmatrix} 0 \\ \phi\left\{ b\left(-\lambda_{\perp}\sum_{c}\Delta f_{2,c,i}^{eq}\right)_{j+ \frac{1}{2},k},\left(-\lambda_{\perp}\sum_{c}\Delta f_{2,c,i}^{eq}\right)_{j+ \frac{3}{2},k}\right\} \\ 
\phi\left\{ b\left(-\lambda_{\perp}\sum_{c}\Delta f_{3,c,i}^{eq}\right)_{j+ \frac{1}{2},k},\left(-\lambda_{\perp}\sum_{c}\Delta f_{3,c,i}^{eq}\right)_{j+ \frac{3}{2},k}\right\} \end{bmatrix} \\
+& \frac{1}{3} \begin{bmatrix} \phi\left\{ b\left(\lambda_{\perp}\sum_{c}\Delta f_{1,c,i}^{eq}\right)_{j- \frac{1}{2},k},\left(\lambda_{\perp}\sum_{c}\Delta f_{1,c,i}^{eq}\right)_{j+ \frac{1}{2},k}\right\} \\0 \\0 \end{bmatrix} \\ 
-& \frac{1}{3} \begin{bmatrix} 0 \\ \phi\left\{ b\left(-\lambda_{\perp}\sum_{c}\Delta f_{2,c,i}^{eq}\right)_{j+ \frac{3}{2},k},\left(-\lambda_{\perp}\sum_{c}\Delta f_{2,c,i}^{eq}\right)_{j+ \frac{1}{2},k}\right\} \\ \phi\left\{ b\left(-\lambda_{\perp}\sum_{c}\Delta f_{3,c,i}^{eq}\right)_{j+ \frac{3}{2},k},\left(-\lambda_{\perp}\sum_{c}\Delta f_{3,c,i}^{eq}\right)_{j+ \frac{1}{2},k}\right\} \end{bmatrix}
\end{split}
\label{eq:2d_HO_2}
\end{equation}
The macroscopic flux, obtained by taking moments of equation \eqref{eq:2d_HO_2}, is given by 
\begin{equation}
\begin{split}
\left(G_{\perp,i}\right)_{j+\frac{1}{2},k, 3O} =& \textbf{P}_{i}\left(\sum_{c}\textbf{h}_{\perp,c,i}\right)_{j+\frac{1}{2},k, 3O} \\ 
=& \left(G_{\perp,i}\right)_{j+\frac{1}{2},k} \\
+&\frac{1}{6}\phi\left\{ b\left(\lambda_{\perp}\sum_{c}\Delta f_{1,c,i}^{eq}\right)_{j+ \frac{1}{2},k},\left(\lambda_{\perp}\sum_{c}\Delta f_{1,c,i}^{eq}\right)_{j- \frac{1}{2},k}\right\} \\
+&\frac{1}{3}\phi\left\{ b\left(\lambda_{\perp}\sum_{c}\Delta f_{1,c,i}^{eq}\right)_{j- \frac{1}{2},k},\left(\lambda_{\perp}\sum_{c}\Delta f_{1,c,i}^{eq}\right)_{j+ \frac{1}{2},k}\right\}\\
-&\frac{1}{6}\phi\left\{ b\left(-\lambda_{\perp}\sum_{c}\Delta f_{2,c,i}^{eq}\right)_{j+ \frac{1}{2},k},\left(-\lambda_{\perp}\sum_{c}\Delta f_{2,c,i}^{eq}\right)_{j+ \frac{3}{2},k}\right\} \\ 
-&\frac{1}{6} \phi\left\{ b\left(-\lambda_{\perp}\sum_{c}\Delta f_{3,c,i}^{eq}\right)_{j+ \frac{1}{2},k},\left(-\lambda_{\perp}\sum_{c}\Delta f_{3,c,i}^{eq}\right)_{j+ \frac{3}{2},k}\right\} \\
-&\frac{1}{3}\phi\left\{ b\left(-\lambda_{\perp}\sum_{c}\Delta f_{2,c,i}^{eq}\right)_{j+ \frac{3}{2},k},\left(-\lambda_{\perp}\sum_{c}\Delta f_{2,c,i}^{eq}\right)_{j+ \frac{1}{2},k}\right\} \\ 
-&\frac{1}{3} \phi\left\{ b\left(-\lambda_{\perp}\sum_{c}\Delta f_{3,c,i}^{eq}\right)_{j+ \frac{3}{2},k},\left(-\lambda_{\perp}\sum_{c}\Delta f_{3,c,i}^{eq}\right)_{j+ \frac{1}{2},k}\right\}
\end{split}
\label{eq:2d_HO_3}
\end{equation}
At this stage, we introduce an approximation inspired by Kumar \& Dass's \cite{kumar} work, which involved approximating the molecular velocity integral of a limiter function of two variables by the limiter function of the integral of those variables in continuous velocity space. In our discrete framework, we substitute summations for integrals. As a result, we approximate \eqref{eq:2d_HO_3} using the following expression.
\begin{equation}
\begin{split}
\left(G_{\perp,i}\right)_{j+\frac{1}{2},k, 3O} =& \left(G_{\perp,i}\right)_{j+\frac{1}{2},k} \\
+&\frac{1}{6}\phi\left\{ b\left(\lambda_{\perp}\sum_{c}\Delta f_{1,c,i}^{eq}\right)_{j+ \frac{1}{2},k},\left(\lambda_{\perp}\sum_{c}\Delta f_{1,c,i}^{eq}\right)_{j- \frac{1}{2},k}\right\} \\ 
+&\frac{1}{3}\phi\left\{ b\left(\lambda_{\perp}\sum_{c}\Delta f_{1,c,i}^{eq}\right)_{j- \frac{1}{2},k},\left(\lambda_{\perp}\sum_{c}\Delta f_{1,c,i}^{eq}\right)_{j+ \frac{1}{2},k}\right\}\\
-&\frac{1}{6}\phi\left\{ b\left(-\lambda_{\perp}\sum_{c}\Delta (f_{2,c,i}^{eq}+f_{3,c,i}^{eq})\right)_{j+ \frac{1}{2},k},\left(-\lambda_{\perp}\sum_{c}\Delta (f_{2,c,i}^{eq}+f_{3,c,i}^{eq})\right)_{j+ \frac{3}{2},k}\right\}\\
-&\frac{1}{3}\phi\left\{ b\left(-\lambda_{\perp}\sum_{c}\Delta (f_{2,c,i}^{eq}+f_{3,c,i}^{eq})\right)_{j+ \frac{3}{2},k},\left(-\lambda_{\perp}\sum_{c}\Delta (f_{2,c,i}^{eq}+f_{3,c,i}^{eq})\right)_{j+ \frac{1}{2},k}\right\} 
\end{split}
\label{eq:2d_HO_4}
\end{equation}
Finally, using definitions \eqref{eq:2d_euler_12c} and \eqref{eq:2d_euler_12d}, equation \eqref{eq:2d_HO_4} can be rewritten as 
\begin{equation}
\begin{split}
\left(G_{\perp,i}\right)_{j+\frac{1}{2},k, 3O} =& \left(G_{\perp,i}\right)_{j+\frac{1}{2},k} \\ 
+& \frac{1}{6}\phi\left\{ b\left(\Delta G^{+}_{\perp,i}\right)_{j+ \frac{1}{2},k},\left(\Delta G^{+}_{\perp,i}\right)_{j- \frac{1}{2},k}\right\}- \frac{1}{6}\phi\left\{ b\left(\Delta G^{-}_{\perp,i}\right)_{j+ \frac{1}{2},k},\left(\Delta G^{-}_{\perp,i}\right)_{j+ \frac{3}{2},k}\right\}\\
+&\frac{1}{3}\phi\left\{ b\left(\Delta G^{+}_{\perp,i}\right)_{j- \frac{1}{2},k},\left(\Delta G^{+}_{\perp,i}\right)_{j+ \frac{1}{2},k}\right\}- \frac{1}{3}\phi\left\{ b\left(\Delta G^{-}_{\perp,i}\right)_{j+ \frac{3}{2},k},\left(\Delta G^{-}_{\perp,i}\right)_{j+ \frac{1}{2},k}\right\}
\end{split}
\label{eq:2d_HO_5}
\end{equation}
We note that as a consequence of the approximation in \eqref{eq:2d_HO_4}, our flux for third order accuracy simplifies and becomes independent of $\lambda_{\parallel}$, thus becoming locally one-dimensional. Finally, temporal derivative is approximated using SSPRK method, as described in \eqref{eq:1d_HO_8}.

\section{Results and Discussion}

\subsection{Experimental Order of Convergence}
To determine the experimental order of accuracy of the proposed numerical scheme, we have solved a simple 1D Euler problem with two components for which the analytical solution is known. Following are the initial conditions for the problem.
\begin{subequations}
\label{eq:EOC_1}
\begin{equation}
\rho_{j}(x,0)= 0.5+ 0.1 sin(\pi x), \ x\in [0,2], \ j=1,2 \ \text{and} \ \gamma_{j}=1.4 
\end{equation}
\begin{equation}
u(x,0)= 0.1, \ p(x,0)= 0.5
\end{equation}
\end{subequations}
The pressure and velocity are thus initially constant, whereas initial density for each component is a sinusoidal perturbation in space. Periodic conditions exist at $x=0$ and $x=2$ boundaries. The exact solution for this test case is known, and is given by,
\begin{subequations}
\label{eq:EOC_2}
\begin{equation}
\rho_{1}(x,t)=\rho_{2}(x,t)= 0.5+ 0.1 sin\left\{\pi (x- 0.1 t)\right\}, \ \rho(x,t)= \rho_{1}+ \rho_{2}= 1+ 0.2 sin\left\{\pi (x- 0.1 t)\right\},
\end{equation}
\begin{equation}
u(x,t)= 0.1, \ p(x,t)= 0.5
\end{equation}
\end{subequations}
The numerical solution for this problem at time $t$=0.5 is compared with the analytical solution. The numerical solution is obtained for varying grid sizes, {\em i.e.}, $Nx$ (=$\frac{2}{\Delta x}$)= 40, 80, 160, .. and so on. Then, the $L_{1}$ and $L_{2}$ errors in solution are computed as follows.  
\begin{subequations}
\label{eq:EOC_3}
\begin{equation}
\left\|\varepsilon_{Nx}\right\|_{L_{1}}= \Delta x \sum^{Nx}_{i=1}|\rho^{i}- \rho^{i}_{exact}|
\end{equation}
\begin{equation}
\left\|\varepsilon_{Nx}\right\|_{L_{2}}= \sqrt{\Delta x \sum^{Nx}_{i=1}(\rho^{i}- \rho^{i}_{exact})^{2}}
\end{equation}
\end{subequations}
Here, $\rho^{i}$ and $\rho^{i}_{exact}$ are the numerical and exact solutions for the $i^{th}$ cell. Now, for a $p^{th}$ order accurate scheme,
\begin{subequations}
\label{eq:EOC_4}
\begin{equation}
\left\|\varepsilon_{Nx}\right\|= C \Delta x^{p}+ O(\Delta x^{p+1}) 
\end{equation} 
Similarly, 
\begin{equation}
\left\|\varepsilon_{Nx/2}\right\|= C (2 \Delta x)^{p}+ O(\Delta x^{p+1})\text{, }(Nx\propto\frac{1}{\Delta x})
\end{equation}
\end{subequations}

\begin{table}[!t] 
\centering
\begin{tabular}{ |c|c|c|c|c|c| }
\hline
Nx& $\Delta$x& $L_{1}$ Error & $(EOC)_{L_{1}}$ & $L_{2}$ Error & $(EOC)_{L_{2}}$\\
\hline
40 & 0.05 & 0.0126783829 & & 0.0099907146 & \\
80 & 0.025 & 0.0064327953 & 0.978853 & 0.0050635125 & 0.980449\\
160 & 0.0125 & 0.0032432732 & 0.987995 & 0.0025539572 & 0.987404\\
320 & 0.00625 & 0.0016302454 & 0.992361 & 0.0012839750 & 0.992117\\
640 & 0.003125 & 0.0008162854 & 0.997944 & 0.0006429518 & 0.997834\\
1280 & 0.0015625 & 0.0004084316 & 0.998979 & 0.0003217150 & 0.998928\\
\hline
\end{tabular}
\centering
\caption{EOC using $L_{1}$ and $L_{2}$ error norms for first order accurate scheme} 
\label{table:EOC_1}
\end{table}

\begin{table}[!t] 
\centering
\begin{tabular}{ |c|c|c|c|c|c| }
\hline
Nx& $\Delta$x& $L_{1}$ Error & $(EOC)_{L_{1}}$ & $L_{2}$ Error & $(EOC)_{L_{2}}$\\
\hline
40 & 0.05 & 0.0019782511 & & 0.0019591484 & \\
80 & 0.025 & 0.0005596572 & 1.821610 & 0.0006493198 & 1.593225\\
160 & 0.0125 & 0.0001504354 & 1.895399 & 0.0002135016 & 1.604683\\
320 & 0.00625 & 0.0000403035 & 1.900167 & 0.0000695290 & 1.618560\\
640 & 0.003125 & 0.0000105647 & 1.931650 & 0.0000225196 & 1.626435\\
1280 & 0.0015625 & 0.0000027415 & 1.946217 & 0.0000072674 & 1.631657\\
\hline
\end{tabular}
\centering
\caption{EOC using $L_{1}$ and $L_{2}$ error norms for second order limited scheme}  
\label{table:EOC_2}
\end{table}

\begin{table}[!t] 
\centering
\begin{tabular}{ |c|c|c|c|c|c| }
\hline
Nx& $\Delta$x& $L_{1}$ Error & $(EOC)_{L_{1}}$ & $L_{2}$ Error & $(EOC)_{L_{2}}$\\
\hline
40 & 0.05 & 0.0003851743 & & 0.0004956926 & \\
80 & 0.025 & 0.0000763896 & 2.334063 & 0.0001315550 & 1.913779\\
160 & 0.0125 & 0.0000140669 & 2.441073 & 0.0000337252 & 1.963765\\
320 & 0.00625 & 0.0000027134 & 2.374152 & 0.0000084261 & 2.000888\\
640 & 0.003125 & 0.0000005213 & 2.379938 & 0.0000020656 & 2.028337\\
1280 & 0.0015625 & 0.0000000955 & 2.448876 & 0.0000004998 & 2.047051\\
\hline
\end{tabular}
\centering
\caption{EOC using $L_{1}$ and $L_{2}$ error norms for third order limited scheme}  
\label{table:EOC_3}
\end{table}

\begin{table}[!t] 
\centering
\begin{tabular}{ |c|c|c|c|c|c| }
\hline
Nx& $\Delta$x& $L_{1}$ Error & $(EOC)_{L_{1}}$ & $L_{2}$ Error & $(EOC)_{L_{2}}$\\
\hline
40 & 0.05 & 0.0000546167 & & 0.0000439340 & \\
80 & 0.025 & 0.0000068813 & 2.988596 & 0.0000055439 & 2.986367\\
160 & 0.0125 & 0.0000008608 & 2.998898 & 0.0000006938 & 2.998378\\
320 & 0.00625 & 0.0000001076 & 2.999712 & 0.0000000867 & 2.999570\\
640 & 0.003125 & 0.0000000135 & 2.998719 & 0.0000000108 & 2.998536\\
1280 & 0.0015625 & 0.0000000018 & 2.924790 & 0.0000000014 & 2.919507\\
\hline
\end{tabular}
\centering
\caption{EOC using $L_{1}$ and $L_{2}$ error norms for third order unlimited scheme}   
\label{table:EOC_3u}
\end{table}
Thus,
\begin{equation}
\frac{\left\|\varepsilon_{Nx/2}\right\|}{\left\|\varepsilon_{Nx}\right\|}= 2^{p}+ O(\Delta x) \Rightarrow log_{2}\left(\frac{\left\|\varepsilon_{Nx/2}\right\|}{\left\|\varepsilon_{Nx}\right\|}\right)= p+ O(\Delta x)
\label{eq:EOC_5}
\end{equation}
The experimental order of convergence (EOC) of the scheme is then given by
\begin{equation}
\textrm{EOC} = log_{2}\left(\frac{\left\|\varepsilon_{Nx/2}\right\|}{\left\|\varepsilon_{Nx}\right\|}\right)
\label{eq:EOC_6}
\end{equation}  
\begin{figure}
\centering
\begin{tabular}{cc}
\includegraphics[width=0.45\textwidth]{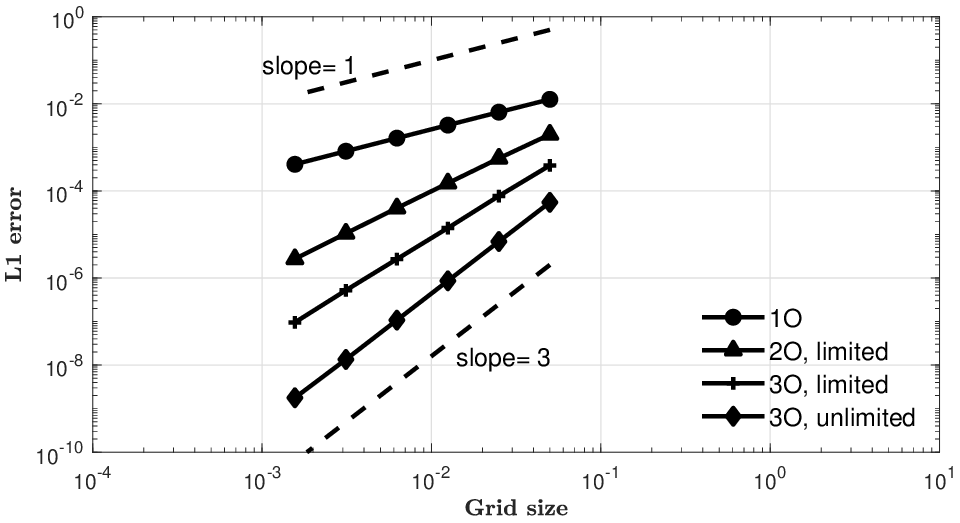} & \includegraphics[width=0.45\textwidth]{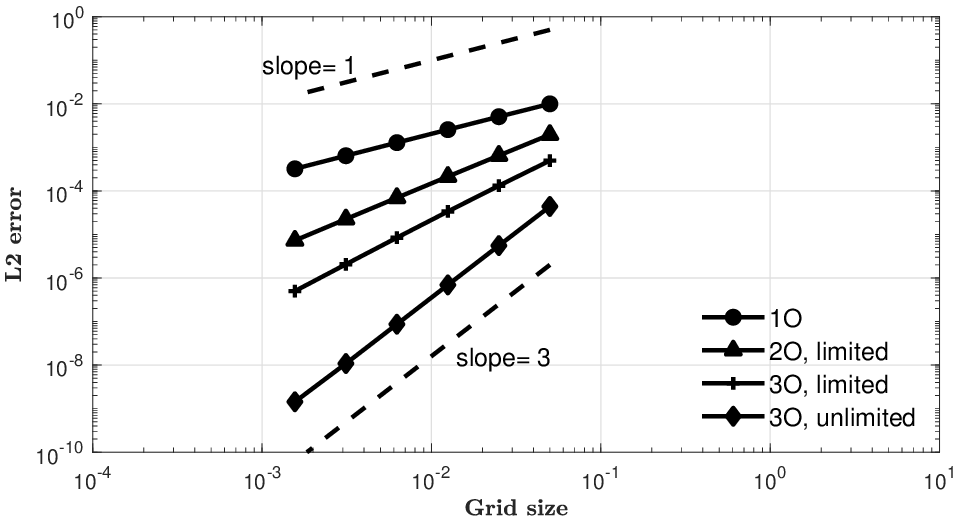}
\end{tabular}
\caption{(a) $L_{1}$ error norm vs grid size, (b) $L_{2}$ error norm vs grid size}  
\label{fig:EOC_1}
\end{figure} 
For the present finite volume approach, numerical solution in each cell corresponds to integral-averaged values. Therefore, for EOC computation, cell integral-averaged value of the exact solution is used for initializing the domain as well as for computing error norms at later times. The computed $L_{1}$ and $L_{2}$ error norms of the present scheme for first order accuracy are tabulated in Table \ref{table:EOC_1}. The error norms for second and third order limited schemes are tabulated in Tables \ref{table:EOC_2} and \ref{table:EOC_3} respectively, whereas those for the unlimited ({\em i.e.}, $\phi$= 1) third order scheme are tabulated in Table \ref{table:EOC_3u}. The log-log plots comparing the EOC with slopes 1 and 3 are shown in Figure \ref{fig:EOC_1}. EOCs for the third order limited scheme are higher than the second order scheme, as expected. However, a loss of accuracy is observed for the limited schemes. This is because for our flux-limited approach, the flux-reconstruction is limited not only at discontinuities but at local extrema as well  (highlighted in \cite{10.1007/978-1-4613-8689-6_9}). Thus, for the present problem with a sinusoidal solution, the reconstruction is limited at the maxima and minima, leading to clipping errors at the two extrema. Since the $L_2$ error norm punishes these localized errors more severely than the $L_1$ norm, the computed $L_2$ error norms are higher. That being said, as the proceeding sections show, the third order scheme generates reasonably accurate results for benchmark multi-component Euler test cases, thereby serving as a suitable high resolution tool.  

\subsection{1D Euler tests}
We have solved some standard multi-component 1D Euler test cases to assess the performance of our numerical scheme. For all the test cases described in the following subsections, the computational domain is $x\in$$[0,1]$, with an initial discontinuity at $x$=0.5. Transmissive boundary conditions are applied at $x$=0 and $x$=1. A CFL number $\sigma$= 0.8 is used for all test cases. Furthermore, except for the test case involving a moving contact discontinuity with different gamma (\ref{moving_contact_diff_gamma}), where results are shown for varying cell sizes, for all other tests $N_{x}$= 200 cells are used.  

\subsubsection{Test case: Steady contact discontinuity}\label{steady_contact}
\begin{figure}[!h] 
\centering
\resizebox{\textwidth}{!}{
\includegraphics[width=15cm]{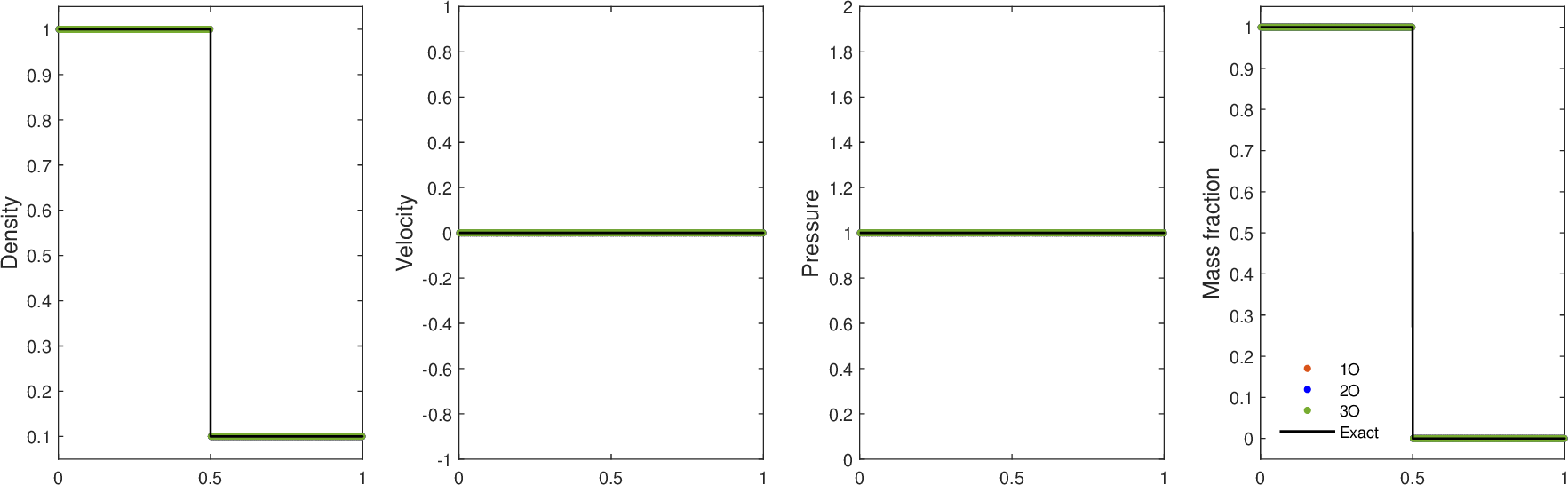}}
\caption{\label{fig:1d_euler_1} 1D test case: Steady contact discontinuity separating two gases with different $\gamma$'s}
\end{figure}

This test case consists of a stationary contact discontinuity separating two gases with different ratio of specific heats ($\gamma$) at initial time. The initial conditions for this problem are given by,
\begin{equation}
\begin{split}
\text{Left state: }& \ \rho_{L}= 1, \ W_{L}= 1, \ u_{L}= 0, \ p_{L}= 1, \ \gamma_{L}= 1.6 \\
\text{Right state: }& \ \rho_{R}= 0.1, \ W_{R}= 0, \ u_{R}= 0, \ p_{R}= 1, \ \gamma_{R}= 1.4
\end{split}
\label{eq:1d_euler_tc_1}
\end{equation} 
Further, $\left(c_{v}\right)_{1}$= $\left(c_{v}\right)_{2}$ is assumed. The numerical solution for this test case at time $t$= 0.1 is compared with the analytical solution in Figure \ref{fig:1d_euler_1}.  
Our results show that the contact discontinuity is captured exactly by our numerical scheme for first as well as higher orders of accuracy. Thus, the numerical diffusion in our scheme is optimal for exact capture of a steady contact discontinuity separating two different gases, indicating its low numerical diffusion. 

\subsubsection{Test case: Moving contact discontinuity, same $\gamma$}
This test case comprises a right-traveling material interface separating two different gases which have the same specific heat ratio, $\gamma$. The initial conditions for this test case are as follows.
\begin{equation}
\begin{split}
\text{Left state: }& \ \rho_{L}= 1, \ W_{L}= 1, \ u_{L}= 1, \ p_{L}= 1, \ \gamma_{L}= 1.4 \\
\text{Right state: }& \ \rho_{R}= 0.1, \ W_{R}= 0, \ u_{R}= 1, \ p_{R}= 1, \ \gamma_{R}= 1.4
\end{split}
\label{eq:1d_euler_tc_2}
\end{equation}
\begin{figure}[!h] 
\centering
\resizebox{\textwidth}{!}{
\includegraphics[width=15cm]{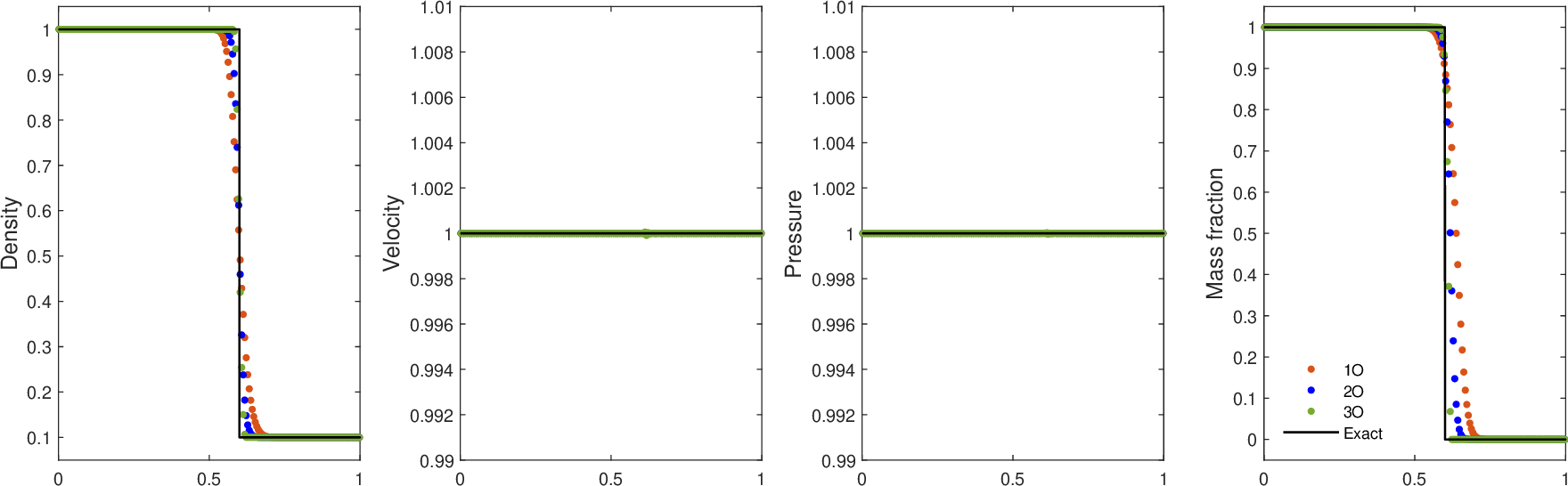}}
\caption{\label{fig:1d_euler_2} 1D test case: Moving contact discontinuity separating two gases with same $\gamma$}
\end{figure}
The results for this test (Figure \ref{fig:1d_euler_2}) show that our numerical scheme accurately captures the moving discontinuity, with some numerical diffusion. Further, oscillations in pressure and velocity are virtually absent. 

\subsubsection{Test case: Moving contact discontinuity, different $\gamma$}\label{moving_contact_diff_gamma}
For this test case, we consider a right-traveling material interface separating gases with different adiabatic constants, $\gamma$ (see \cite{ABGRALL2001594}, \cite{GOUASMI2020112912}). The following are the initial conditions for this problem.
\begin{equation}
\begin{split}
\text{Left state: }& \ \rho_{L}= 1, \ W_{L}= 1, \ u_{L}= 1, \ p_{L}= 1, \ \gamma_{L}= 1.6 \\
\text{Right state: }& \ \rho_{R}= 0.1, \ W_{R}= 0, \ u_{R}= 1, \ p_{R}= 1, \ \gamma_{R}= 1.4
\end{split}
\label{eq:fig:1d_euler_tc_3}
\end{equation} 
Further, it is assumed that $\left(c_{v}\right)_{1}$= $\left(c_{v}\right)_{2}= 1$. 
\begin{figure}[!h] 
\centering
\resizebox{\textwidth}{!}{
\includegraphics[width=15cm]{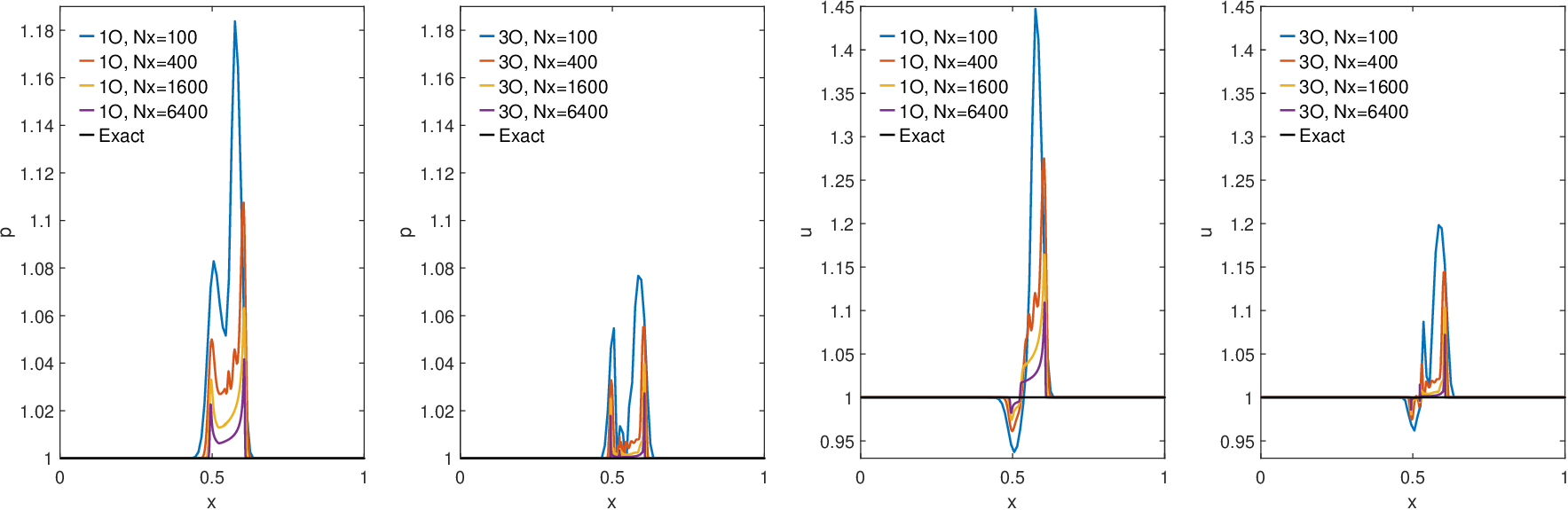}}
\caption{\label{fig:1d_euler_3a} 1D test case: Moving contact discontinuity separating two gases with different $\gamma$. 1O and 3O results for pressure and velocity at time $t$= 0.022}
\end{figure}
\begin{figure}[!h] 
\centering
\resizebox{\textwidth}{!}{
\includegraphics[width=15cm]{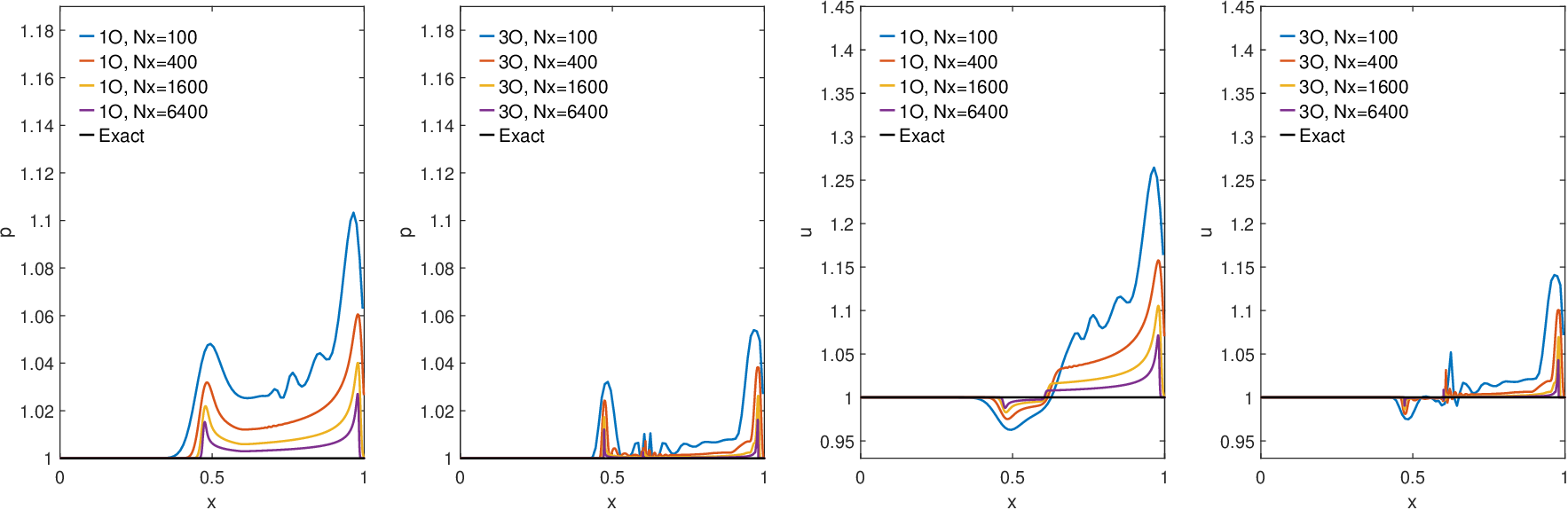}}
\caption{\label{fig:1d_euler_3b} 1D test case: Moving contact discontinuity separating two gases with different $\gamma$. 1O and 3O results for pressure and velocity at time $t$= 0.1}
\end{figure}
The numerical results for this test case at times $t$=0.022 and $t$=0.1 are shown in Figures \ref{fig:1d_euler_3a} and \ref{fig:1d_euler_3b} respectively. Numerical oscillations in pressure and velocity are observed, which are typical for a conservative numerical scheme (see \cite{ABGRALL2001594}).  These oscillations slowly reduce as the number of grid points is increased. Increasing the order of accuracy also reduces the numerical oscillations.   

\subsubsection{Test case: Sod's shock tube problem, same $\gamma$}
This test case consists of a discontinuity separating a high pressure and a low pressure gas at initial time, with both gases having the same $\gamma$ (\cite{LIU2003651}). The initial conditions for this problem are as follows.
\begin{equation}
\begin{split}
\text{Left state: }& \ \rho_{L}= 2, \ W_{L}= 1, \ u_{L}= 0, \ p_{L}= 10, \ \gamma_{L}= 1.4 \\
\text{Right state: }& \ \rho_{R}= 1, \ W_{R}= 0, \ u_{R}= 0, \ p_{R}= 1, \ \gamma_{R}= 1.4
\end{split}
\label{eq:1d_euler_tc_4}
\end{equation}
The solution for this problem for t$>$0 comprises a right-traveling shock, a right-traveling contact discontinuity and left traveling expansion wave. 
\begin{figure}[!h] 
\centering
\resizebox{\textwidth}{!}{
\includegraphics[width=15cm]{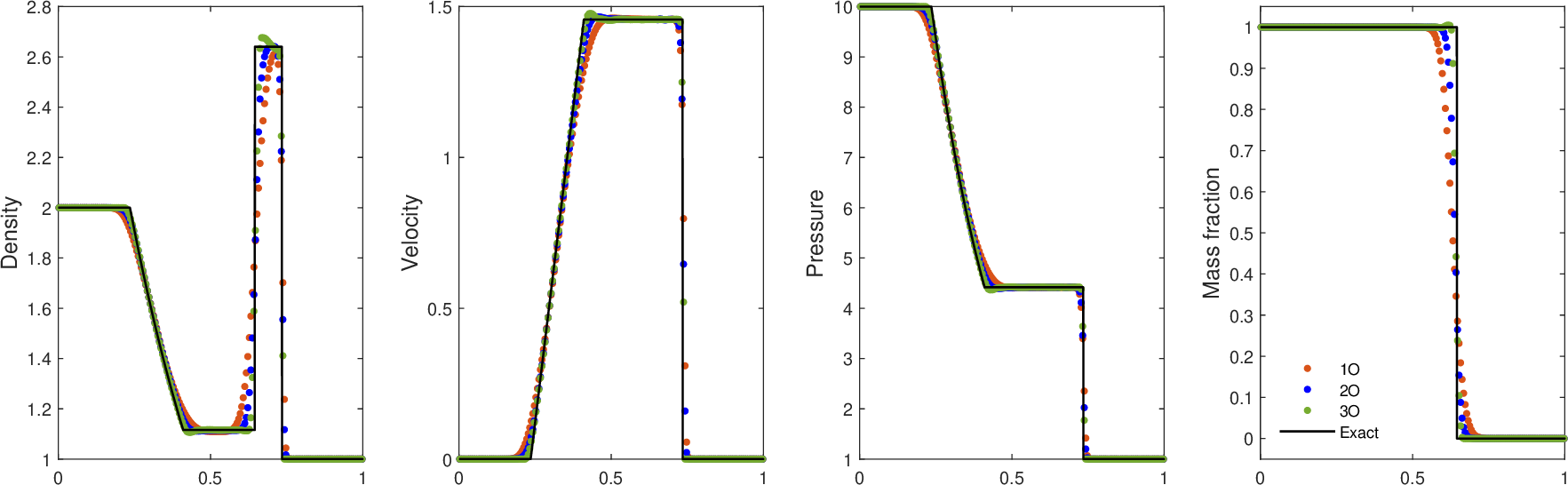}}
\caption{\label{fig:1d_euler_4} 1D test case: Sod's shock tube problem, with the two gases having same $\gamma$}
\end{figure}
Figure \ref{fig:1d_euler_4} shows the numerical result for this problem at $t$= 0.1. The results are reasonably accurate, with no entropy-violating expansion shock being formed. There is a significant increase in the resolution as the order of accuracy of the scheme increases.  

\subsubsection{Test case: Sod's shock tube problem, different $\gamma$}\label{sod_diff_y}
This test case is another version of the multi-component shock tube problem, with the left and right state gases having different $\gamma$ (\cite{LARROUTUROU199159}, \cite{KARNI199431}). The initial conditions for this test case are given below.
\begin{equation}
\begin{split}
\text{Left state: }& \ \rho_{L}= 1, \ W_{L}= 1, \ u_{L}= 0, \ p_{L}= 1, \ \gamma_{L}= 1.4 \\
\text{Right state: }& \ \rho_{R}= 0.125, \ W_{R}= 0, \ u_{R}= 0, \ p_{R}= 0.1, \ \gamma_{R}= 1.2
\end{split}
\label{eq:1d_euler_tc_5}
\end{equation}
\begin{figure}[!h] 
\centering
\resizebox{\textwidth}{!}{
\includegraphics[width=15cm]{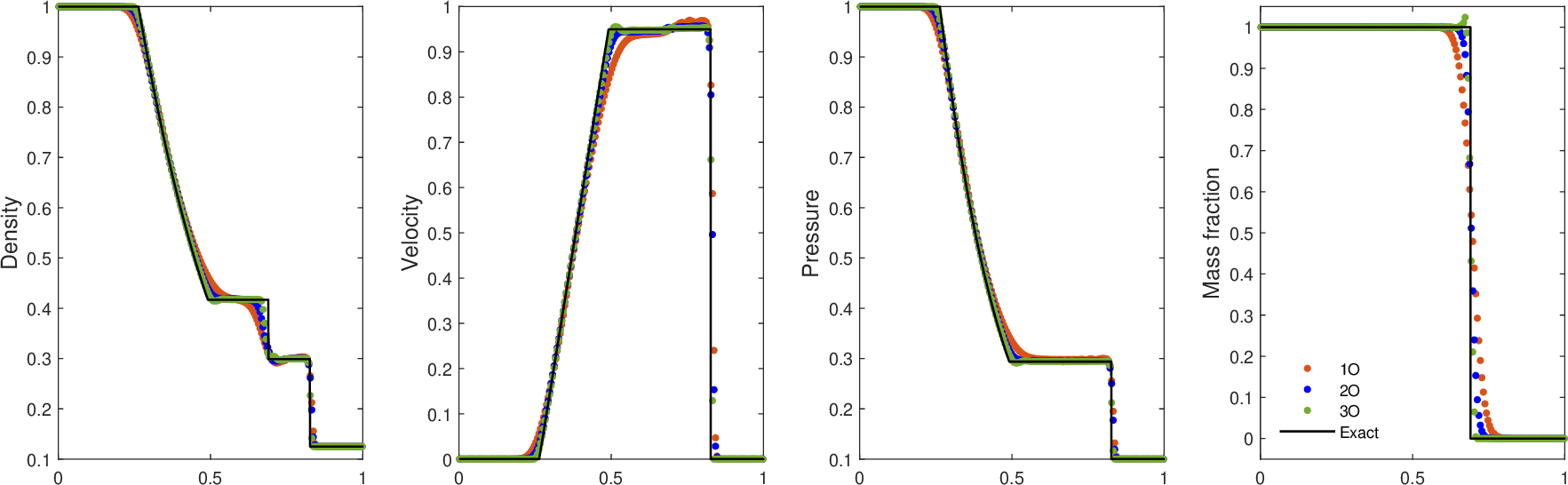}}
\caption{\label{fig:1d_euler_5} 1D test case: Sod's shock tube problem, with the two gases having different $\gamma$}
\end{figure}
Further, $\left(c_{v}\right)_{1}$= $\left(c_{v}\right)_{2}$= 1.  This problem also features a right traveling shock and a  contact discontinuity, along with a left traveling expansion wave. The numerical results at $t$= 0.2 are presented in Figure \ref{fig:1d_euler_5}. While the results are entropic, the third-order solution shows that the mass fraction exceeds the [0,1] range upstream of the material interface, indicating a loss of positivity of density in that region. It underscores the fact that the higher order extension is not necessarily positivity preserving.  

\subsubsection{Test for positivity of mass fraction}
This so called 'positivity of mass fraction problem' is taken from \cite{ABGRALL2001594}. The initial conditions for this test case are given by 
\begin{equation}
 \begin{split}
\text{Left state: }& \ \rho_{L}= 1, \ W_{L}= 1, \ u_{L}= -1, \ H_{L}= 1, \ \gamma_{L}= 1.4 \\
\text{Right state: }& \ \rho_{R}= 1, \ W_{R}= 0, \ u_{R}= 1, \ H_{R}= 5, \ \gamma_{R}= 1.4
\end{split}
\label{eq:1d_euler_tc_6}
\end{equation} 
Here, enthalpy $H$= $E$+ $\frac{p}{\rho}$. The solution for this problem features a left traveling shock and contact discontinuity, and a rarefaction wave with a sonic point. 
\begin{figure}[!h] 
\centering
\resizebox{\textwidth}{!}{
\includegraphics[width=15cm]{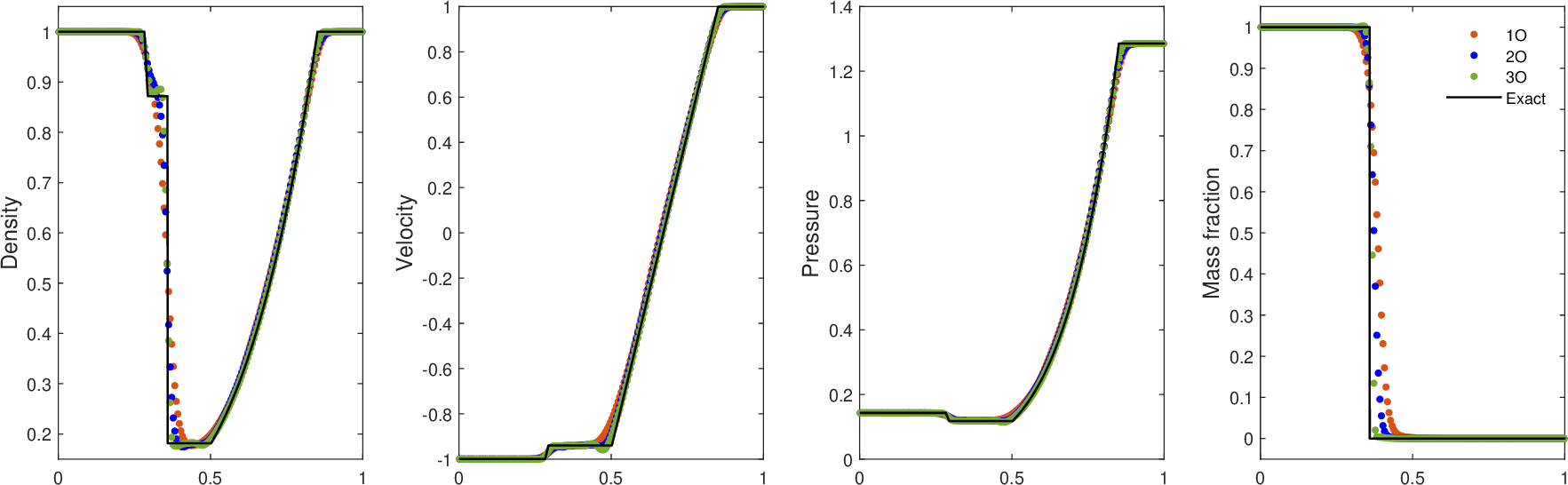}}
\caption{\label{fig:1d_euler_6} 1D test case: Positivity of mass fraction problem}
\end{figure}
The numerical results for this problem at time $t=0.15$ are shown in Figure \ref{fig:1d_euler_6}. The numerical results are entropic and reasonably accurate.

\subsection{2D Euler test case: Triple point problem}
\begin{figure}[h!] 
\centering
\begin{tikzpicture}
\begin{axis}
  [xmin=0,xmax=7,ymin=0,ymax=3,
	xtick={0,1,2,3,4,5,6,7},
	ytick={0,1,2,3},
	unit vector ratio*=1 1 1,
	tick style={draw=none},
	]
\draw [thick](axis cs:0,0) rectangle (axis cs:7,3);
\draw[black,thick] (axis cs:1,0) --(axis cs:1,3);
\draw[black,thick] (axis cs:1,1.5) --(axis cs:7,1.5);
\node[anchor= west] at (axis cs:1.1,1.25){(1,1.5)};
\node at (axis cs:0.45,1.5){\textbf{I}};
\node at (axis cs:4,2.25){\textbf{II}};
\node at (axis cs:4,0.75){\textbf{III}};
\end{axis}
\end{tikzpicture}
\caption{Schematic showing the different regions at initial time for the triple point problem.}
\label{fig:2d_tp_schematic}%
\end{figure}
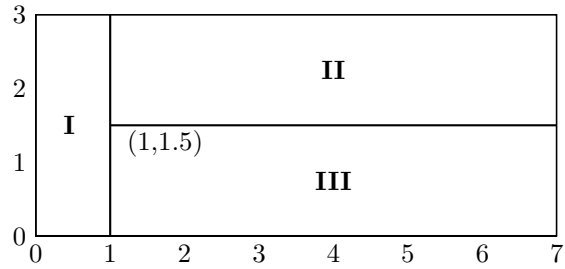

The multi-material triple point problem is a two-component 2D Riemann problem (see \cite{Pan_Cheng_Wang_Xu_2017})  defined on a rectangular domain $[0,7]\times[0,3]$.  The domain is divided into three regions, as illustrated in Figure \ref{fig:2d_tp_schematic}, with each region assigned its own initial conditions. The initial conditions for the three regions are as follows. 

\begin{equation}
 \begin{split}
\text{Region I: }& \ \rho_{\text{I}}= 1, \ W_{\text{I}}= 1, \ u_{\text{I}}= 0, \ v_{\text{I}}= 0, \ p_{\text{I}}= 1, \ \gamma_{\text{I}}= 1.5 \\
\text{Region II: }& \ \rho_{\text{II}}= 0.125, \ W_{\text{II}}= 1, \ u_{\text{II}}= 0, \ v_{\text{II}}= 0, \ p_{\text{II}}= 0.1, \ \gamma_{\text{II}}= 1.5 \\
\text{Region III: }& \ \rho_{\text{III}}= 1, \ W_{\text{III}}= 0, \ u_{\text{III}}= 0, \ v_{\text{III}}= 0, \ p_{\text{III}}= 0.1, \ \gamma_{\text{III}}= 1.4
\end{split}
\label{eq:2d_euler_tp_ic}
\end{equation}
\begin{figure}
\centering
\begin{tabular}{cc}
\includegraphics[width=0.45\textwidth]{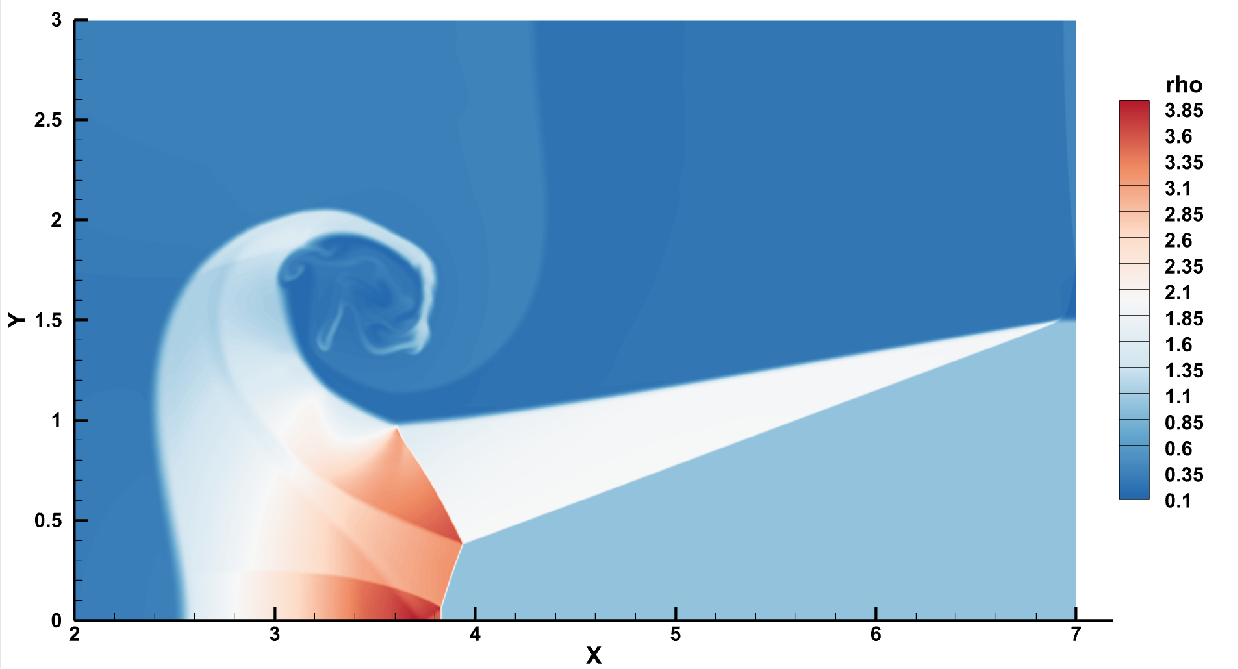} & \includegraphics[width=0.45\textwidth]{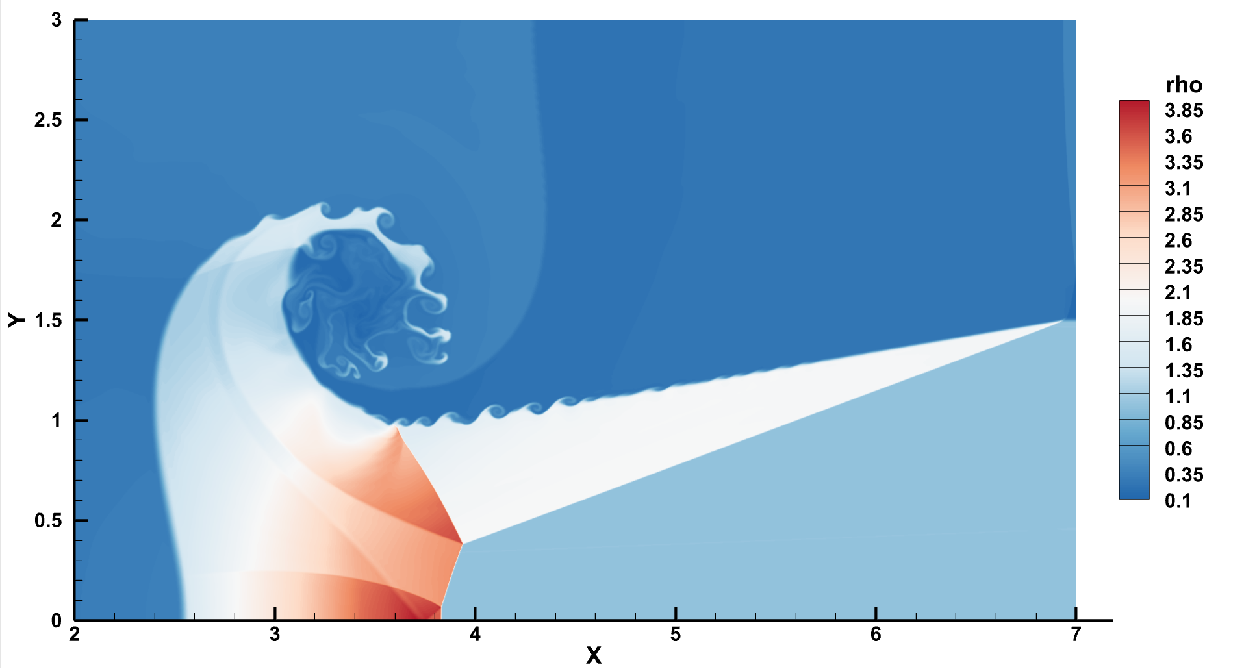}\\
\includegraphics[width=0.45\textwidth]{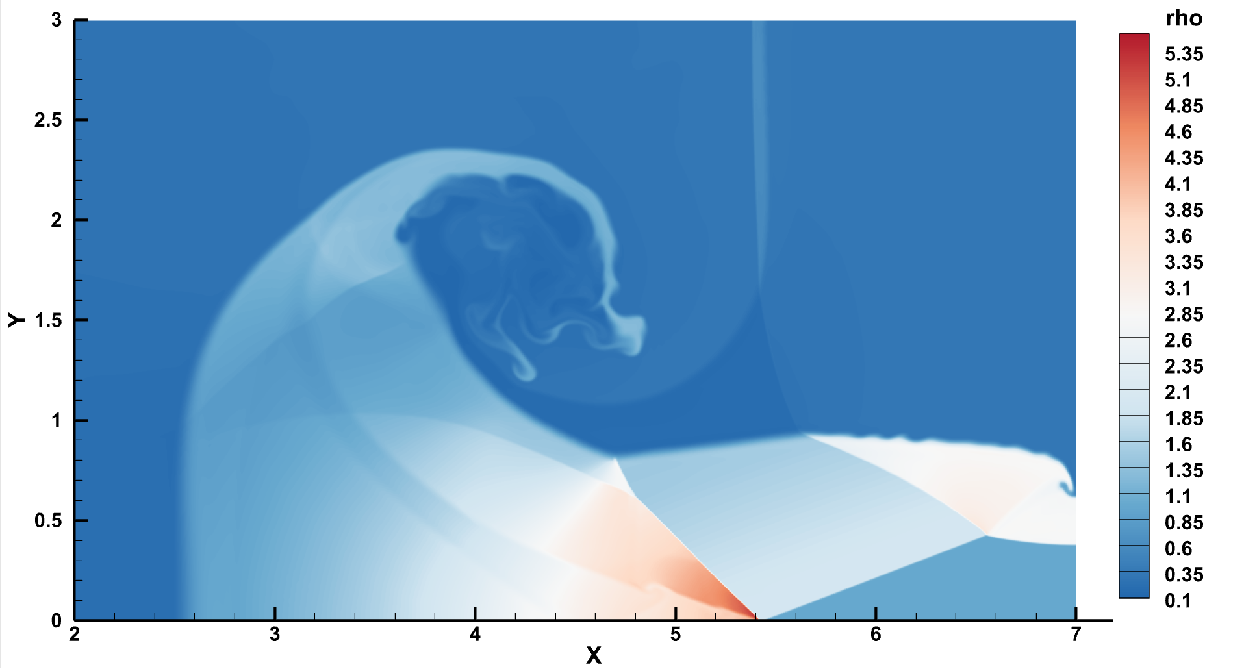} & \includegraphics[width=0.45\textwidth]{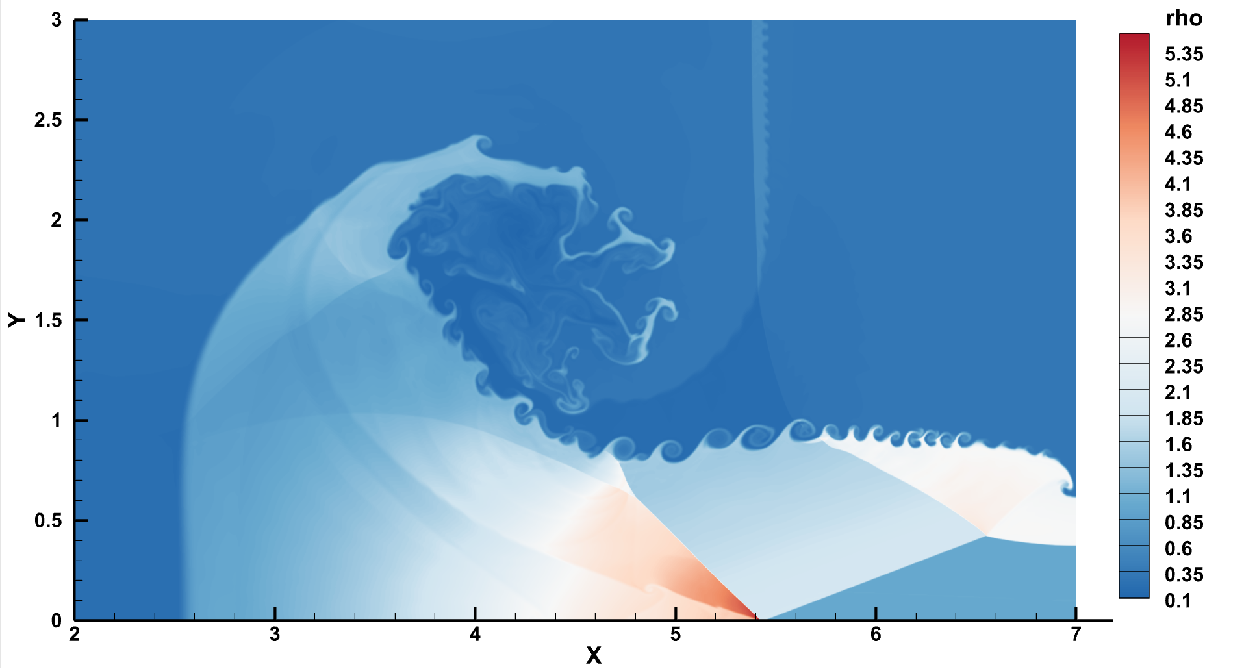}
\end{tabular}
\caption{Triple-point problem, 3O total density contours for $1400\times600$ and $2800\times1200$ grids at: Top) $t$= 3.5,  Bottom) $t$= 5}
\label{fig:2d_euler_tp_contours}
\end{figure}
Flow tangency ({\em i.e.} inviscid wall) conditions are applied at all four boundaries. The computational domain is discretized using Cartesian grid, with equal grid size along the $x$ and $y$ directions ({\em i.e.}, $\Delta x$= $\Delta y$). CFL no. taken is 0.8. For this problem, the high pressure in region I creates a shock wave which moves to the right through regions II and III. As density is lower in region II, the shock travels faster in region II than in region III. As a result, a vortex forms which swirls the components in all three regions around the triple point. Figure \ref{fig:2d_euler_tp_contours} shows the third order accurate total density contours at times $t$=3.5 and 5 for $1400\times600$ and $2800\times1200$ grids. With a finer grid, the Kelvin-Helmholtz instability is more clearly observed.  

\subsection{2D Euler test case: Shock-bubble interaction}
\begin{figure}[h!] 
\centering
\begin{tikzpicture}[scale=1.2]
\small\begin{axis}
 [every axis plot post/.append style={
  mark=none,domain=0:50,samples=50,smooth},
  xmin=0,xmax=50,ymin=0,ymax= 20,
	axis x line*=bottom, 
  axis y line*=left,
	axis line style={draw=none},
	xtick={},
	unit vector ratio*=1 1 1,
	xticklabels={\empty},
	yticklabel={\empty},
	tick style={draw=none},
	]
\draw [thick](axis cs:2, 2) rectangle (axis cs:48,18);
\draw[black,thick,dashed,-] (axis cs:2, 10) --(axis cs:48,10);
\draw[black,thick](axis cs:40, 2) --(axis cs:40,18);
\path (axis cs:30,10) coordinate (A);       
\path (axis cs:33.5,10) coordinate (B);
\draw[thick]  let \p1=(A),\p2=(B), \n1={veclen(\x2-\x1,\y2-\y1)} in (A) circle [radius=\n1];
\node at (axis cs:1, 19){$A$};
\node at (axis cs:49, 19){$B$};
\node at (axis cs:49, 1){$C$};
\node at (axis cs:1, 1){$D$};
\node at (axis cs:49, 10){$C'$};
\node at (axis cs:3.2, 11.2){$D'$};
\node at (axis cs:21, 13){\textbf{I}};
\node at (axis cs:30, 11.5){\textbf{II}};
\node at (axis cs:44, 13){\textbf{III}};
\draw[black,thick,<->] (axis cs:2, 0.5) --(axis cs:48,0.5)node[font=\tiny,midway,fill=white]{445};
\draw[black,thick,<->] (axis cs:2, 16.5) --(axis cs:40,16.5)node[font=\tiny,midway,fill=white]{275};
\draw[black,thick,<->] (axis cs:2, 8.5) --(axis cs:30,8.5)node[font=\tiny,midway,fill=white]{225};
\draw[black,thick,<->] (axis cs:0.5, 2) --(axis cs:0.5,18)node[font=\tiny,midway,sloped,fill=white]{89};
\draw[black,thick,<->] (axis cs:26.5, 15) --(axis cs:33.5,15)node[font=\tiny,midway,fill=white]{50};
\draw[thick,-stealth](axis cs:42, 4) --(axis cs:38,4)node [font=\tiny,anchor=east]{$M_{s}$=1.22};
\end{axis}
\end{tikzpicture}
\caption{Schematic of computational domain for shock-bubble interaction tests (not to scale). Lengths in $mm$}  
\label{fig:2d_shock_bubble_schematic}%
\end{figure}
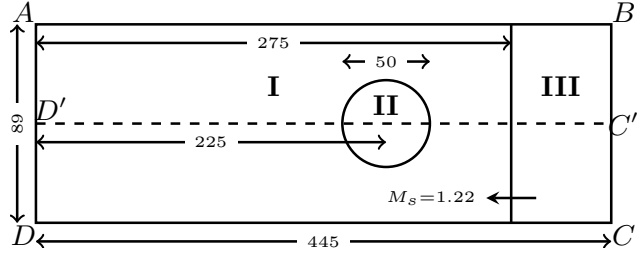

\begin{figure}[!h]
\captionsetup[subfigure]{labelformat=empty}
\centering
\begin{subfigure}{.9\textwidth}
\centering
\resizebox{.9\textwidth}{!}{
\includegraphics[height=3cm]{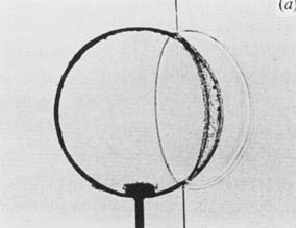}
\quad
\includegraphics[height=3cm]{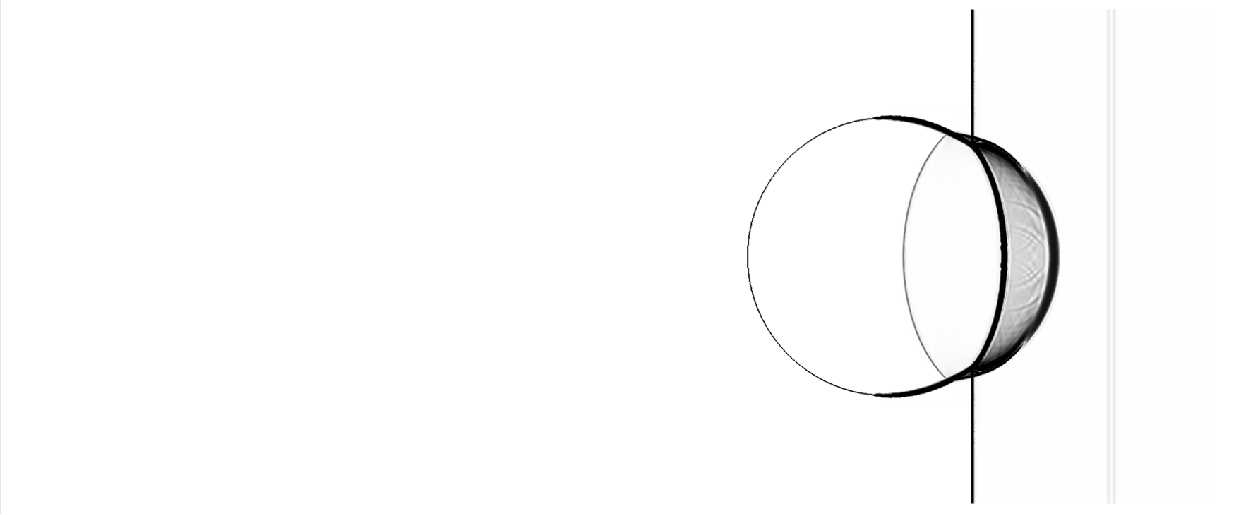}
}
\caption{t= 23 (32) $\mu$s}
\end{subfigure}

\begin{subfigure}{.9\textwidth}
\centering
\resizebox{.9\textwidth}{!}{
\includegraphics[height=3cm]{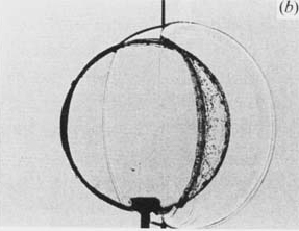}
\quad
\includegraphics[height=3cm]{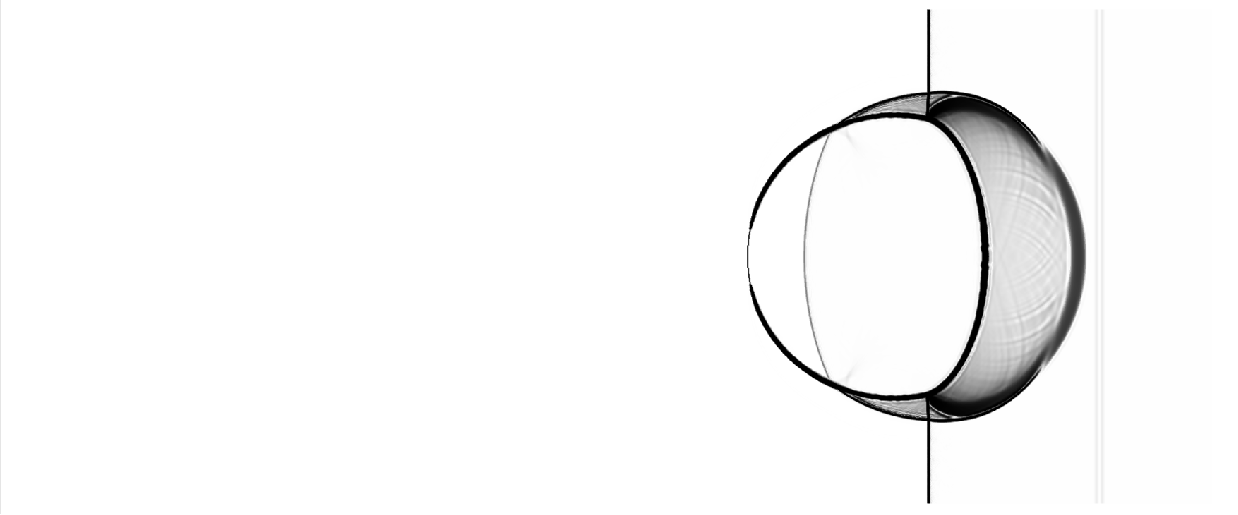}
}
\caption{t= 42 (52) $\mu$s}
\end{subfigure}

\begin{subfigure}{.9\textwidth}
\centering
\resizebox{.9\textwidth}{!}{
\includegraphics[height=3cm]{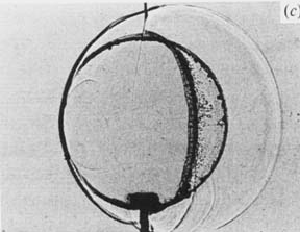}
\quad
\includegraphics[height=3cm]{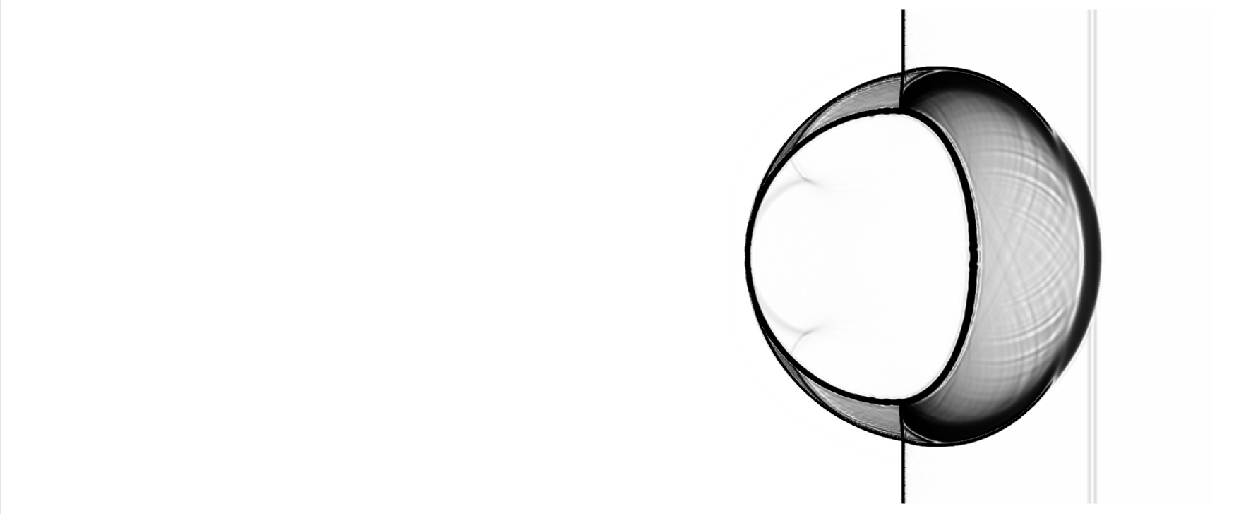}
}
\caption{t= 53 (62) $\mu$s}
\end{subfigure}

\begin{subfigure}{.9\textwidth}
\centering
\resizebox{.9\textwidth}{!}{
\includegraphics[height=3cm]{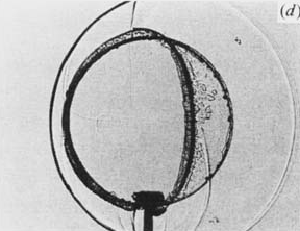}
\quad
\includegraphics[height=3cm]{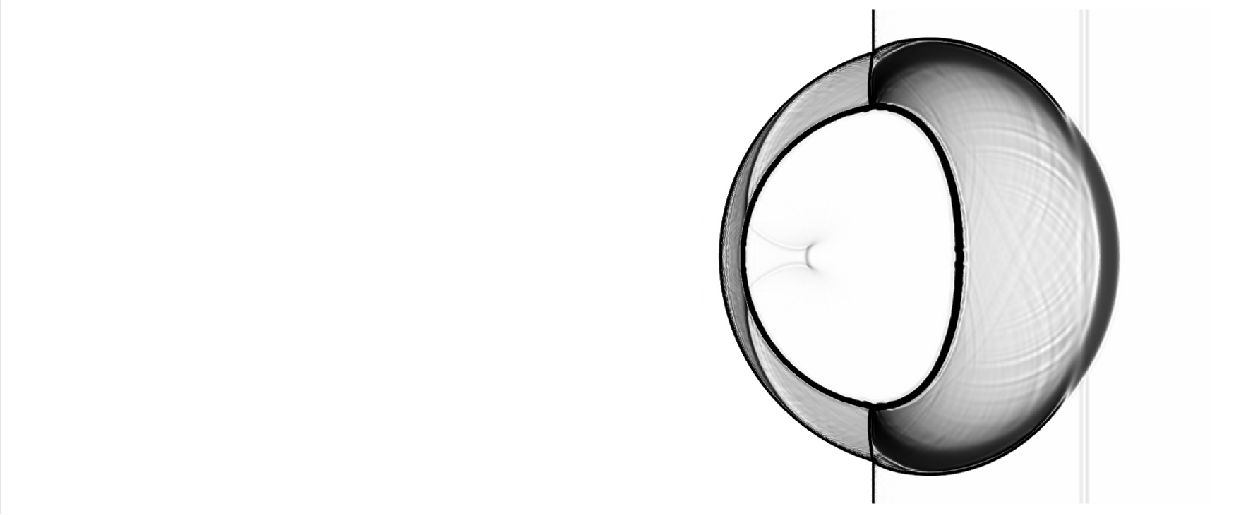}
}
\caption{t= 66 (72) $\mu$s}
\end{subfigure}

\begin{subfigure}{.9\textwidth}
\centering
\resizebox{.9\textwidth}{!}{
\includegraphics[height=3cm]{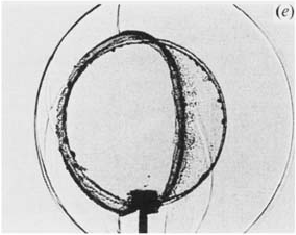}
\quad
\includegraphics[height=3cm]{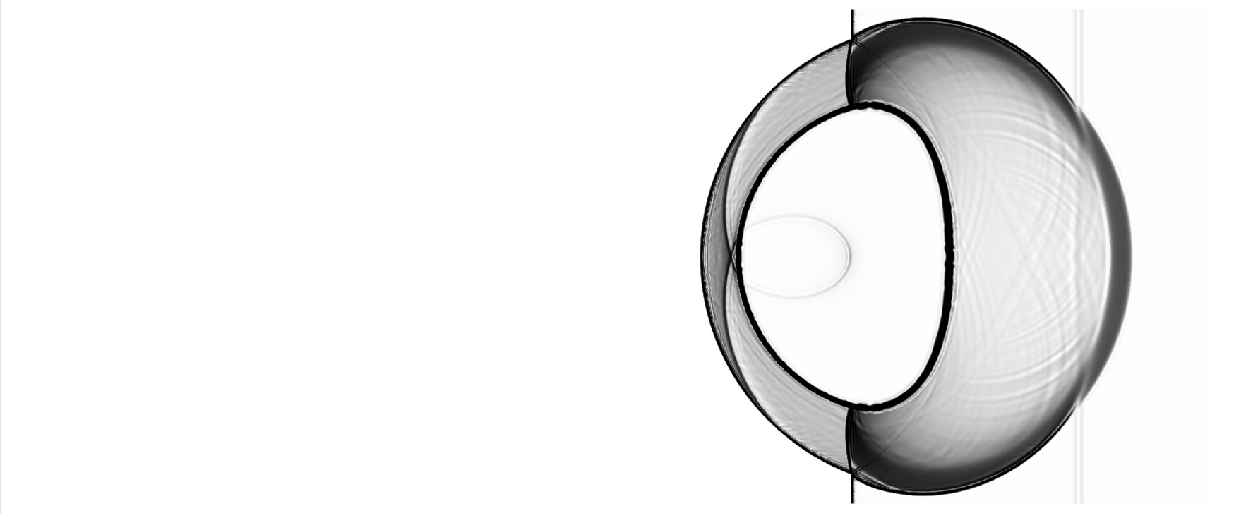}
}
\caption{t= 75 (82) $\mu$s}
\end{subfigure}
\caption{Schlieren-type images of density for the Shock-Helium bubble interaction problem. LEFT: experimental results \cite{Haas_Sturtevant_1987}, RIGHT: $3O$ accurate numerical results. Times in parentheses are the corresponding times in experiment.}
\label{fig:2d_shock_He_bubble_sch1}
\end{figure}

\begin{figure}[!h]
\captionsetup[subfigure]{labelformat=empty}
\centering
\begin{subfigure}{.9\textwidth}
\centering
\resizebox{.9\textwidth}{!}{
\includegraphics[height=3cm]{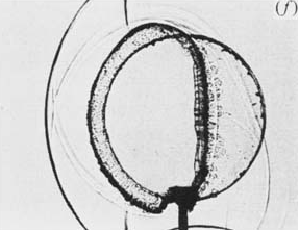}
\quad
\includegraphics[height=3cm]{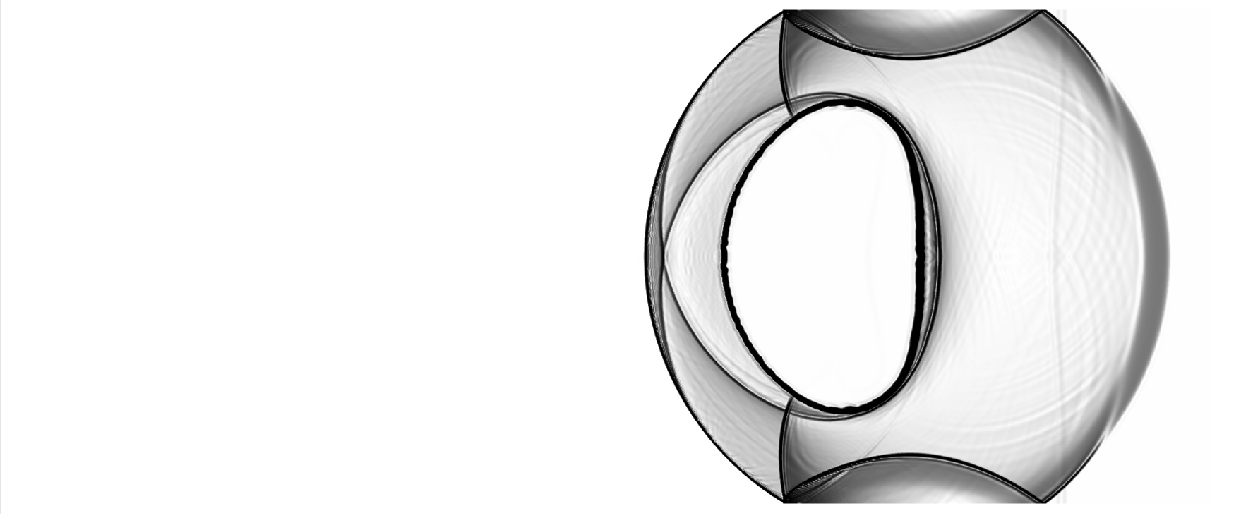}
}
\caption{t= 102 (102) $\mu$s}
\end{subfigure}

\begin{subfigure}{.9\textwidth}
\centering
\resizebox{.9\textwidth}{!}{
\includegraphics[height=3cm]{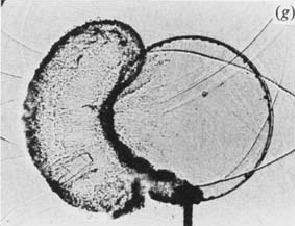}
\quad
\includegraphics[height=3cm]{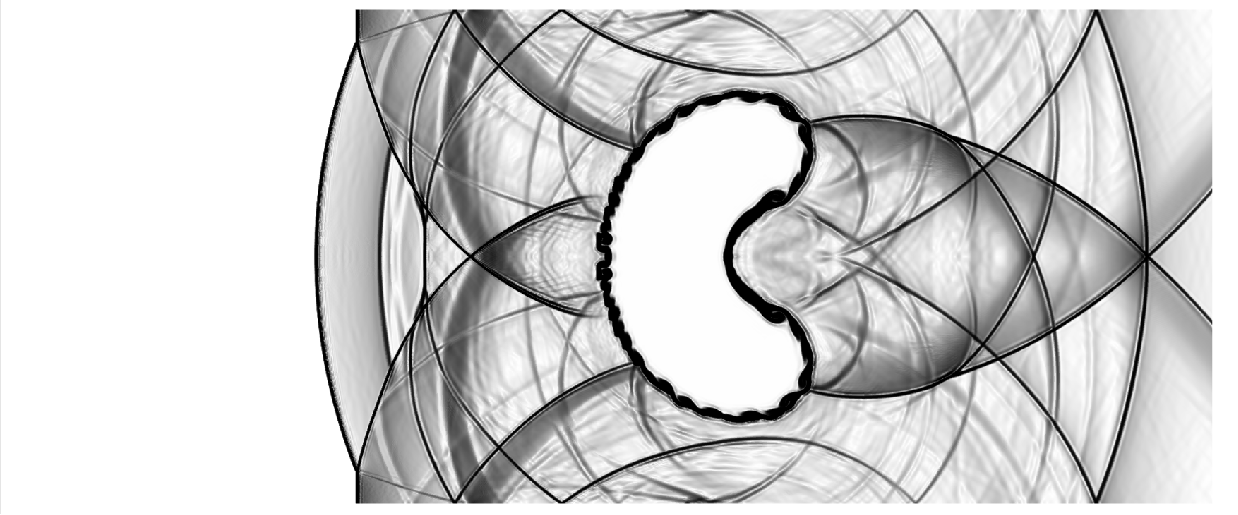}
}
\caption{t= 260 (245) $\mu$s}
\end{subfigure}

\begin{subfigure}{.9\textwidth}
\centering
\resizebox{.9\textwidth}{!}{
\includegraphics[height=3cm]{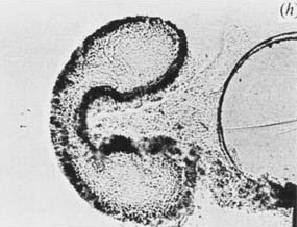}
\quad
\includegraphics[height=3cm]{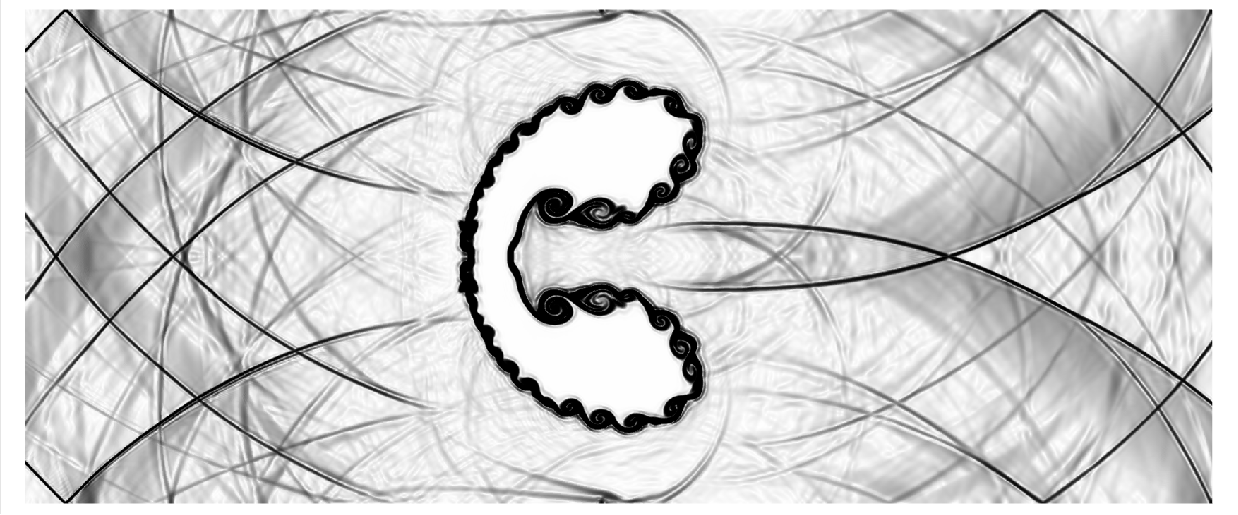}
}
\caption{t= 445 (427) $\mu$s}
\end{subfigure}

\begin{subfigure}{.9\textwidth}
\centering
\resizebox{.9\textwidth}{!}{
\includegraphics[height=3cm]{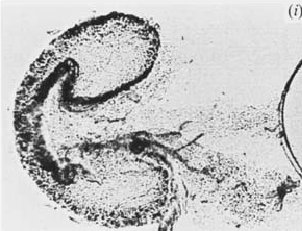}
\quad
\includegraphics[height=3cm]{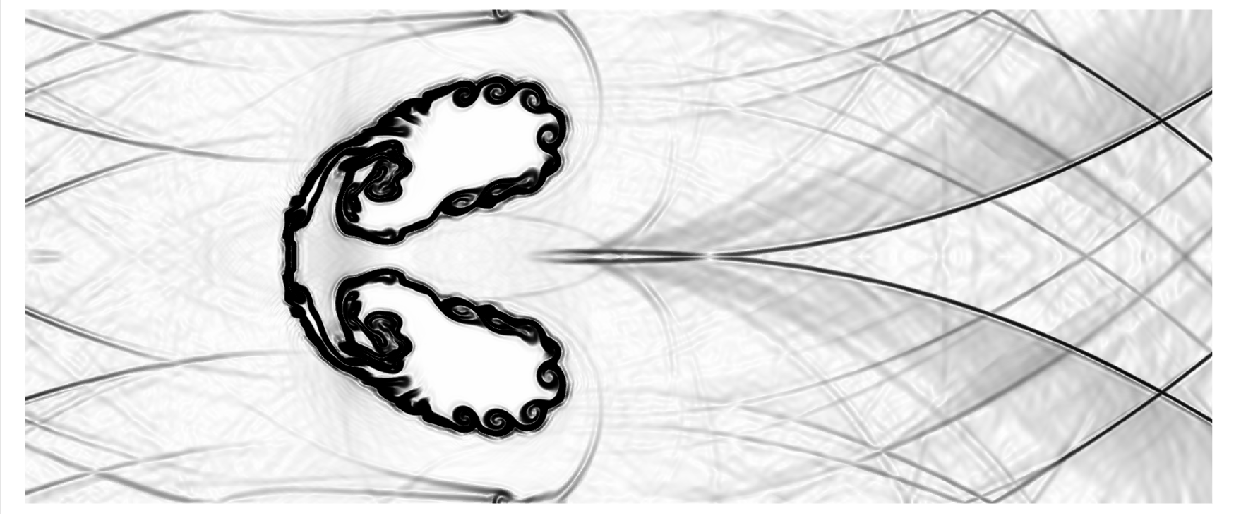}
}
\caption{t= 674 $\mu$s}
\end{subfigure}

\begin{subfigure}{.9\textwidth}
\centering
\resizebox{.9\textwidth}{!}{
\includegraphics[height=3cm]{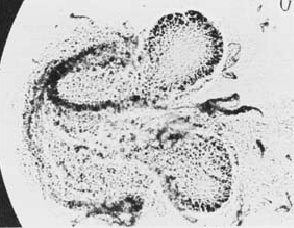}
\quad
\includegraphics[height=3cm]{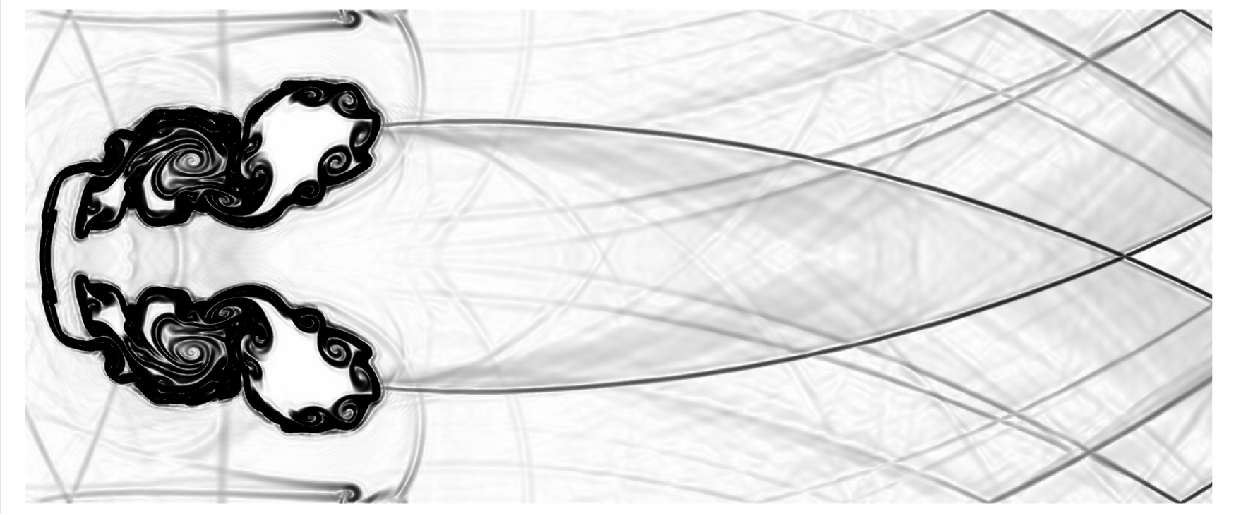}
}
\caption{t= 983 $\mu$s}
\end{subfigure}

\caption{Schlieren-type images of density for the Shock-Helium bubble interaction problem. LEFT: experimental results \cite{Haas_Sturtevant_1987}, RIGHT: $3O$ accurate numerical results. Times in parentheses are the corresponding times in experiment.}
\label{fig:2d_shock_He_bubble_sch2}
\end{figure}

\begin{figure}[h!]
\captionsetup[subfigure]{labelformat=empty}
\centering
\begin{subfigure}{.9\textwidth}
\centering
\resizebox{.9\textwidth}{!}{
\includegraphics[height=3cm]{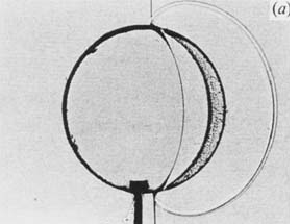}
\quad
\includegraphics[height=3cm]{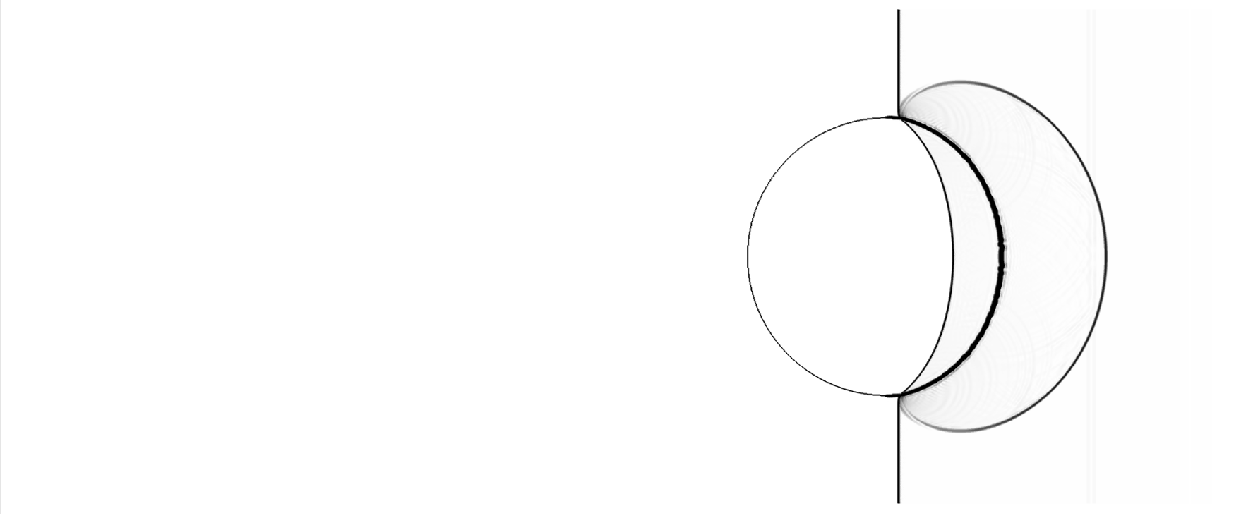}
}
\caption{t= 55 $\mu$s}
\end{subfigure}

\begin{subfigure}{.9\textwidth}
\centering
\resizebox{.9\textwidth}{!}{
\includegraphics[height=3cm]{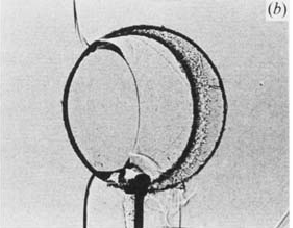}
\quad
\includegraphics[height=3cm]{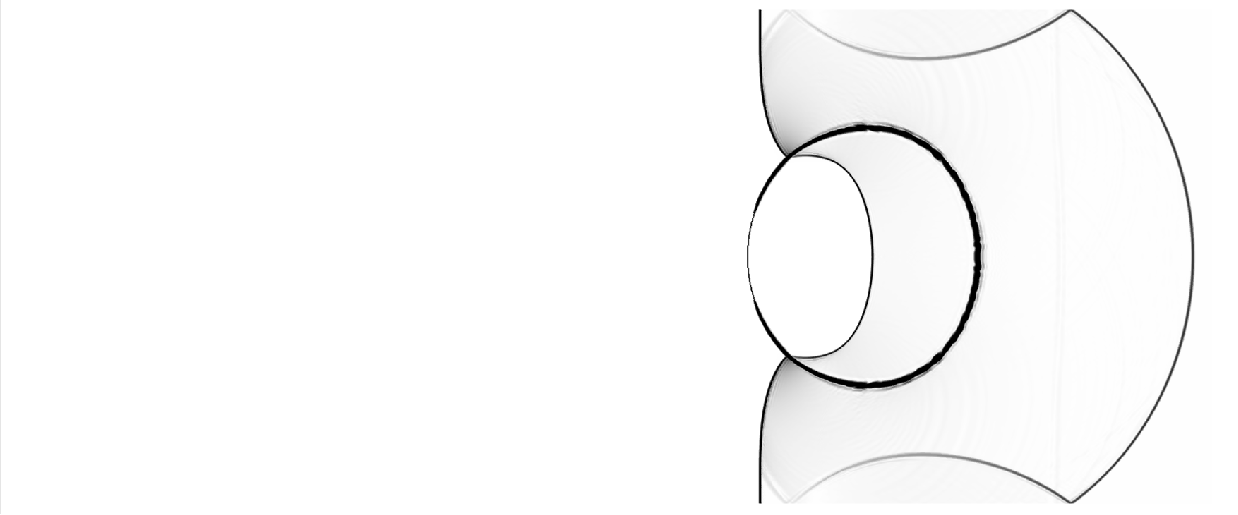}
}
\caption{t= 115 $\mu$s}
\end{subfigure}

\begin{subfigure}{.9\textwidth}
\centering
\resizebox{.9\textwidth}{!}{
\includegraphics[height=3cm]{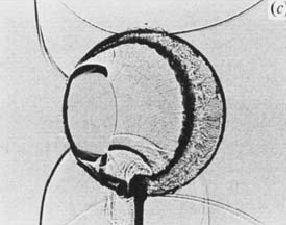}
\quad
\includegraphics[height=3cm]{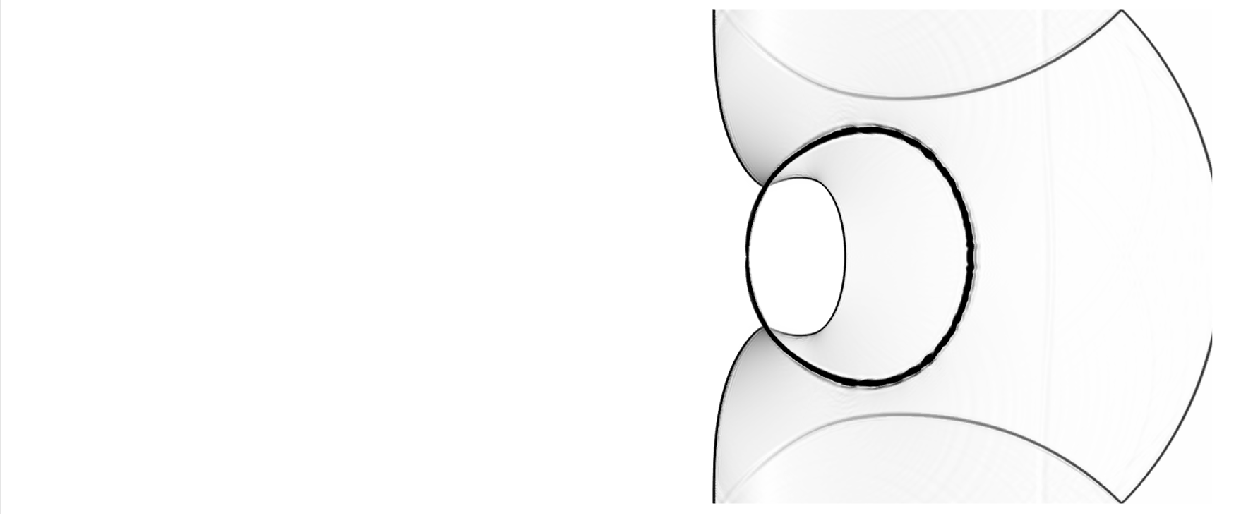}
}
\caption{t= 135 $\mu$s}
\end{subfigure}

\begin{subfigure}{.9\textwidth}
\centering
\resizebox{.9\textwidth}{!}{
\includegraphics[height=3cm]{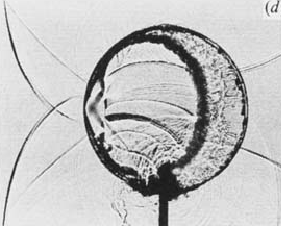}
\quad
\includegraphics[height=3cm]{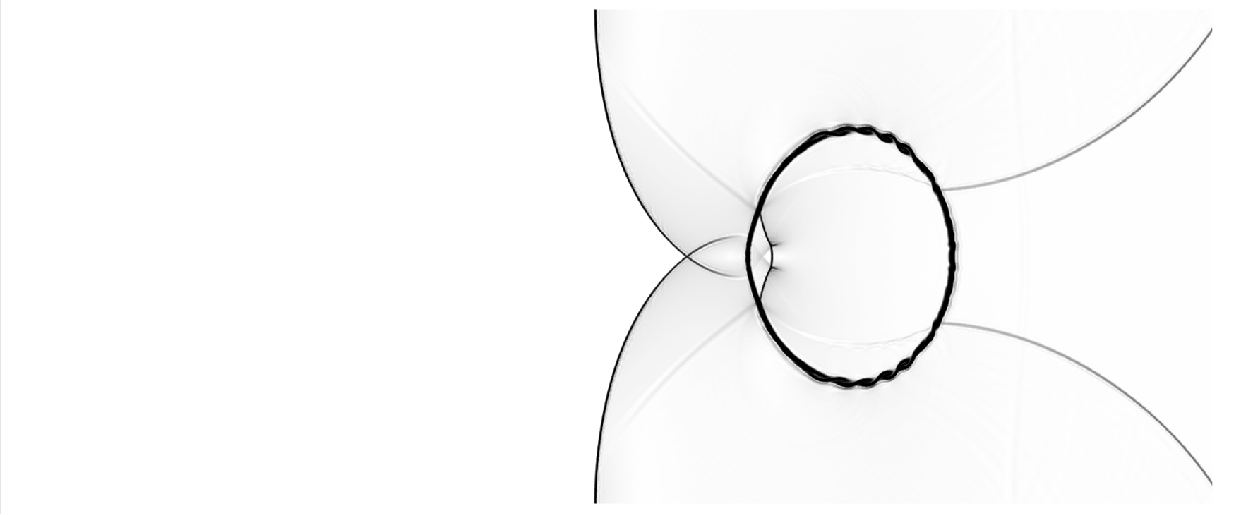}
}
\caption{t= 187 $\mu$s}
\end{subfigure}

\begin{subfigure}{.9\textwidth}
\centering
\resizebox{.9\textwidth}{!}{
\includegraphics[height=3cm]{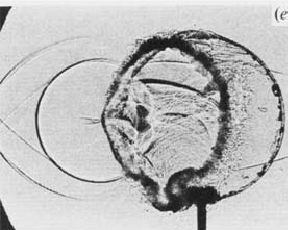}
\quad
\includegraphics[height=3cm]{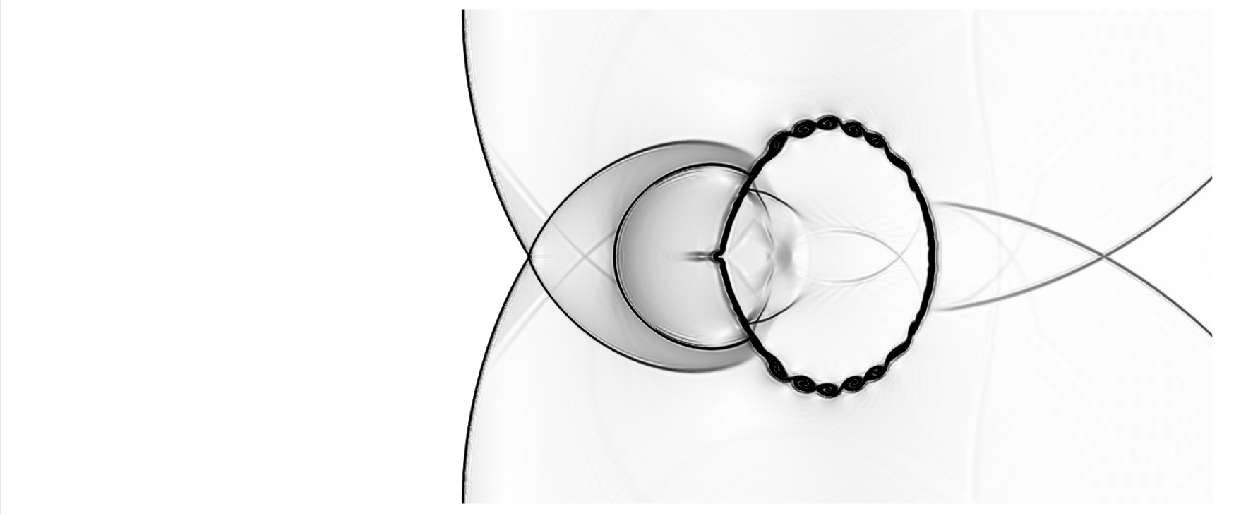}
}
\caption{t= 247 $\mu$s}
\end{subfigure}

\caption{Schlieren-type images of density for the Shock-R22 bubble interaction problem. LEFT: experimental results \cite{Haas_Sturtevant_1987}, RIGHT: $3O$ accurate numerical results.}
\label{fig:2d_shock_R22_bubble_sch1}
\end{figure}

\begin{figure}[h!]
\captionsetup[subfigure]{labelformat=empty}
\centering
\begin{subfigure}{.9\textwidth}
\centering
\resizebox{.9\textwidth}{!}{
\includegraphics[height=3cm]{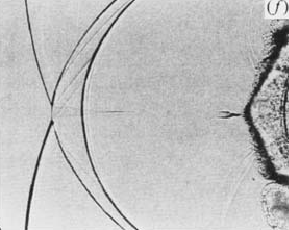}
\quad
\includegraphics[height=3cm]{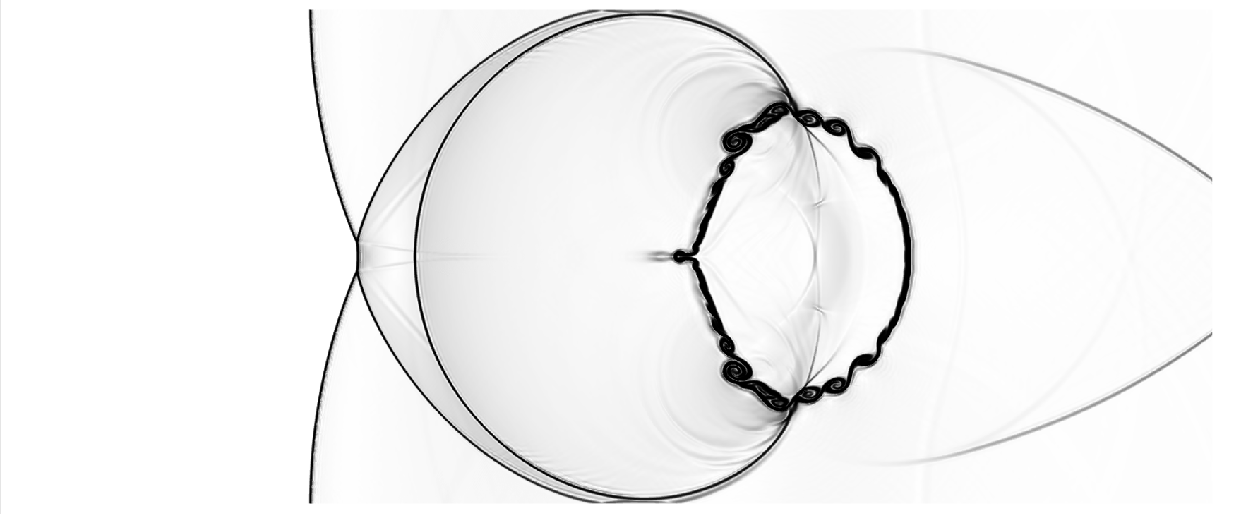}
}
\caption{t= 318 $\mu$s}
\end{subfigure}

\begin{subfigure}{.9\textwidth}
\centering
\resizebox{.9\textwidth}{!}{
\includegraphics[height=3cm]{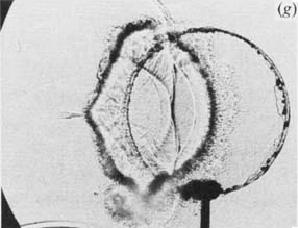}
\quad
\includegraphics[height=3cm]{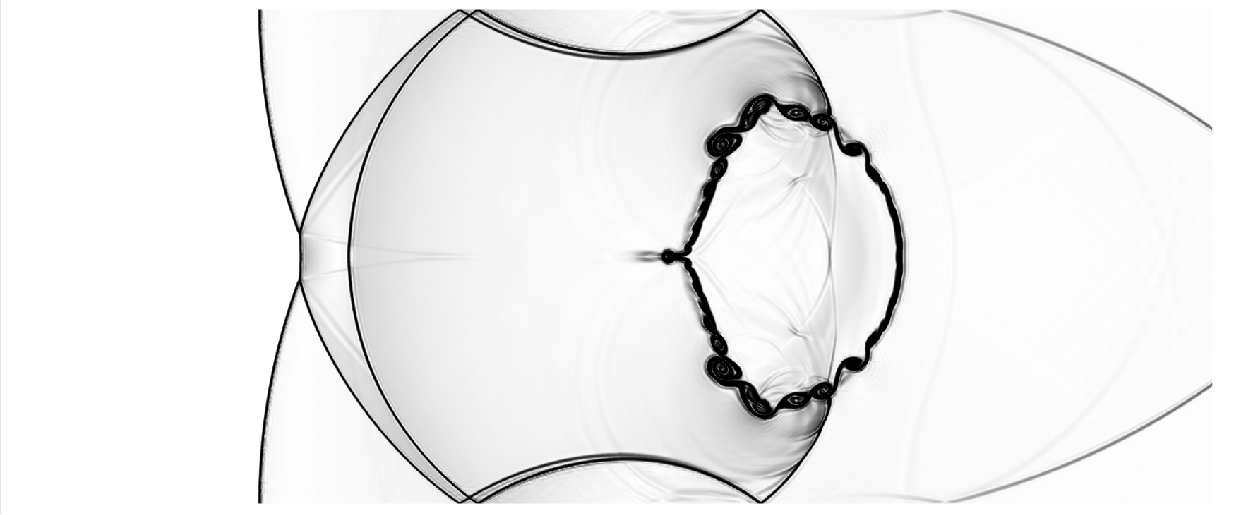}
}
\caption{t= 342 $\mu$s}
\end{subfigure}

\begin{subfigure}{.9\textwidth}
\centering
\resizebox{.9\textwidth}{!}{
\includegraphics[height=3cm]{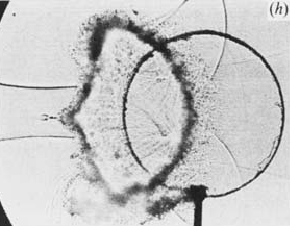}
\quad
\includegraphics[height=3cm]{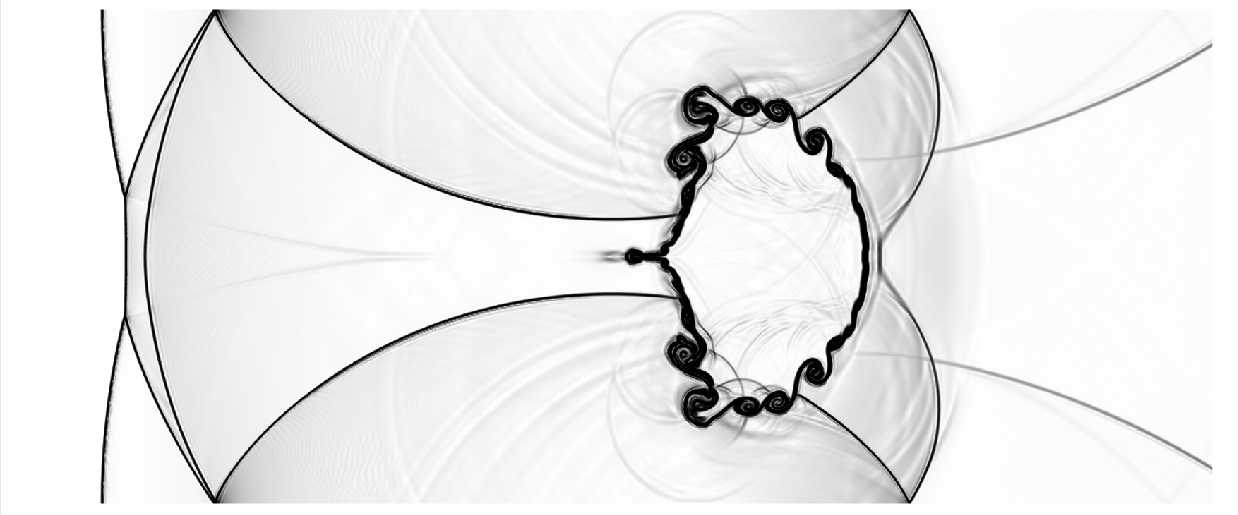}
}
\caption{t= 417 $\mu$s}
\end{subfigure}

\begin{subfigure}{.9\textwidth}
\centering
\resizebox{.9\textwidth}{!}{
\includegraphics[height=3cm]{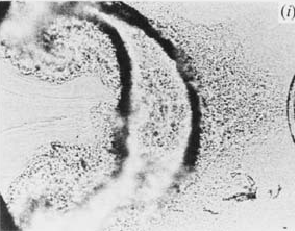}
\quad
\includegraphics[height=3cm]{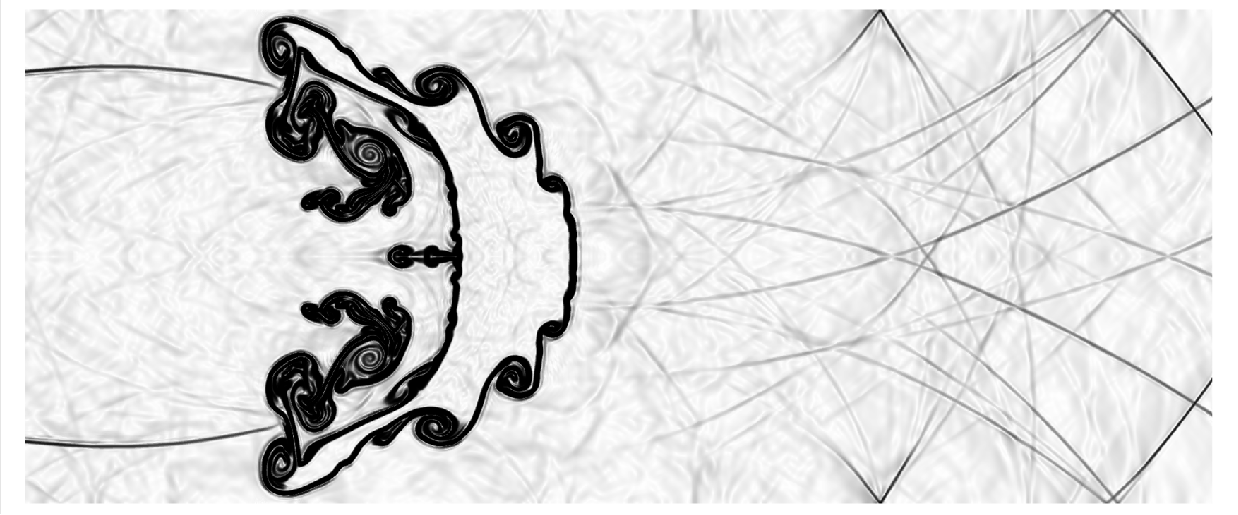}
}
\caption{t= 1020 $\mu$s}
\end{subfigure}

\caption{Schlieren-type images of density for the Shock-R22 bubble interaction problem. LEFT: experimental results \cite{Haas_Sturtevant_1987}, RIGHT: 3O accurate numerical results.}
\label{fig:2d_shock_R22_bubble_sch2}
\end{figure}

Numerical tests are performed for two 2D problems which simulate the interaction of a planar weak shock with a cylindrical inhomogeneity. Specifically, these numerical tests are based on the experiments performed by Haas and Sturtevant \cite{Haas_Sturtevant_1987}, which comprise a planar Mach 1.22 shock moving through air and impinging on a cylindrical bubble. The bubble consists of either Helium or Refrigerant 22. The gas constant $R$ and adiabatic constant $\gamma$ for air, Helium, Helium+ 28 $\%$ air and R22 are tabulated in Table \ref{table:gas_prop}.   
\begin{table}%
\centering
\begin{tabular}{ |c|c|c| }
\hline
Gas component & R (KJ/K/mol) & $\gamma$ \\
\hline
air & 0.2867 & 1.4\\
He & 2.0768 &1.667\\
72$\%$ He+ 28$\%$ air & 1.5768& 1.645\\
R22& 0.0914& 1.249\\
\hline
\end{tabular}
\caption{Gas properties used for computations.}
\label{table:gas_prop}
\end{table}
Figure \ref{fig:2d_shock_bubble_schematic} shows a schematic of the computational domain for the two tests. 
The rectangular domain $ABCD$ with dimensions of 445 $mm$ $\times$ 89 $mm$ has a planar Mach 1.22 shock located at $x_{0}$= 275 $mm$ at initial time traveling left into quiescent air (region I). For region I, we take pressure $p_{\text{I}}$= 101325 $Pa$, density $\rho_{\text{I}}$= 1.225 $kg/m^{3}$, and thus a sound speed of 340.3 $m/s$. The flow conditions in region III are computed using compressible flow relations for a moving normal shock. A cylindrical bubble (region II) with a radius of 25 $mm$ is located at $x$= 225 $mm$ along the center-line $C'D'$. The bubble consists of either refrigerant R22 ($CHClF_{2}$) or Helium ($He$) with 28 $\%$ air contamination by mass. At initial time, the bubble (region II) is at rest and is assumed to be in thermal and mechanical equilibrium with the surrounding air (region I), meaning both regions share the same temperature $T$ and pressure $p$. Thus, at initial time,
\begin{equation}
\rho_{\text{I}}R_{\text{I}}T= \rho_{\text{II}}R_{II}T \Rightarrow\rho_{\text{II}}= \rho_{\text{I}}\frac{R_{\text{I}}}{R_{\text{II}}}
\label{eq:2d_euler_sb_1}
\end{equation}  
Since flow is symmetric about $C'D'$, only upper half of the domain ({\em i.e.}, $ABC'D'$) is considered for computational domain and is discretized using a Cartesian grid into $4000\times400$ square cells. Reflecting conditions are applied at the top and bottom boundaries. Extrapolation condition is applied at the left and right boundaries. CFL no ($\sigma$) of 0.8 is used. Third order accurate numerical results are obtained at different times $t$ from when the shock first arrives at the upstream end of the bubble. The contours of numerical Schlieren of total density are plotted alongside Haas and Sturtevant's experimental images in Figures \ref{fig:2d_shock_He_bubble_sch1}, \ref{fig:2d_shock_He_bubble_sch2}, \ref{fig:2d_shock_R22_bubble_sch1} and \ref{fig:2d_shock_R22_bubble_sch2}. The numerical schlieren $\phi$ is computed using the gradient of density by applying the following formula (see \cite{GOUASMI2020112912}).
\begin{subequations}
\label{eq:2d_euler_sb_sch}
\begin{equation}
\phi= exp\left(-K|\nabla \rho|/|\nabla \rho|_{max}\right)\text{, where}
\end{equation}
\begin{equation}
\nabla \rho = \frac{\partial \rho}{\partial x_{1}}\widehat{e}_{1}+ \frac{\partial \rho}{\partial x_{2}}\widehat{e}_{2}, \ \ K= 10 \ W+ 150 \ (1-W)
\end{equation}
\end{subequations}

\subsubsection{Shock-Helium bubble interaction}
The incident shock striking the upstream interface of the Helium bubble leads to the formation of a reflected wave outside the bubble and a refracted shock wave inside the bubble. The divergent refracted shock travels downstream faster than the incident shock outside, due to the higher speed of sound inside the bubble than in the air outside. In the experiments of Haas and Sturtevant, the speed of the refracted shock and the speed of sound inside the bubble is observed to be lower than that expected from one-dimensional gas dynamics for pure Helium. This leads them to estimate that the Helium bubble has 28 $\%$ air contamination by mass. However, even with this correction, a mismatch in the corresponding experimental and numerical times is noted by Marquina and Mulet \cite{MARQUINA2003120}. They attribute this mismatch to a possible non-uniform contamination of the Helium bubble by air.

\begin{figure}%
\centering
\begin{minipage}{0.55\textwidth}
\centering
\includegraphics[width=0.9\linewidth]{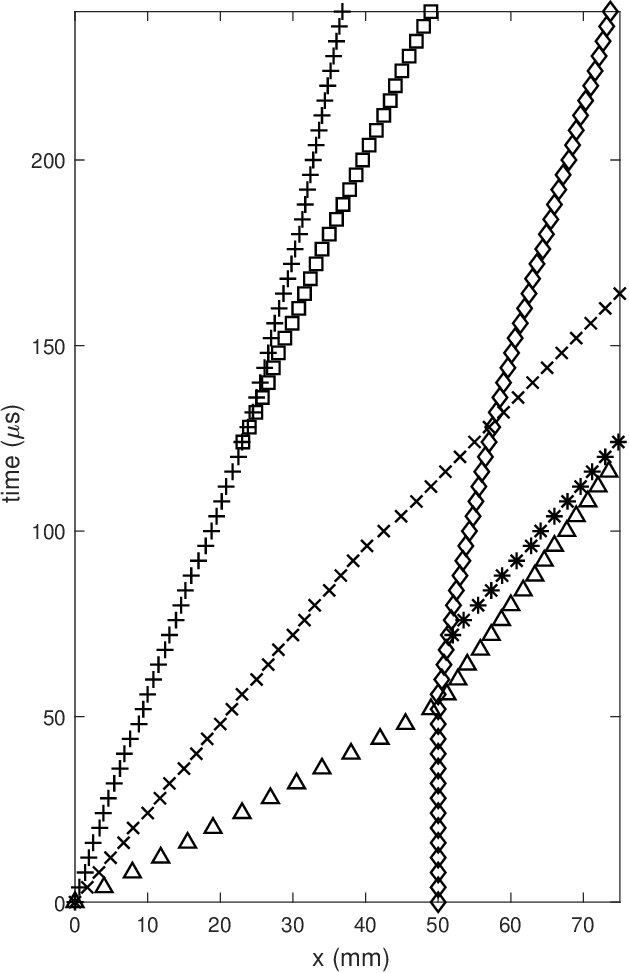}
\end{minipage}
\begin{minipage}{0.4\textwidth}
\centering
\includegraphics[width=0.9\linewidth]{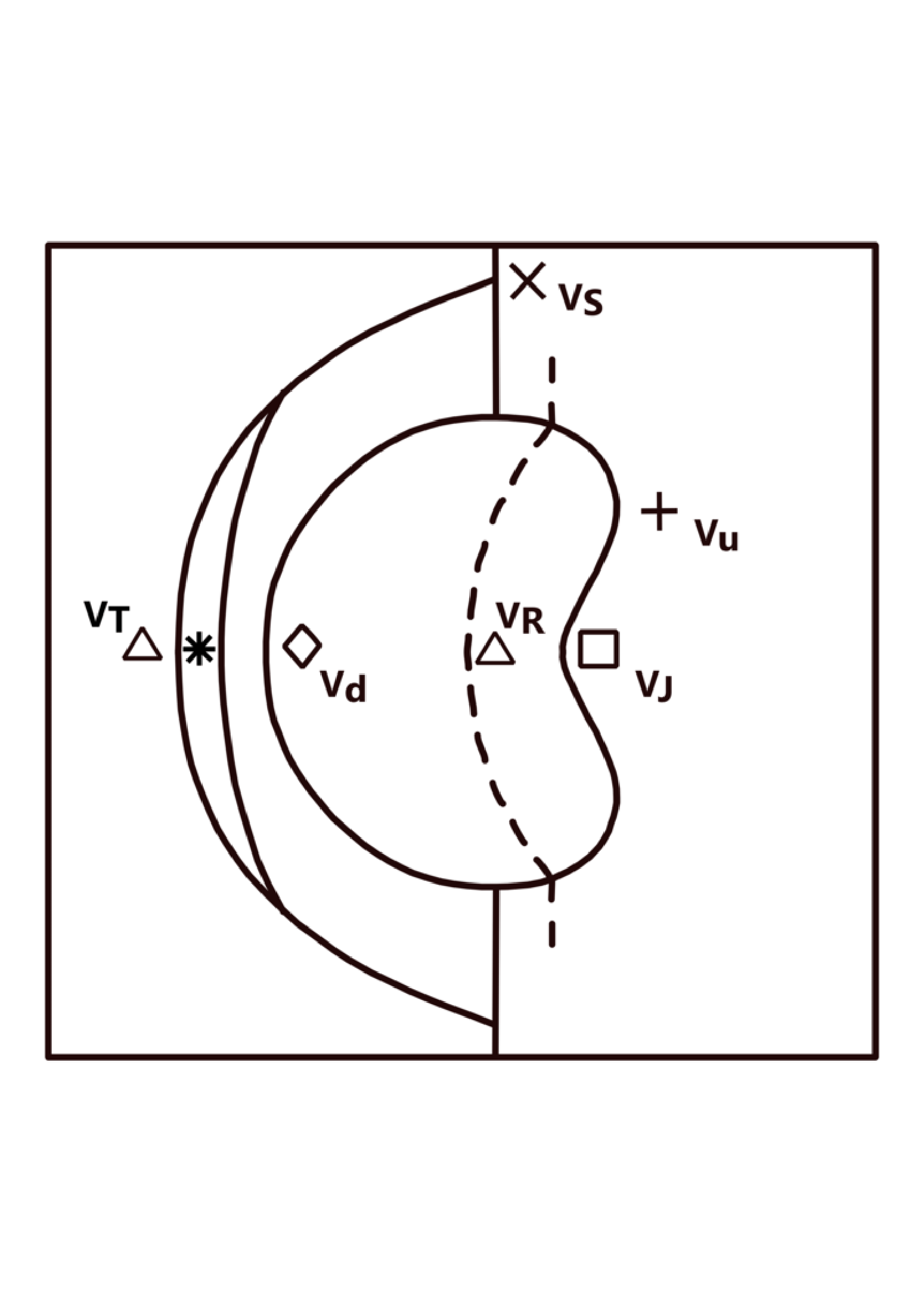}
\end{minipage}
\caption{Shock-Helium bubble interaction problem: t-x diagram (LEFT) of the features shown in the schematic diagram (RIGHT). $V_{S}$: incident shock, $V_{R}$: refracted shock, $V_{T}$: transmitted shock, $V_{u}$: upstream edge of the bubble, $V_{d}$: downstream edge of the bubble, $V_{J}$: air jet head.}%
\label{fig:2d_shock_He_bubble_tx}
\end{figure}

As the refracted shock travels downstream inside the bubble, it also interacts with the bubble interface, leading to internally reflected waves forming inside the bubble. By time $t$= 53 $\mu s$, the refracted shock has reached the downstream interface of the bubble. It then emerges out of the bubble as transmitted shock. Meanwhile, the upstream interface of the bubble is flattening. By $t$= 260 $\mu s$, the bubble has taken a kidney shape. The positions of the incident shock, refracted/ transmitted shock and upstream and downstream ends of the bubble are recorded at regular time intervals. The $t$-$x$ plot for these features is shown in Figure \ref{fig:2d_shock_He_bubble_tx}. Using linear regression, the average velocity of these features is computed and tabulated in Table \ref{table:velocities_He}. The results match well with the experimental data as well as with some popular numerical results.

\begin{table}[!h] 
\centering
\begin{tabular}{ |c|c|c|c|c|c|c|c| }
\hline
 & $V_{S}$ & $V_{R}$ & $V_{T}$ & $V_{ui}$ & $V_{uf}$ & $V_{d}$ & $V_{J}$ \\
\hline
Our Kinetic Model & 415 & 942 & 375 & 180 & 112 & 149 & 225\\
Marquina and Mulet \cite{MARQUINA2003120} & 414 & 943 & 373 & 176 & 111 & 153 & 229\\
Quirk and Karni \cite{Quirk_Karni_1996} & 422 & 943 & 377 & 178 & & 146 & 227\\
Haas and Sturtevant \cite{Haas_Sturtevant_1987} & 410 & 900 & 393 & 170 & 113 & 145 & 230\\
\hline
\end{tabular}
\caption{Velocities of features described in Figure \ref{fig:2d_shock_He_bubble_tx}. The time interval for computing each velocity are: $V_{S}$[0,60], $V_{R}$[0,52], $V_{T}$[52,240], $V_{ui}$[0,52], $V_{uf}$[140,240], $V_{d}$[140,240], $V_{J}$[140,240].}
\label{table:velocities_He}
\end{table}


\subsubsection{Shock-R22 bubble interaction}
\begin{figure}%
\centering
\begin{minipage}{0.55\textwidth}
\centering
\includegraphics[width=0.9\linewidth]{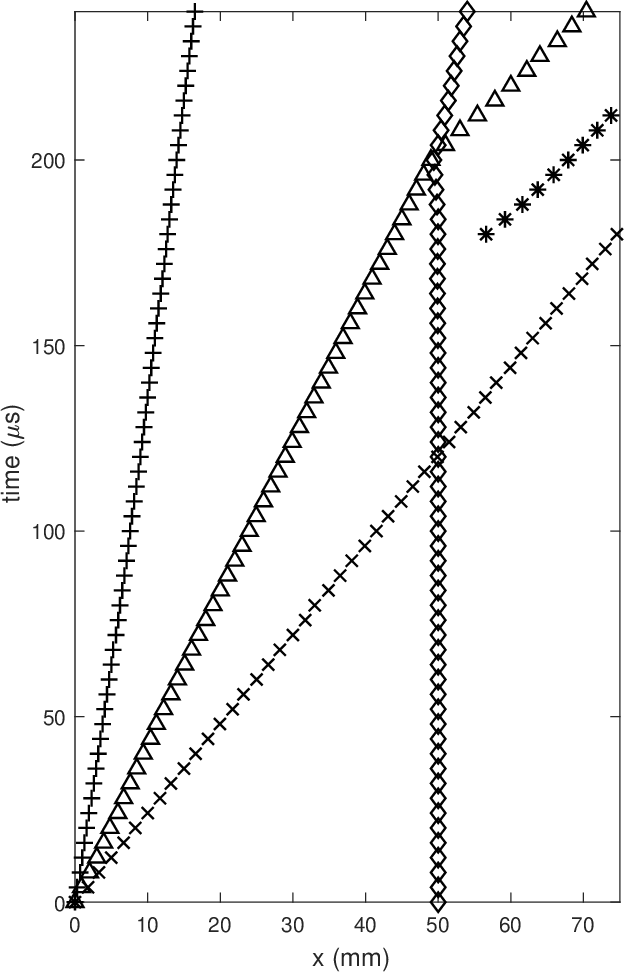}
\end{minipage}
\begin{minipage}{0.4\textwidth}
\centering
\includegraphics[width=0.9\linewidth]{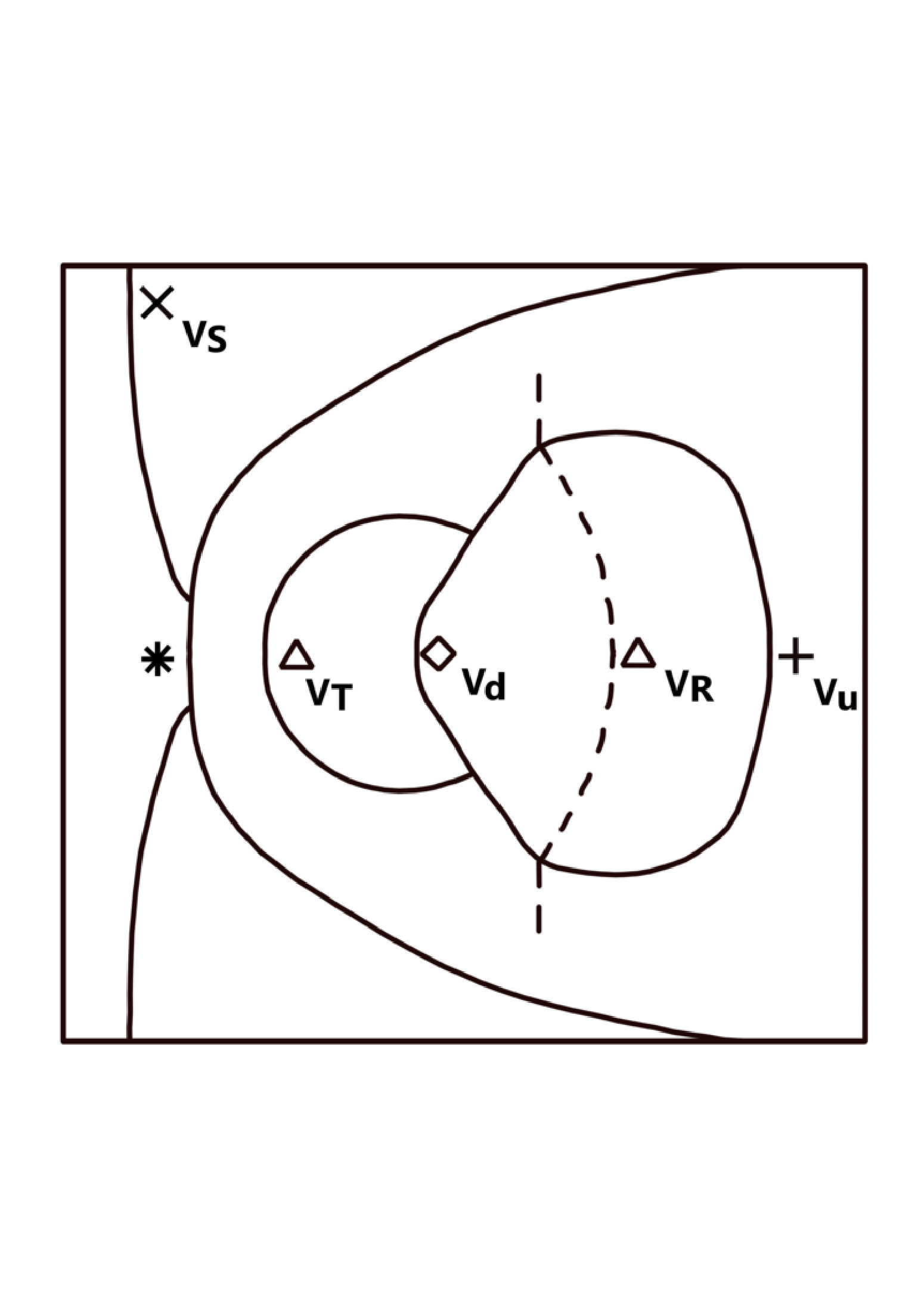}
\end{minipage}
\caption{Shock-R22 bubble interaction problem: t-x diagram (LEFT) of the features shown in the schematic diagram (RIGHT). $V_{S}$: incident shock, $V_{R}$: refracted shock, $V_{T}$: transmitted shock, $V_{u}$: upstream edge of the bubble, $V_{d}$: downstream edge of the bubble.}%
\label{fig:2d_shock_R22_bubble_tx}%
\end{figure}

After the planar shock strikes the upstream interface of the R22 bubble, a convergent refracted shock is formed inside the bubble. This refracted shock travels slower than the incident shock outside, due to the lower speed of sound inside the bubble than that outside. In their experiments, Haas and Sturtevant observed that the experimentally obtained speed of sound inside the bubble is very close to the expected value from gas dynamics. Thus, air contamination for the R22 bubble case is neglected (see also \cite{Quirk_Karni_1996}). As the refracted shock propagates downstream, it curves inward more. By $t$= 200 $\mu s$, the refracted shock reaches the downstream interface of the bubble, and then emerges out of the bubble as transmitted shock. The $t$-$x$ plot in Figure \ref{fig:2d_shock_R22_bubble_tx} shows the propagation of the incident shock, refracted/transmitted shock, and the upstream and downstream interfaces of the bubble. The computed mean velocities of these features have been tabulated in Table \ref{table:velocities_R22}. Our results have a good match with the experimental results of Haas and Sturtevant, as well as with the numerical results of Quirk \& Karni \cite{Quirk_Karni_1996}.

\begin{table}[p!]%
\centering
\begin{tabular}{ |c|c|c|c|c|c| }
\hline
 & $V_{S}$ & $V_{R}$ & $V_{T}$ & $V_{ui}$ & $V_{d}$ \\
\hline
Our Kinetic Model & 415 & 245 & 543 & 76 & 109\\
Quirk and Karni \cite{Quirk_Karni_1996} & 420 & 254 & 560 & 70 & 116\\
Haas and Sturtevant \cite{Haas_Sturtevant_1987} & 415 & 240 & 540 & 73 & 78(N/A)\\
\hline
\end{tabular}
\caption{Velocities of features described in Figure \ref{fig:2d_shock_R22_bubble_tx}. The time interval for computing each velocity are: $V_{S}$[0,180], $V_{R}$[0,200], $V_{T}$[204,240], $V_{ui}$[0,52], $V_{d}$[208,240].}
\label{table:velocities_R22}
\end{table}

\section{Conclusions}
We have presented kinetic models in 1D and 2D for the multi-component Euler equations that employ flexible velocities determined by local flow properties. These velocities are defined to satisfy conditions for preservation of positivity of both the component densities and the overall pressure under a CFL-like time step restriction for the first-order accurate scheme. Furthermore, the velocity definitions have been refined to ensure exact capture of steady multi-material contact discontinuities. To enhance accuracy and reduce pressure oscillations across moving contact discontinuities, the basic scheme is extended to third-order accuracy using a Chakravarthy-Osher type flux-limited approach coupled with Strong Stability Preserving Runge-Kutta (SSPRK) method. Since the resulting scheme does not strongly rely on the macroscopic eigenstructure, it remains simple and computationally inexpensive. Benchmark test cases, including shock-bubble interaction problems, have been solved and compared with results from the literature, demonstrating that the proposed scheme accurately captures the relevant flow features.   

\section*{CRediT author statement}
\textbf{Shashi Shekhar Roy}: Conceptualization, Methodology, Investigation, Software, Validation, Formal analysis, Writing - Original draft, Revised draft.\\

\textbf{S. V. Raghurama Rao}: Conceptualization, Supervision, Resources, Writing - Review \& Editing.

\section*{Declaration of competing interest}
The authors declare that they have no known financial interests or personal relationships with any other people or organizations that could influence the work presented here.



\appendix
\section{Positivity condition}
\label{appendix:a0}
We consider the requirement of positivity of the term $\left(\lambda\right)_{j+\frac{1}{2}}\textbf{U}_{j+1} - \textbf{G}_{j+1}$ for two- component 1D Euler equations. That is,
\begin{equation}
\left(\lambda\right)_{j+\frac{1}{2}}\textbf{U}_{j+1} - \textbf{G}_{j+1}=  \begin{bmatrix} G_{1} & G_{2} & G_{3} & G_{4} \end{bmatrix}^{T} \in \textbf{W}
\label{eq:a0_1}
\end{equation}
Since $\lambda\geq$0, the condition \eqref{eq:a0_1} can be restated as
\begin{equation}
\textbf{U}_{j+1} - \textbf{G}_{j+1}/\left(\lambda\right)_{j+\frac{1}{2}}=  \begin{bmatrix} U_{1} & U_{2} & U_{3} & U_{4}\end{bmatrix}^{T} \in \textbf{W}
\label{eq:a0_2}
\end{equation}
Now, the requirement of non-negativity of component densities and overall pressure in \eqref{eq:a0_2} implies that,
\begin{equation}
U_{1}\geq 0, U_{2}\geq 0 \text{ and } 2 U_{2}U_{4}- U^{2}_{3}\geq 0 \ \left(\text{equivalently, }G_{1}\geq 0, G_{2}\geq 0 \text{ and } 2 G_{2}G_{4}- G^{2}_{3}\geq 0\right)
\label{eq:a0_3}
\end{equation}
The condition $G_{1}\geq 0$ gives us
\begin{equation}
\rho_{j+1}W_{j+1}\left(\lambda\right)_{j+\frac{1}{2}} -(\rho W u)_{j+1}\geq 0 \Rightarrow \left(\lambda\right)_{j+\frac{1}{2}} \geq u_{j+1}
\label{eq:a0_4}
\end{equation}
Similarly, the condition $G_{2}\geq 0$ gives us
\begin{equation}
\rho_{j+1}\left(\lambda\right)_{j+\frac{1}{2}} -(\rho u)_{j+1}\geq 0 \Rightarrow \left(\lambda\right)_{j+\frac{1}{2}} \geq u_{j+1}
\label{eq:a0_4a}
\end{equation}
which is the same condition as \eqref{eq:a0_4}. The condition $2 G_{2}G_{4}- G^{2}_{3}\geq 0$ leads to
\begin{eqnarray}
&&2\left[\rho_{j+1}\left(\lambda\right)_{j+\frac{1}{2}} -(\rho u)_{j+1}\right]\left[(\rho E)_{j+1}\left(\lambda\right)_{j+\frac{1}{2}} -\left\{\left(\rho E+ p\right)u\right\}_{j+1}\right] - \left[(\rho u)_{j+1}\left(\lambda\right)_{j+\frac{1}{2}} - (\rho u^{2}+ p)_{j+1}\right]^{2} \geq 0 \nonumber\\
&&\Rightarrow p_{j+1} \left[ \frac{2}{\gamma_{j+1} -1}\rho_{j+1}\left\{\left(\lambda\right)_{j+\frac{1}{2}}- u_{j+1}\right\}^{2} -p_{j+1}\right]\geq 0\text{ (on simplifying)}\nonumber\\
&&\Rightarrow \left(\lambda\right)_{j+\frac{1}{2}} \geq u_{j+1}+ \sqrt{\frac{\gamma_{j+1} -1}{2 \gamma_{j+1}}} a_{j+1}
\label{eq:a0_5}
\end{eqnarray}
Here $a = \sqrt{\frac{\gamma p}{\rho}}$ is the speed of sound. We note that the condition \eqref{eq:a0_5} automatically satisfies \eqref{eq:a0_4}. Thus, the positivity condition \eqref{eq:a0_1} leads to a limitation on $\lambda$ as specified in \eqref{eq:a0_5}.


\begin{thebibliography}{10}
\expandafter\ifx\csname url\endcsname\relax
  \def\url#1{\texttt{#1}}\fi
\expandafter\ifx\csname urlprefix\endcsname\relax\def\urlprefix{URL }\fi
\expandafter\ifx\csname href\endcsname\relax
  \def\href#1#2{#2} \def\path#1{#1}\fi

\bibitem{chu1965kinetic}
C.~Chu, Kinetic-theoretic description of the formation of a shock wave, Physics
  of Fluids 8~(1) (1965) 12--22.

\bibitem{sanders1974possible}
R.~Sanders, K.~H. Prendergast, The possible relation of the 3-kiloparsec arm to
  explosions in the galactic nucleus, The Astrophysical Journal 188 (1974)
  489--500.

\bibitem{pullin1980direct}
D.~Pullin, Direct simulation methods for compressible inviscid ideal-gas flow,
  Journal of Computational Physics 34~(2) (1980) 231--244.

\bibitem{reitz1981one}
R.~D. Reitz, One-dimensional compressible gas dynamics calculations using the
  {Boltzmann} equation, Journal of Computational Physics 42~(1) (1981)
  108--123.

\bibitem{deshpande1986second}
S.~M. Deshpande, A second-order accurate kinetic-theory-based method for
  inviscid compressible flows, Tech. Rep. NASA TP 2613 (1986).

\bibitem{deshpande1986kinetic}
S.~M. Deshpande, Kinetic theory based new upwind methods for inviscid
  compressible flows, in: AIAA $24^{th}$ Aerospace Sciences Meeting, no.
  AIAA-86-0275, 1986.

\bibitem{mandal1994kinetic}
J.~C. Mandal, S.~M. Deshpande, Kinetic {Flux} {Vector} {Splitting} for {Euler}
  equations, Computers \& Fluids 23~(2) (1994) 447--478.

\bibitem{kaniel1988kinetic}
S.~Kaniel, A kinetic model for the compressible flow equations, Indiana
  University Mathematics Journal 37~(3) (1988) 537--563.

\bibitem{perthame1990boltzmann}
B.~Perthame, Boltzmann type schemes for gas dynamics and the entropy property,
  SIAM Journal on Numerical Analysis 27~(6) (1990) 1405--1421.

\bibitem{prendergast1993numerical}
K.~H. Prendergast, K.~Xu, Numerical hydrodynamics from gas-kinetic theory,
  Journal of Computational Physics 109~(1) (1993) 53--66.

\bibitem{raghurama1995peculiar}
S.~V. Raghurama~Rao, S.~M. Deshpande, Peculiar velocity based upwind method for
  inviscid compressible flows, Computational Fluid Dynamics Journal 3 (1995)
  415--432.

\bibitem{ROY2023106016}
S.~S. Roy, S.~V. Raghurama~Rao, A kinetic scheme with variable velocities and
  relative entropy, Computers \& Fluids 265 (2023) 106016.

\bibitem{roy2024kineticschemebasedpositivity}
S.~S. Roy, S.~V. Raghurama~Rao, A kinetic scheme based on positivity
  preservation with exact shock capture (2025).
\newblock \href {http://arxiv.org/abs/2403.14794} {\path{arXiv:2403.14794}}.

\bibitem{10.1007/978-3-663-13975-1_1}
R.~Abgrall, Extension of roe's approximate riemann solver to equilibrium and
  nonequilibrium flows, in: Proceedings of the Eighth GAMM-Conference on
  Numerical Methods in Fluid Mechanics, Vieweg+Teubner Verlag, Wiesbaden, 1990,
  pp. 1--10.

\bibitem{Fernandez1989}
G.~Fernandez, B.~Larrouturou, Hyperbolic Schemes for Multi-Component Euler
  Equations, Vieweg+Teubner Verlag, Wiesbaden, 1989, pp. 128--138.

\bibitem{ABGRALL2001594}
R.~Abgrall, S.~Karni, Computations of compressible multifluids, Journal of
  Computational Physics 169~(2) (2001) 594--623.

\bibitem{LARROUTUROU199159}
B.~Larrouturou, How to preserve the mass fractions positivity when computing
  compressible multi-component flows, Journal of Computational Physics 95~(1)
  (1991) 59--84.

\bibitem{KARNI199431}
S.~Karni, Multicomponent flow calculations by a consistent primitive algorithm,
  Journal of Computational Physics 112~(1) (1994) 31--43.

\bibitem{LeFloch_1994}
T.~Y. Hou, P.~G.~L. Floch, Why nonconservative schemes converge to wrong
  solutions: Error analysis, Mathematics of Computation 62~(206) (1994)
  497--530.

\bibitem{GOUASMI2020112912}
A.~Gouasmi, K.~Duraisamy, S.~M. Murman, Formulation of entropy-stable schemes
  for the multicomponent compressible euler equations, Computer Methods in
  Applied Mechanics and Engineering 363 (2020) 112912.

\bibitem{bhatnagar1954model}
P.~L. Bhatnagar, E.~P. Gross, M.~Krook, A model for collision processes in
  gases. i. small amplitude processes in charged and neutral one-component
  systems, Physical Review 94~(3) (1954) 511--525.

\bibitem{10.1007/BFb0083869}
B.~Larrouturou, L.~Fezoui, On the equations of multi-component perfect of real
  gas inviscid flow, in: Nonlinear Hyperbolic Problems, Springer Berlin
  Heidelberg, Berlin, Heidelberg, 1989, pp. 69--98.

\bibitem{shrinath2023kinetic}
K.~Shrinath, N.~Maruthi, S.~V. Raghurama~Rao, V.~Vasudev~Rao, A kinetic flux
  difference splitting method for compressible flows, Computers \& Fluids 250
  (2023) 105702.

\bibitem{natalini1998discrete}
R.~Natalini, A discrete kinetic approximation of entropy solutions to
  multidimensional scalar conservation laws, Journal of differential equations
  148~(2) (1998) 292--317.

\bibitem{aregba2000discrete}
D.~Aregba-Driollet, R.~Natalini, Discrete kinetic schemes for multidimensional
  systems of conservation laws, SIAM Journal on Numerical Analysis 37~(6)
  (2000) 1973--2004.

\bibitem{bouchut1999construction}
F.~Bouchut, Construction of {BGK} models with a family of kinetic entropies for
  a given system of conservation laws, Journal of Statistical Physics 95~(1)
  (1999) 113--170.

\bibitem{10.1007/978-1-4613-8689-6_9}
S.~Osher, S.~Chakravarthy, Very high order accurate tvd schemes, in:
  Oscillation Theory, Computation, and Methods of Compensated Compactness,
  Springer New York, New York, NY, 1986, pp. 229--274.

\bibitem{gottlieb2001strong}
S.~Gottlieb, C.-W. Shu, E.~Tadmor, Strong stability-preserving high-order time
  discretization methods, SIAM review 43~(1) (2001) 89--112.

\bibitem{kumar}
R.~Kumar, A.~K. Dass, A new flux-limiting approach–based kinetic scheme for
  the euler equations of gas dynamics, International Journal for Numerical
  Methods in Fluids 90~(1) (2019) 22--56.

\bibitem{LIU2003651}
T.~Liu, B.~Khoo, K.~Yeo, Ghost fluid method for strong shock impacting on
  material interface, Journal of Computational Physics 190~(2) (2003) 651--681.

\bibitem{Pan_Cheng_Wang_Xu_2017}
L.~Pan, J.~Cheng, S.~Wang, K.~Xu, A two-stage fourth-order gas-kinetic scheme
  for compressible multicomponent flows, Communications in Computational
  Physics 22~(4) (2017) 1123–1149.

\bibitem{Haas_Sturtevant_1987}
J.-F. Haas, B.~Sturtevant, Interaction of weak shock waves with cylindrical and
  spherical gas inhomogeneities, Journal of Fluid Mechanics 181 (1987) 41–76.

\bibitem{MARQUINA2003120}
A.~Marquina, P.~Mulet, A flux-split algorithm applied to conservative models
  for multicomponent compressible flows, Journal of Computational Physics
  185~(1) (2003) 120--138.

\bibitem{Quirk_Karni_1996}
J.~J. Quirk, S.~Karni, On the dynamics of a shock–bubble interaction, Journal
  of Fluid Mechanics 318 (1996) 129–163.

\end{thebibliography}

\end{document}